\documentclass{article}
\usepackage[utf8]{inputenc}
\usepackage[T1]{fontenc}
\usepackage{lmodern}
\usepackage{microtype}
\usepackage{booktabs}
\usepackage{threeparttable}
\usepackage{geometry}
\usepackage{setspace}
\usepackage{amsmath,amssymb,amsthm}  
\usepackage{bm}                    
\usepackage{graphicx}             
\usepackage{caption}
\usepackage{subcaption}             
\usepackage{booktabs}                
\usepackage{hyperref}      
\usepackage{algorithm} 
\usepackage{algpseudocode}
\hypersetup{
    colorlinks=true,
    linkcolor=blue,
    citecolor=blue,
    urlcolor=blue
}
\usepackage{natbib}
\usepackage{xcolor,mathtools}      
\usepackage{ulem}

\mathtoolsset{showonlyrefs=true}

\newcommand{\eqlnostar}[2]{\begin{align}\label{#1}#2\end{align}}
\newcommand{\eqstar}[1]{\begin{align*}#1\end{align*}}
\newcommand{\eq}[1]{\ifthenelse{\equal{#1}{*}}
  {\eqstar}
  {\eqlnostar{#1}}
 }

\usepackage{todonotes}

\usepackage{appendix}

\title{Forecasting implied volatility surface with generative diffusion models}
\author{
  Chen JIN\thanks{Department of Statistical and Actuarial Sciences, University of Western Ontario, London, ON, Canada. Email: cjin94@uwo.ca; aagarw93@uwo.ca} 
  \and 
  Ankush AGARWAL\footnotemark[2]
}

\begin{document}

\maketitle

\begin{abstract}
\noindent We introduce a conditional Denoising Diffusion Probabilistic Model (DDPM) for generating one-day-ahead approximately arbitrage-free implied volatility surfaces. To capture the path-dependent nature of volatility dynamics, we condition our model on a set of market variables, including exponentially weighted moving averages (EWMAs) of historical vol-surfaces, returns and squared returns of the underlying asset, and scalar risk indicators associated with the underlying asset. A key challenge is that the historical data used for training often contain arbitrage opportunities, particularly in the earlier part of the sample, which conflicts with the goal of generating arbitrage-free surfaces. We address this by using a weighting scheme based on the signal-to-noise ratio (SNR) to incorporate the arbitrage penalty into the loss function. The scheme dynamically adjusts the penalty strength across the diffusion process.  Through numerical experiments using market data on S\&P 500 options, we demonstrate the superior performance of our proposed model in volatility forecasting compared to existing modern and classical approaches.  

\smallskip


\smallskip


\smallskip

\end{abstract}

\section{Introduction}

The implied volatility surface (vol-surface) is a vital tool in finance as it provides information on market expectations of future asset price fluctuations implied from option prices across different maturities and strikes. Financial traders and risk managers rely on this information daily to price complex (exotic) derivatives, manage portfolio risk, and calibrate their valuation models. In this work, we develop a conditional generative diffusion model for forecasting the one-day-ahead vol-surface of the S\&P 500 (SPX) index options. In particular, we design a novel loss function which is able to handle arbitrage violations in the historical data of option prices while producing volatility surface forecasts. As a benchmark, we use the recently proposed generative machine learning technique of generative adversarial networks (GANs) to illustrate the accuracy of our approach, and also perform exhaustive comparison studies against classical time-series, stochastic and machine learning models.

A classical approach to modelling the vol-surface using market data is through the local volatility model of \citet{Dupire_1994} where we fit option implied volatility to the volatility function. Even though the approach calibrates a static surface with great accuracy, its predicted future dynamics are often unrealistic. Specifically, in this local volatility approach, the volatility smile flattens as the asset price increases. Stochastic volatility models such as the stochastic alpha beta rho (SABR) model of \citet{Hagan_2002} treat volatility as a separate random process correlated with the asset and successfully capture some of the market observed phenomena, but often struggle to capture the complex, idiosyncratic shapes observed in vol-surfaces. Forecasting of vol-surfaces for risk management purposes, however, is a totally different challenge. Here, the modelling task is about capturing the dynamics of the surface instead of calibrating to it statically.

A large literature models the dynamics of the implied volatility surface directly. \citet{dumas1998implied} showed that deterministic implied-volatility functions fitted cross-sectionally are unstable over time. \citet{skiadopoulos2000dynamics} and \citet{Cont_Da_Fonseca_2002} found that a few principal components explain most of the daily variation of the S\&P 500 surface and modelled their dynamics as low-order autoregressive processes, and \citet{fengler2007semiparametric} proposed a semiparametric factor model. \citet{goncalves2006predictable} and \citet{bernales2014can} forecast the coefficients of a parametric surface with vector autoregressions and documented statistically significant predictability. These approaches produce point forecasts and do not impose absence of arbitrage on the forecast surface.

More recently, the availability of large datasets has motivated data-driven machine learning approaches that model surface dynamics and then subsequently produce simulated surfaces which can be used for scenario analysis and stress testing. For example, \citet{Bergeron_2021} employed Variational Autoencoders (VAEs) trained on historical S\&P 500 options data. Their primary application was to learn the underlying probability distribution of implied volatilities, facilitating the generation of synthetic surfaces that reflect empirical market structures. In the recent works of \cite{choudhary2024funvol} and \cite{ning2023arbitrage}, neural stochastic differential equations (SDEs) and VAEs, respectively, were used to create arbitrage-free vol-surface simulators without necessarily focusing on forecasting the vol-surface using the time series data. Another recent example of an arbitrage-free vol-surface simulator is the work by \cite{buehler2026dysanos}. In the pioneering work of \citet{Vuletic_Cont_2024}, the authors introduced a Generative Adversarial Network (GAN) framework named VolGAN, which is trained on SPX implied volatility time series. This model learns the joint dynamics of the volatility surface and the underlying asset allowing for day-ahead vol-surface forecasting. Another work in this same strand of literature is the paper of \cite{LSTMIV}, who used a long short-term memory (LSTM) network with an attention mechanism. 

Static fitting, or smoothing, constructs an arbitrage-free surface from a single day's quotes. Related neural approaches include \citet{ackerer2020deep}, \citet{zheng2021incorporating} and \citet{wiedemann2025operator}. \citet{zheng2021incorporating}, in particular, build no-arbitrage and shape restrictions into the network design and training loss, an aim closely related to our arbitrage penalty, but their model maps moneyness and maturity to implied volatility on a given day rather than forecasting the next day's surface. Forecasting, the task considered here and in \citet{Vuletic_Cont_2024} and \cite{LSTMIV}, predicts the distribution of the next day's surface given the information available today.

Despite recent advances in machine-learning-based techniques, most existing generative methods are designed as simulators of realistic vol-surface samples rather than forecasting by conditioning on the current market state. The confidence intervals produced by VolGAN \citep{Vuletic_Cont_2024} frequently miss the realized surface, while the LSTM model of \citet{LSTMIV} produces only point forecasts of the vol-surface. Thus, in this work, we are specifically interested in proposing an alternative generative modelling approach that produces more accurate vol-surface forecasts using the options implied volatility time series data. Denoising Diffusion Probabilistic Models (DDPMs) have emerged as a distinct class of generative models, offering an alternative to GANs and VAEs. The comparative performance of these architectures has been a subject of active research; for instance, \citet{Dhariwal_2021} demonstrated that diffusion models can achieve image generation quality superior to GANs, a finding further supported in diverse domains such as audio synthesis \citep{Kong_2021} and time-series forecasting \citep{Rasul_2021}. The diffusion framework generates data by learning to reverse a gradual noise corruption process. This method was first introduced by \citet{Sohl_2015} and was later linked to matching data gradients \citep{Song_2019}, making the training process stable and reliable. The ability of diffusion models to capture complex high-dimensional distributions has led to their adoption in financial applications, where accurate density estimation is critical for risk modelling.

In this work, we make the following contributions:

\begin{enumerate}
\item We develop a conditional diffusion framework for one-day-ahead vol-surface forecasting based on a Variance-Preserving SDE. Motivated by the path-dependent volatility framework of \cite{Gazzani_Guyon_2025}, our model is conditioned on the current surface, short- and long-horizon exponentially weighted moving averages (EWMAs) of past surfaces, EWMAs of underlying returns and squared returns, and the VIX return. We pass the surface-valued conditions to the network by channel-stacking them with the noisy target, and the scalar conditions through feature-wise linear modulation (FiLM), which allows a compact U-Net to exploit both sources of information without spending spatial capacity on spatially constant inputs. Unlike the generative simulators discussed above, the model produces a predictive distribution of the next day's surface given today's market state.
\item 
We introduce an arbitrage regularization scheme for diffusion training that allows the model to learn arbitrage-free surfaces from historical data that often contain arbitrage opportunities. We do this by designing a composite loss function that augments the standard score-matching objective with an explicit financial arbitrage penalty. The penalty is evaluated on the surface implied by the network's denoising estimate at each sampled diffusion step and weighted by a capped signal-to-noise ratio following the Min-SNR strategy of \citet{Hang2023minSNR}. Our approach ensures that the arbitrage penalty is suppressed when the estimate is noise-dominated and enforces it strictly in the low-noise refinement stages, preventing instability in the training procedure when the no-arbitrage constraint is applied uniformly across the diffusion process. 

\item We provide evidence on the calibration of the forecast distribution across ten random seeds, verify the proxy penalty against the exact Durrleman condition, and study the role of each conditioning variable in an ablation study.
\end{enumerate}

 We perform a comprehensive numerical evaluation of our model using SPX options data, benchmarking it against VolGAN, VolGAN with our conditioning set, Vector Autoregressive (VAR) model of order one, SABR model, a principal-component factor model with autoregressive dynamics (PCA-AR(1)), and an attention-augmented long short-term memory (LSTM) model. Our model achieves the lowest overall MAPE across all benchmarks (3.00\% versus 3.32\% for VAR(1), 3.84\% for VolGAN and 3.90\% for PCA-AR(1)), and the Diebold--Mariano statistics are significant at the 0.1\% level against every competitor. The 90\% confidence intervals of the forecasts are well calibrated in the liquid regions of the surface, with breach rates of 11.1\% at 1-month ATM and 8.5\% at 1-month ITM, compared with 26.7\% and 55.0\% for VolGAN. Across ten random seeds, the overall MAPE of the diffusion model is $2.996\% \pm 0.055\%$, and every seed outperforms the best of ten VolGAN runs. Finally, our SNR-weighted penalty reduces the mean arbitrage violation of generated surfaces by roughly a factor of 3.5 relative to a diffusion model without the penalty, with no loss in forecasting accuracy. In addition, we perform an extensive ablation study which reveals contributions of different conditioning variables. The main finding is that the surface conditions, in particular the current surface, drive forecast accuracy, whereas the return, squared-return and short-horizon surface conditions mainly affect the calibration of the ATM confidence intervals.

The remainder of this paper is organized as follows. Section \ref{sec:background} provides a brief background on the concepts of implied volatility surface and sets the stage for the forecasting problem. In Section \ref{sec:conditional_diffusion}, we provide a high-level overview of how the diffusion model is able to learn the underlying distributional dynamics from data. We introduce the loss function used to train our generative diffusion model in Section \ref{subsec:score_matching}. Exact model implementation details, especially the market conditioning set, and neural network architecture are discussed in Section \ref{sec:framework}. The market-data results are presented in Section~\ref{sec:experiments}, together with several diagnostic studies and ablation studies. We conclude with Section \ref{sec:conclusion}. Appendices \ref{subsec:heston_results}--\ref{sec:ablation_appendix} contain additional experimental results and model hyperparameters.

\section{Background and Problem Formulation}
\label{sec:background}
As is well-known, implied volatility, $\sigma_{\text{imp}}$, is the value of the model parameter $\sigma$ (volatility) in the Black-Scholes model \citep{Black_Scholes_1973} which makes the model European call and put option prices equal to their market prices. Call and put options are financial derivatives that grant their holders the right to buy or sell the underlying asset at a predetermined strike price at maturity.  
The normalized Black-Scholes call and put prices, $c$ and $p$, are expressed as:
\eq{eq:optionprice}{
    c(m, \tau, \sigma) := \mathcal{N}(d_1) - m \mathcal{N}(d_2), \qquad p(m, \tau, \sigma) := m \mathcal{N}(-d_2) - \mathcal{N}(-d_1),
}
where $\mathcal{N}(\cdot)$ is defined as the cumulative distribution function of the normal distribution, and the terms $d_1$ and $d_2$ as:
\eqstar{
    d_1 := \frac{-\ln(m) + \frac{1}{2}\sigma^2\tau}{\sigma\sqrt{\tau}}, \qquad d_2 := d_1 - \sigma\sqrt{\tau},
}
with moneyness $m:=K/S$ as the standardized strike ($K$ is the strike price and $S$ is the underlying spot price) and time-to-maturity $\tau$ (in years). For simplicity, we assume that the risk-free interest rate is $r = 0$ and that the underlying pays no dividends, so that $m$ coincides with forward moneyness.

For a given underlying asset, multiple options are concurrently traded across a range of strike prices and maturities. By extracting $\sigma_{\text{imp}}$ across this entire grid of observed market prices, we can construct an implied volatility surface. Under the perfect Black-Scholes framework, this surface would be flat; however, actual surfaces exhibit different time-varying shapes. In this work, we consider the implied volatility surface (vol-surface) over a fixed grid with moneyness levels $\mathbf{m}  \in \mathbb{R}^{N_m}$ and time-to-maturities 
$\boldsymbol{\tau} \in \mathbb{R}^{N_\tau}$.
We denote the vol-surface as a matrix $\boldsymbol{\sigma} \in \mathbb{R}^{N_m \times N_\tau}$, where each element $\sigma_{i,j} := \sigma_{\text{imp}}(m_i, \tau_j)$, $i=1\dots N_m, j=1\dots N_\tau$, $m_i$ and $\tau_j$ as the $i$-th and $j$-th elements of $\mathbf{m}$ and $\boldsymbol{\tau}$, respectively. 

Not every arbitrary matrix $\boldsymbol{\sigma} \in \mathbb{R}^{N_m \times N_\tau}$ represents a valid vol-surface due to the arbitrage constraints. Calendar arbitrage occurs when an option with a longer time-to-maturity is priced cheaper than a shorter option at the same moneyness level. This is equivalent to saying that we require the total implied variance $w(m_i, \tau_j) := \sigma_{i,j}^2 \tau_j$ to be non-decreasing in $\tau$ for a valid vol-surface.  This condition, stated at fixed forward moneyness, can be found in \citet[Sec.~2.2.2.1, Eq.~(2.15), p.~32]{bergomi2016stochastic}. Thus, we quantify the calendar arbitrage penalty $\Phi_{\tau}(\boldsymbol{\sigma})$ as:
\eqlnostar{eq:arb pen calendar}{
    \Phi_{\tau}(\boldsymbol{\sigma}) := \sum_{i,j} \max\left(0, w(m_i, \tau_j) - w(m_i, \tau_{j+1}) \right)^2,
}
where $i \in \{1, \dots, N_m\}$ and $j \in \{1, \dots, N_\tau-1\}$.

Additionally, butterfly arbitrage occurs when, for a fixed maturity, the call price fails to be a convex function of the strike. Since $m = K/S$, convexity in $K$ is equivalent to convexity in $m.$ See \citet[Section 2.2.2]{bergomi2016stochastic}.
Alternatively, instead of imposing the butterfly arbitrage condition on option prices, it is
often numerically more convenient to work directly with implied volatilities. This leads to the so-called Durrleman condition on the total implied variance
(see, for example, \citet{roper2010arbitrage} or  \citet[Lemma~2.2]{gatheral2014arbitrage}). Written in terms of the
implied volatility $\sigma = \sigma(m,\tau_j)$, the condition reads
\eqlnostar{eq:durrleman-m-sigma}{
g(m) := \left( 1 - \frac{m \log m \, \partial_m \sigma}{\sigma} \right)^2
- \frac{m^2 \tau_j^2 \sigma^2 (\partial_m \sigma)^2}{4}
+ m\, \tau_j\, \sigma \left( \partial_m \sigma + m\, \partial_{mm} \sigma \right) \;\geq\; 0.
}
In our implementation, we regularize butterfly arbitrage through the convexity of the implied
volatility smile in moneyness,
\eqlnostar{eq:arb pen butterfly}{
    \Phi_{b}(\boldsymbol{\sigma}) = \sum_{i,j} \max\left(0, -\left[ \sigma_{i-1,j} - 2\sigma_{i,j} + \sigma_{i+1,j} \right] \right)^2,
}
where $i \in \{2, \dots, N_m-1\}$ and $j \in \{1, \dots, N_\tau\}$. This penalty targets the
curvature term $m^2 \tau_j \sigma\, \partial_{mm}\sigma$ in \eqref{eq:durrleman-m-sigma}. This is the only term in \eqref{eq:durrleman-m-sigma} that involves the second derivative $\partial_{mm}\sigma$. Penalizing it avoids evaluating the full expression at every diffusion step. However, this is only a proxy. In particular, it is overly restrictive at
short maturities where \eqref{eq:durrleman-m-sigma} permits substantial negative curvature of
the smile. We therefore also verify the exact condition \eqref{eq:durrleman-m-sigma} on the generated
surfaces in Section~\ref{subsec:arb_comparison}. 
The total structural arbitrage penalty for any given surface is then defined as the sum of these two terms:
\eq{eq: total_arb}{
    \Phi(\boldsymbol{\sigma}) := \Phi_{\tau}(\boldsymbol{\sigma}) + \Phi_{b}(\boldsymbol{\sigma}).
}
We focus on the modelling task of learning the distribution of the vol-surface, which is approximately supported on a manifold $\mathcal{M} \subset \mathbb{R}_{>0}^{N_m \times N_\tau}$ such that $\Phi(\boldsymbol{\sigma})$ is ``almost zero''. $\mathbb{R}_{>0}$ denotes the strictly positive half of the real line.

Specifically, we aim to use generative diffusion modelling to perform this task.
In the diffusion model, the basic approach is to first gradually corrupt data with the standard Gaussian noise $\mathcal{N}(\mathbf{0}, \mathbf{I})$ and then learn to reverse this corruption process in order to estimate the original data distribution. The typical values of implied volatility seen in the market are much smaller than 100\%, whereas standard Gaussian noise lies between $-3$ and $3$ with high probability. Due to this difference in magnitude between actual data and Gaussian noise, the corruption procedure in the diffusion model would not be gradual but instead the real data will start to look like Gaussian noise very quickly. For this reason, the vol-surfaces need to be standardized before being corrupted by the Gaussian noise. If we use the vol-surface data directly in the diffusion model, the financial signal would get overwhelmed by the injected noise due to this magnitude mismatch. The positivity of the generated vol-surfaces from the diffusion model is also not guaranteed. To resolve these issues, we define an element-wise transformation mapping $\mathcal{T}: \mathbb{R}_{>0}^{N_m \times N_\tau} \to \mathbb{R}^{N_m \times N_\tau},$ which ensures that the vol-surface data used in the model is standardized and the generated vol-surfaces are positive. For a vol-surface $\boldsymbol{\sigma} \in \mathbb{R}_{>0}^{N_m \times N_\tau}$, we denote the standardized vol-surface $\mathbf{x}_0 := \mathcal{T}(\boldsymbol{\sigma})$ as:
\eqlnostar{eq:standardization transform}{
    \mathbf{x}_{0, ij} := \mathcal{T}(\boldsymbol{\sigma}_{ij}) =  \frac{\log(\sigma_{ij}) - \mu_{ij}}{s_{ij}}, \quad i = 1, \ldots,N_m, j = 1,\ldots, N_{\tau},
} 
where $\mu_{ij}$ and $s_{ij}$ represent the sample mean and sample standard deviation of $\log(\sigma_{ij})$ computed from the data sample used for training the diffusion model (training set). The variability of implied volatility differs markedly across the surface: short-maturity implied volatilities fluctuate much more than long-maturity ones (in our training data, the standard deviation of daily changes in log implied volatility at $\tau = 1/252$ is 2.4 times that at $\tau = 1$), and the wings behave differently from the centre. A single scaling would therefore leave some grid points with variance far above one and others far below one. The point-wise standardization \eqref{eq:standardization transform} gives every grid point zero mean and unit variance over the training set, so that the injected Gaussian noise has a comparable effect at all grid points. Note that \eqref{eq:standardization transform} normalizes only the marginal means and variances. The strong correlations between grid points are left intact, so the standardized data do not have identity covariance. This is not required by the diffusion model, which learns the joint distribution, including these correlations, through the score. The original data distribution used in the subsequent sections, denoted by $p_{\text{data}},$ is based on $\mathbf{x}_0$ defined above. 
To recover the vol-surface from the trained diffusion model, the generated samples are mapped back using the inverse transformation $\mathcal{T}^{-1}$.

\section{Conditional Score-Based Generative Modelling}
\label{sec:conditional_diffusion}
With the training surfaces standardized grid point by grid point, we model the dynamics of the surface using a conditional diffusion process. We begin with the unconditional forward and backward process, then introduce how we guide the model using a conditioning set $\mathcal{Y}$ that encapsulates the market information and directly integrate it into the architecture.

\subsection{Unconditional Forward-Backward Process}

Diffusion models were originally popularized in the domain of computer vision \citep{Ho_Jain_Abbeel_2020, Nichol_Dhariwal_2021}, where they were designed to generate images by learning the static marginal distribution of a training dataset. In contrast, our application requires dynamically forecasting path-dependent financial time series. Since our conditional forecasting relies on the foundational mechanics of stochastic data corruption and score matching, we first outline the standard unconditional framework for illustration purposes. This serves as the theoretical prerequisite before we introduce the market conditions required to adapt the model for conditional time-series prediction. The forward diffusion process $\{\mathbf{x}_t: t \in [0,T]\}$ is constructed by gradually corrupting the observed market surface $\mathbf{x}_0 \sim p_{\text{data}}$ with noise. The idea here is to start with $\mathbf{x}_0$ and gradually add Gaussian noise continuously over $t \in [0,T]$ such that at the final horizon $T$, the surface is completely transformed into $\mathbf{x}_T \sim p_T$, where $p_T$ is a known prior, usually taken to be the Gaussian distribution. This forward diffusion process is typically modelled as an SDE:
\eqlnostar{eq: forward diffusion}{
    d\mathbf{x}_t = \mathbf{f}(\mathbf{x}_t, t)dt + g(t)d\mathbf{w}_t, \quad \mathbf{x}_0 \sim p_{\text{data}},
}
where $\mathbf{w}_t$ is the standard Brownian motion. The term $\mathbf{f}(\cdot, t): \mathbb{R}^{N_m \times N_{\tau}} \to \mathbb{R}^{N_m \times N_{\tau}}$ is a vector-valued function called the drift coefficient, and $g(\cdot): \mathbb{R} \to \mathbb{R}$ is a scalar function known as the diffusion coefficient. To control the behaviour of the diffusion, the drift is coupled to the added noise by defining $\mathbf{f}(\mathbf{x}_t, t) = -\frac{1}{2} g(t)^2 \mathbf{x}_t$. This specific formulation acts as a mean-reverting force that pulls the data toward the origin at a rate strictly proportional to the variance injected by $g(t)$. The overall variance of the process then remains bounded rather than exploding toward infinity. This configuration is formally known in the literature as a Variance-Preserving (VP) SDE. Under this balanced setting, the marginal probability density $p_t$ of the solution $\mathbf{x}_t$ in \eqref{eq: forward diffusion} converges exponentially quickly to the Gaussian prior at the terminal time $T$. Note that $p_t$ is unknown since we do not know $p_{\text{data}}$; however, the forward transition density $\mathbf{x}_t | \mathbf{x}_0 \sim p_{t|0}$ is Gaussian and known in closed form since $\mathbf{f}(\mathbf{x},\cdot)$ and $g$ are deterministic functions. 

To generate new vol-surfaces, we reverse the diffusion process \eqref{eq: forward diffusion} by starting from noise $\mathbf{x}_T \sim p_T$ at $t=T$. This process runs backward in time such that at $t=0$, it recovers the underlying distribution of the data $\mathbf{x}_0 \sim p_{\text{data}}$. The reverse process is also modelled as an SDE. Flowing backwards from $T$ to $0$, the generative dynamics are governed by the reverse-time SDE, as derived by \citet{Anderson_1982}:
\eqlnostar{eq: reverse diffusion}{
    d\mathbf{x}_t = \left[ \mathbf{f}(\mathbf{x}_t, t) - g(t)^2 \nabla \log p_t(\mathbf{x}_t) \right] dt + g(t)d\mathbf{\bar{w}}_t, \quad \mathbf{x}_T \sim p_T,
}
where $\mathbf{\bar{w}}_t$ is a standard Brownian motion running backwards. Substituting the VP drift coefficient $\mathbf{f}(\mathbf{x}_t, t)$ in \eqref{eq: reverse diffusion}, the equation simplifies to:
\eq{eq:unconditional_backward}{
    d\mathbf{x}_t = -g(t)^2 \left[ \frac{1}{2}\mathbf{x}_t + \nabla \log p_t(\mathbf{x}_t) \right] dt + g(t)d\mathbf{\bar{w}}_t.
}
The drift term above depends on the time-dependent score function $\nabla \log p_t(\mathbf{x}_t)$, which represents the gradient of the log marginal probability density. However, the exact score function is intractable because $p_{\text{data}}$ is unknown. We approximate this theoretical score using a neural network estimator. By substituting the estimated score into the reverse equation, we can numerically simulate the time-reversed SDE to sample new, unconditional volatility surfaces from the prior distribution.

However, simulating \eqref{eq:unconditional_backward} merely samples vol-surfaces from the historical data distribution. As such, this procedure cannot be used to generate a day-ahead forecast of the vol-surface given the history. For time-series forecasting, we adapt this generative framework by explicitly conditioning the reverse diffusion process on the current market state. By injecting historical market context into the score estimator, we constrain the generative trajectory to pinpoint the specific expected surface for the upcoming trading day.

\subsection{The Conditional Forward and Reverse SDEs}
\label{subsec:conditional_sdes}

The original score-based generative framework assumes that the target data is independent and identically distributed (i.i.d.) from a single distribution, $p_{\text{data}}(\cdot)$. While this assumption holds for domains such as image generation, it structurally fails in the context of quantitative finance. Financial time series are inherently non-stationary and the distribution of valid future vol-surfaces is highly dependent on the market state, volatility clustering, and price dynamics observed today. Generating a vol-surface unconditionally would simply sample from a static historical average. To overcome this limitation, we formulate the generating process as a sequence of state-dependent conditional distributions. Specifically, we assume that the target surface for the upcoming trading day $k+1$ is drawn from a distribution governed entirely by the current conditioning set, denoted as $p_{\text{data}}(\cdot \mid \mathcal{Y}^{(k)})$, where $\mathcal{Y}^{(k)}$ is the conditioning set of historical market states observed up to the current trading day $k$. This shift forces the model to learn a series of dynamic distributions of probable future surfaces rather than a single historical baseline.

To adapt to this conditional diffusion process \citep{cao2026diffusion, Tang_Lin_Yang_2024, baldassari2023conditional}, let the arbitrage-free target surface for the upcoming trading day $k+1$ be denoted as $\mathbf{x}_0^{(k+1)} \sim p_{\text{data}}(\cdot \mid \mathcal{Y}^{(k)})$. The forward diffusion mechanism remains mathematically identical to the unconditional case \eqref{eq: forward diffusion}, since the forward transition density $\mathbf{x}_t | \mathbf{x}_0$ is still Gaussian and independent of the market condition $\mathcal{Y}^{(k)}$. By applying the forward SDE \eqref{eq: forward diffusion} to $\mathbf{x}_0^{(k+1)}$, the process transforms the conditional data distribution $p_{\text{data}}(\cdot \mid \mathcal{Y}^{(k)})$ into the tractable Gaussian prior $p_T(\cdot) \approx \mathcal{N}(\mathbf{0}, \mathbf{I})$, where $\mathbf{I}$ is the identity matrix.

To achieve generative forecasting, we reverse the conditional forward diffusion process. The generative dynamics are governed by the conditional reverse-time SDE:
\eq{eq:reverse_sde}{
    d\mathbf{x}_t^{(k+1)} = \left[ \mathbf{f}(\mathbf{x}_t^{(k+1)}, t) - g(t)^2 \nabla\log p_t(\mathbf{x}_t^{(k+1)} \mid \mathcal{Y}^{(k)}) \right] dt + g(t) d\mathbf{\bar{w}}_t, \quad \mathbf{x}_T^{(k+1)} \sim p_T,
}
where $\mathbf{\bar{w}}_t$ represents the standard Wiener process moving backwards in time from $T$ to $0$. 
Here we substitute the unconditional score in \eqref{eq: reverse diffusion} with the conditional score function $\nabla \log p_t(\mathbf{x}_t^{(k+1)} \mid \mathcal{Y}^{(k)})$, where $p_t(\cdot \mid \mathcal{Y}^{(k)})$ is the marginal density of the solution $\mathbf{x}_t^{(k+1)}$ of \eqref{eq:reverse_sde}. We enforce the score function to be conditional with respect to $\mathcal{Y}^{(k)}$ since we are aiming to learn $p_{\text{data}}(\cdot \mid \mathcal{Y}^{(k)})$ from the reverse-time SDE. By conditioning on $\mathcal{Y}^{(k)}$, the reverse trajectory is constrained to sample from the distribution of expected vol-surfaces for the next trading day.

\section{Conditional Score Estimation via Regularized Loss Function}
\label{subsec:score_matching}
As can be seen from \eqref{eq:reverse_sde}, training the conditional diffusion model requires estimating the intractable score function $\nabla \log p_t(\mathbf{x}_t^{(k+1)} \mid \mathcal{Y}^{(k)})$. Directly optimizing a neural network to match this marginal score will require knowing the true probability density function, which is not possible. Instead, we employ the Denoising Score Matching (DSM) loss function $\mathcal{L}_{\text{DSM}}$\ in the following way:
\eq{eq:DSM}{
    \mathcal{L}_{\text{DSM}}(\theta) := \mathbb{E}_{t, \mathbf{x}_0^{(k+1)}, \mathbf{x}_t^{(k+1)}\mid \mathbf{x}_0^{(k+1)}} \left[ \left\| \mathbf{s}_\theta \left( \mathbf{x}_t^{(k+1)}, t, \mathcal{Y}^{(k)} \right) - \nabla \log p_t(\mathbf{x}_t^{(k+1)} \mid \mathbf{x}_0^{(k+1)}) \right\|^2 \right],
}
where $\mathbf{s}_\theta$ represents the conditional score estimator. A neural network, represented by its parameter vector $\theta$, is used to calculate $\mathbf{s}_\theta$, as shown later. Since the terminal noise is purely random, it is inherently independent of $\mathcal{Y}^{(k)}$. Therefore, during the reverse process, we enforce the neural network to take $\mathcal{Y}^{(k)}$ as an input. Note that we can write the following using the law of total probability and the conditional independence of $\mathbf{x}_t^{(k+1)}$ and $\mathcal{Y}^{(k)}$ given $\mathbf{x}_0^{(k+1)}$
\eqstar{
p_t(\mathbf{x}_t^{(k+1)} \mid \mathcal{Y}^{(k)}) &= \int p_t(\mathbf{x}_t^{(k+1)} \mid \mathbf{x}_0^{(k+1)}) p_{\text{data}}(\mathbf{x}_0^{(k+1)} \mid \mathcal{Y}^{(k)}) d\mathbf{x}_0^{(k+1)}.
}
Using the above interpretation of the unknown conditional density $p_t(\mathbf{x}_t^{(k+1)} \mid \mathcal{Y}^{(k)})$, \citet{Vincent2011} showed that the DSM loss \eqref{eq:DSM} is mathematically equivalent to matching the intractable marginal score $\nabla\log p_t(\mathbf{x}_t^{(k+1)} \mid \mathcal{Y}^{(k)})$. 

The expectation in \eqref{eq:DSM} is taken over the distribution of $t$ which is uniform over $[0, T]$, the target surface $\mathbf{x}_0^{(k+1)} \sim p_{\text{data}}(\cdot \mid \mathcal{Y}^{(k)})$, and the noisy state $\mathbf{x}_t^{(k+1)} \mid \mathbf{x}_0^{(k+1)}$. The transition density $p_t(\mathbf{x}_t^{(k+1)} \mid \mathbf{x}_0^{(k+1)})$ is Gaussian with mean $\mu_t \mathbf{x}_0^{(k+1)}$ and variance $v_t \mathbf{I}$, where 
\eqlnostar{eq: mean and variance}{\mu_t := \exp\Bigl(-\frac{1}{2} \int_0^t g(s)^2 ds\Bigr), \qquad v_t:= 1 - \exp\Bigl(-\int_0^t g(s)^2 ds\Bigr).}
The score of this transition density is known and is given as
\eqlnostar{eq: score transition density}{
    \nabla \log p_t(\mathbf{x}_t^{(k+1)} \mid \mathbf{x}_0^{(k+1)}) = -\frac{\boldsymbol{\epsilon}}{\sqrt{v_t}},
}
where $\boldsymbol{\epsilon} \sim \mathcal{N}(\mathbf{0}, \mathbf{I})$ is the standard Gaussian noise added at time $t$. From the score in \eqref{eq: score transition density}, we can see that the loss function \eqref{eq:DSM} targets the added noise.  
Following the methodology of \citet{Ho_Jain_Abbeel_2020}, we train the neural network with parameter $\theta$, with inputs $\left( \mathbf{x}_t^{(k+1)}, t, \mathcal{Y}^{(k)} \right)$, to predict the injected noise $\epsilon$ rather than the conditional score directly. The predicted noise from the neural network, denoted by $\boldsymbol{\epsilon}_\theta\left( \mathbf{x}_t^{(k+1)}, t, \mathcal{Y}^{(k)} \right)$, is then used to compute the conditional score estimator following \eqref{eq: score transition density} as
\eq{eq:estimator_reparameterization}{
    \mathbf{s}_\theta \left( \mathbf{x}_t^{(k+1)}, t, \mathcal{Y}^{(k)} \right) = -\frac{1}{\sqrt{v_t}} \boldsymbol{\epsilon}_\theta \left( \mathbf{x}_t^{(k+1)}, t, \mathcal{Y}^{(k)} \right).
}
Using representations \eqref{eq: score transition density} and \eqref{eq:estimator_reparameterization} in the loss function \eqref{eq:DSM} yields a new weighted noise-matching loss function given as
\eqlnostar{eq: gaussian loss}{
    \mathcal{L}(\theta) := \mathbb{E}_{t, \mathbf{x}_0^{(k+1)}, \boldsymbol{\epsilon}} \left[ \frac{1}{v_t} \left\| \boldsymbol{\epsilon} - \boldsymbol{\epsilon}_\theta \left( \mathbf{x}_t^{(k+1)}, t, \mathcal{Y}^{(k)} \right) \right\|^2 \right].
}
The above loss function is preferable to the $\mathcal{L}_{\text{DSM}}$ loss \eqref{eq:DSM} since the term $\nabla \log p_t(\mathbf{x}_t^{(k+1)} \mid \mathbf{x}_0^{(k+1)})$ diverges as $t \to 0$, making it less effective when  estimating $\mathbf{s}_\theta$. By shifting the target to $\boldsymbol{\epsilon}$, the network only needs to predict standard Gaussian noise, ensuring its outputs maintain a consistent and stable scale across all $t$. In practice, omitting the time-dependent weighting factor $\frac{1}{v_t}$ in \eqref{eq: gaussian loss} leads to greater training stability and improved empirical performance, as it implicitly downweights the loss at very small noise scales where the score matching is numerically unstable. We therefore use the simplified unweighted mean squared error:
\eq{eq:noiseloss}{
    \mathcal{L}_{\text{simple}}(\theta) := \mathbb{E}_{t, \mathbf{x}_0^{(k+1)}, \boldsymbol{\epsilon}} \left[ \left\| \boldsymbol{\epsilon} - \boldsymbol{\epsilon}_\theta \left( \mathbf{x}_t^{(k+1)}, t, \mathcal{Y}^{(k)} \right) \right\|^2 \right].
}
By minimizing \eqref{eq:noiseloss} across $t$, the neural network with parameter $\theta$ learns the conditional score estimator in \eqref{eq:DSM} which is used to simulate the reverse SDE and forecast the day-ahead vol-surface as discussed in Section \ref{subsec:conditional_sdes}. 

To ensure that the vol-surface forecasts satisfy no-arbitrage conditions, we augment the loss function $\mathcal{L}_{\text{simple}}$ with a differentiable penalty. Notice that in \eqref{eq:noiseloss}, the neural network is tasked to predict the noise that was added to $\mathbf{x}_t^{(k+1)}$; however, the arbitrage penalties \eqref{eq:arb pen calendar} and \eqref{eq:arb pen butterfly} must be evaluated on vol-surfaces. Since we know the transition density $p_t(\mathbf{x}_t^{(k+1)} \mid \mathbf{x}_0^{(k+1)})$ is Gaussian, we can generate an estimate of $\mathbf{x}_0^{(k+1)}$ using the predicted noise and the noisy state as 
\eqlnostar{eq:predicted surface}{
    \hat{\mathbf{x}}_0^{(k+1)} = \frac{\mathbf{x}_t^{(k+1)} - \sqrt{v_t} \boldsymbol{\epsilon}_\theta(\mathbf{x}_t^{(k+1)}, t, \mathcal{Y}^{(k)})}{\mu_t}.
}
The arbitrage constraints can only apply to the vol-surface $\boldsymbol{\sigma}$, so we need to recover the corresponding vol-surface estimate of trading day $k+1$ from $\hat{\mathbf{x}}_0^{(k+1)}$, which is computed 
as $\hat{\boldsymbol{\sigma}}^{(k+1)} = \mathcal{T}^{-1}(\hat{\mathbf{x}}_0^{(k+1)})$, with $\mathcal{T}$ defined in \eqref{eq:standardization transform}. With this vol-surface estimate, we enforce the arbitrage constraints as $\Phi(\hat{\boldsymbol{\sigma}}^{(k+1)})$ by applying $\hat{\boldsymbol{\sigma}}^{(k+1)}$ to \eqref{eq: total_arb}. However, because the initial projection $\hat{\mathbf{x}}_0^{(k+1)}$ (and consequently $\hat{\boldsymbol{\sigma}}^{(k+1)}$) becomes highly inaccurate at early stages of the reverse process (when $t$ is large and the surface is dominated by noise), directly penalizing the surface across all timesteps introduces severe training instability. To resolve this, we scale the arbitrage penalty using the Min-SNR (signal-to-noise ratio) weighting strategy \citep{Hang2023minSNR}, which suppresses the regularization by capping the SNR at a predefined maximum threshold $\gamma$. In the VP framework of the diffusion model, the continuous SNR is calculated as $\text{SNR}(t) = \frac{\mu_t^2}{v_t}$. The final loss function used to optimize the diffusion model is then defined as
\eqlnostar{eq:final loss function}{
    \mathcal{L}_{\text{total}}(\theta) = \mathcal{L}_{\text{simple}}(\theta) + \lambda \cdot \mathbb{E}_{t, \mathbf{x}_0^{(k+1)}, \boldsymbol{\epsilon}} \left[ \min(\text{SNR}(t), \gamma) \cdot \Phi(\hat{\boldsymbol{\sigma}}^{(k+1)}) \right],
}
where $\lambda$ is a hyperparameter controlling the strength of the arbitrage regularization. The hyperparameter $\lambda$ is tuned with Optuna \citep{optuna_2019} by minimizing the validation loss in two round of searches with 20 trials each. In each trial, $\lambda$ is sampled from a log-uniform distribution, so that every order of magnitude of the search range is explored equally. The first round explores a wide range, $\lambda \in [0.1, 50]$, and the second round narrows down the search to $\lambda \in [1, 15]$.
In Section \ref{subsec:arb_comparison}, we evaluate the performance of the arbitrage regularization by comparing a model trained with loss function $\mathcal{L}_{\text{total}}$ against a baseline model trained with loss function $\mathcal{L}_{\text{simple}}$.  The weight $\min(\text{SNR}(t), \gamma)$ is the Min-SNR-$\gamma$ weight of \citet{Hang2023minSNR} expressed in $\mathbf{x}_0$-space. Since $\hat{\mathbf{x}}_0^{(k+1)} - \mathbf{x}_0^{(k+1)} = \text{SNR}(t)^{-1/2}\big(\boldsymbol{\epsilon} - \boldsymbol{\epsilon}_\theta(\mathbf{x}_t^{(k+1)}, t, \mathcal{Y}^{(k)})\big)$, the denoised estimate on which $\Phi$ is evaluated becomes increasingly noisy as $\text{SNR}(t)$ decreases. The cap $\gamma$ prevents the nearly noise-free steps, where $\text{SNR}(t) \to \infty$, from dominating the gradient. Unlike \citet{Hang2023minSNR}, who apply this weight to the denoising loss itself (equivalently, the weight $\min(\text{SNR}(t), \gamma)/\text{SNR}(t)$ on the $\boldsymbol{\epsilon}$-prediction loss), we keep the unweighted objective \eqref{eq:noiseloss} of \citet{Ho_Jain_Abbeel_2020} for the denoising term, because weighting it as well generated surfaces which do not look like vol-surfaces but are rather jagged, in our setting. We choose the SNR cap $\gamma$ by observing how adding noise changes the arbitrage of a vol-surface. We take an arbitrarily chosen vol-surface from the dataset and for each diffusion timestep $t$, add the amount of noise prescribed by the noise schedule and compute its arbitrage penalty. When $t$ is small, $\text{SNR}(t)$ is high and the arbitrage penalty remains low. As $t$ increases, the arbitrage penalty eventually increases sharply. We set $\gamma$ to the value of $\text{SNR}(t)$ at the timestep where a sharp increase in arbitrage penalty is observed. As a result, all timesteps before this point, where the arbitrage penalty is still low, receive the same weight $\gamma$. As in both of these works, a single timestep $t$ is sampled uniformly for each training example. 

\section{Model Architecture and Implementation Details}
\label{sec:framework}
So far the market conditioning set $\mathcal{Y}^{(k)}$ mentioned in Section \ref{subsec:conditional_sdes} has not been defined. Therefore, this section first details the components of $\mathcal{Y}^{(k)}$, drawing on established results in the volatility modelling literature to capture path-dependent market dynamics. After that, we
outline the conditional neural network architecture designed to integrate these market conditions.
Finally, we consolidate the market conditioning set and the network architecture by presenting complete algorithms used for model training and one-day-ahead surface generation.

\subsection{Market Conditioning Set}
\label{subsec:conditions}

The conditioning set $\mathcal{Y}^{(k)}$ includes two sources of information: a high-dimensional component capturing the history of the vol-surface, and a low-dimensional scalar component capturing the market conditions. By integrating these two sources, the model ensures the forecast for day $k+1$ represents information available up to the current trading day $k$. 
The high-dimensional component consists of the 5-day (short-term trend) and 20-day (long-term trend) exponentially weighted moving averages (EWMAs) of past surfaces, denoted as $\mathbf{M}_{\text{short}}^{(k)}$ and $\mathbf{M}_{\text{long}}^{(k)}$, respectively. These moving averages are updated recursively as:
\eq{*}{
\mathbf{M}_{\text{short}}^{(k)} &= \alpha_{\text{short}} \mathbf{x}_0^{(k)} + (1 - \alpha_{\text{short}}) \mathbf{M}_{\text{short}}^{(k-1)}, \\
\mathbf{M}_{\text{long}}^{(k)} &= \alpha_{\text{long}} \mathbf{x}_0^{(k)} + (1 - \alpha_{\text{long}}) \mathbf{M}_{\text{long}}^{(k-1)},
}
where the decay factors $\alpha_{\text{short}}$ and $\alpha_{\text{long}}$ correspond to the short and long window spans. These sequences are initialized with the first observed surface such that $\mathbf{M}_{\text{short}}^{(0)} = \mathbf{M}_{\text{long}}^{(0)} = \mathbf{x}_0^{(0)}$. The high-dimensional component also includes the vol-surface observed on the current trading day $\mathbf{x}_0^{(k)}$. 

The market conditions are represented by a 
vector $\mathbf{c}^{(k)} \in \mathbb{R}^5$. The selection of these variables is motivated by the path-dependent volatility framework proposed by \citet{Gazzani_Guyon_2025}, who demonstrated that volatility exhibits a strong linear dependence on historical EWMAs of both asset returns and squared returns. We denote the underlying asset return on trading day $k$ as $r_k=\frac{S_k-S_{k-1}}{S_{k-1}}$, where $S_k$ is the daily closing price of the underlying on trading day $k$ (the S\&P 500 index in our case). The short-term and long-term EWMAs of the asset returns are denoted as $m_r^{S,(k)}$ and $m_r^{L,(k)}$, respectively, and the EWMAs of the squared returns as $m_\sigma^{S,(k)}$ and $m_\sigma^{L,(k)},$ respectively. These conditions are calculated recursively for $k \geq 1$:
\eq{*}{
    m_r^{i,(k)} &= \alpha_r^i r_k + (1 - \alpha_r^i) m_r^{i,(k-1)}, \quad i \in \{S, L\}, \\
    m_\sigma^{j,(k)} &= \alpha_\sigma^j r_k^2 + (1 - \alpha_\sigma^j) m_\sigma^{j,(k-1)}, \quad j \in \{S, L\},
}
where the decay factors are set to capture both the short (S) and long (L) terms, and volatility clustering. These sequences are initialized on the first day of the dataset using the realized observations, such that $m_r^{i,(0)} = r_0$ and $m_\sigma^{j,(0)} = r_0^2$. To provide more information in the conditioning set, we also include the daily return of the VIX index, $r_{\text{vix}}^{(k)}$, for $k \geq 1$. The complete scalar conditioning vector evaluated at day $k$ is thus defined as:
\eq{*}{
    \mathbf{c}^{(k)} = \left[ m_r^{S,(k)}, m_r^{L,(k)}, m_\sigma^{S,(k)}, m_\sigma^{L,(k)}, r_{\text{vix}}^{(k)} \right]^\top.
}
Therefore, the conditioning set $\mathcal{Y}^{(k)}$ ($k \geq 1$) is defined as:
\eq{*}{
    \mathcal{Y}^{(k)} \coloneqq \left\{ \mathbf{x}_0^{(k)}, \mathbf{M}_{\text{short}}^{(k)}, \mathbf{M}_{\text{long}}^{(k)}, \mathbf{c}^{(k)} \right\}.
}

\subsection{Conditional Network Architecture}
\label{subsec:architecture}
\begin{figure}[htbp]
    \centering
    \includegraphics[width=0.9\textwidth]{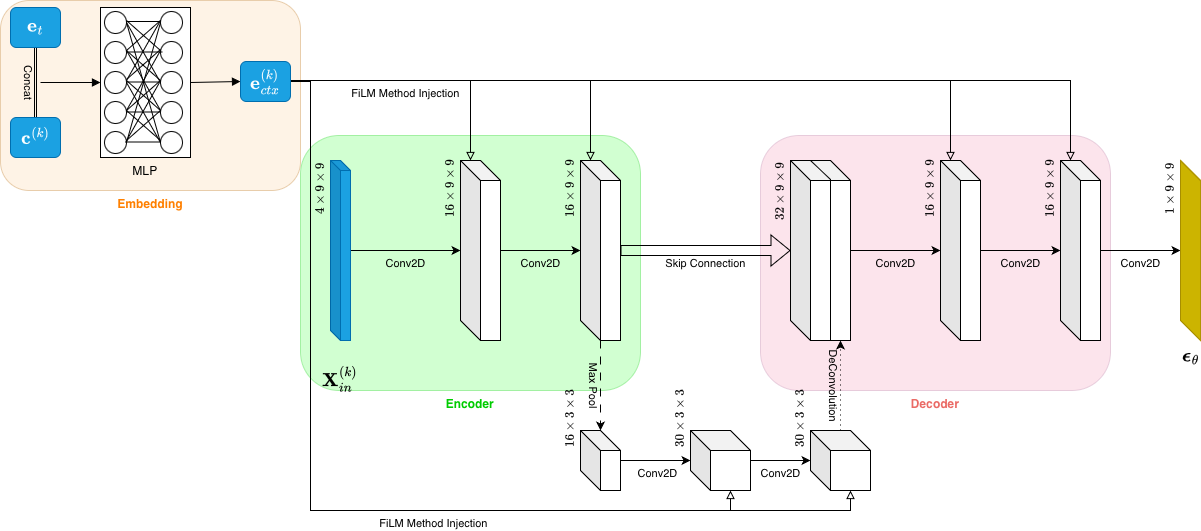}
    \caption{Architecture of the conditional noise-predicting U-Net. The model inputs are highlighted in blue and the output is highlighted in yellow. White blocks represent the convolutional blocks of the U-Net, with their corresponding tensor dimensions $C_l \times H_l \times W_l$ indicated at the top-left corner. The shaded background areas show each of the architecture components: the time and condition embedding by MLP (orange), the U-Net encoder (green), and the U-Net decoder (pink).}
    \label{fig:arch}
\end{figure}

We implement the noise-predicting neural network, parameterized by $\theta,$ by using a conditional U-Net architecture \citep{Ronneberger_Fischer_Brox_2015}, which minimizes the loss function     $\mathcal{L}_{\text{total}}$ \eqref{eq:final loss function}. The neural network follows an encoder-decoder design as shown in Figure \ref{fig:arch}. The encoder consists of a series of convolutional blocks, which are effective at extracting structural features from relationships across adjacent moneyness and maturity. The convolutional blocks are then followed by a max-pooling layer which downsamples the data and forces the subsequent convolutional blocks to focus on the concentrated features of the entire vol-surface. The decoder mirrors this process, using convolutional blocks and transposed convolutions to upsample the features back to the original dimension. Meanwhile, a skip connection is applied to reintegrate the discarded information from the max-pooling layer so that the neural network can combine both the structural and concentrated features for noise prediction. It is implemented by concatenating the output of each encoder block directly with the corresponding decoder block. Since the high-dimensional market conditions $\mathbf{x}_0^{(k)}, \mathbf{M}_{\text{short}}^{(k)},$ $\mathbf{M}_{\text{long}}^{(k)}$ defined in Section \ref{subsec:conditions} and noisy target state $\mathbf{x}_t^{(k+1)}$ are matrices that share the same dimensions, we introduce an additional axis, called channel, to concatenate those matrices into a 3-dimensional tensor. By stacking $\mathbf{x}_0^{(k)}, \mathbf{M}_{\text{short}}^{(k)}, \mathbf{M}_{\text{long}}^{(k)}$ with the noisy target state $\mathbf{x}_t^{(k+1)}$ along the channel, we construct an augmented multi-channel input tensor $\mathbf{X}_{\text{in}}^{(k)} \in \mathbb{R}^{4 \times N_m \times N_\tau}$. The convolutional blocks can evaluate this information simultaneously by processing $\mathbf{X}_{\text{in}}^{(k)}$. As previously discussed in Section \ref{subsec:score_matching}, the neural network requires the tensor $\mathbf{X}_{\text{in}}^{(k)}$, the diffusion time $t$, and the vector $\mathbf{c}^{(k)}$ to predict the injected noise. The convolutional layers of the encoder process $\mathbf{X}_{\text{in}}^{(k)}$ directly, allowing the neural network to extract the underlying spatial features of the noisy vol-surfaces. Since $t$ is a scalar and $\mathbf{c}^{(k)}$ is a vector, they cannot fit into the model architecture directly. We resolve this by first projecting $t$ into a high-dimensional feature space with fixed sinusoidal embeddings \citep{Vaswani_Shazeer_Parmar_Uszkoreit_Jones_Gomez_Kaiser_Polosukhin_2017}. For a specified embedding dimension $d_{\text{emb}} \geq 2$, the explicit components of the time vector $\mathbf{e}_t \in \mathbb{R}^{d_{\text{emb}}}$ are given by:
\eq{*}{
\mathbf{e}_t[2j] &= \sin\left(\frac{t}{10000^{2j/d_{\text{emb}}}}\right), \\
\mathbf{e}_t[2j+1] &= \cos\left(\frac{t}{10000^{2j/d_{\text{emb}}}}\right),
}
for $j = 0, \dots, \frac{d_{\text{emb}}}{2}-1$. The resulting time embedding $\mathbf{e}_t$ is then concatenated with $\mathbf{c}^{(k)}$ and processed by a multi-layer perceptron (MLP) consisting of two hidden layers with SiLU activations \citep{Elfwing_Uchibe_Doya_2018}. This produces a joint context embedding:
\eq{*}{
\mathbf{e}_{\text{ctx}}^{(k)} = \text{MLP}\left( \text{Concat}(\mathbf{e}_t, \mathbf{c}^{(k)}) \right).
}
This embedding $\mathbf{e}_{\text{ctx}}^{(k)}$ encodes information from both $\mathbf{e}_t$ and $\mathbf{c}^{(k)}$. The neural network 
$\theta$ integrates $\mathbf{e}_{\text{ctx}}^{(k)}$ into the convolutional blocks via Feature-wise Linear Modulation (FiLM) \citep{Perez_Strub_Vries_Dumoulin_Courville_2017}. In each FiLM operation, the intermediate output of the $l$-th convolutional block $\mathbf{F}_l$ is combined with the embedding 
$\mathbf{e}_{\text{ctx}}^{(k)}$ in the following way:
\eqstar{
    \text{FiLM}(\mathbf{F}_l \mid \mathbf{e}_{\text{ctx}}^{(k)}) = \boldsymbol{\kappa}_l(\mathbf{e}_{\text{ctx}}^{(k)}) \odot \mathbf{F}_l + \boldsymbol{\beta}_l(\mathbf{e}_{\text{ctx}}^{(k)}).
}
This affine transformation is performed across individual channels. In the above, $\mathbf{F}_l \in \mathbb{R}^{C_l \times H_l \times W_l},$ where $C_l$ represents the channel depth, and $H_l$ and $W_l$ denote the dimensions of the height and width at convolutional step $l$. The scaling vector $\boldsymbol{\kappa}_l \in \mathbb{R}^{C_l}$ and the shifting vector $\boldsymbol{\beta}_l \in \mathbb{R}^{C_l}$ are learned linear projections of $\mathbf{e}_{\text{ctx}}^{(k)}$. By applying this method after every block in both the encoder and decoder, $\mathbf{c}^{(k)}$ can shift and scale the intermediate feature outputs to dynamically shape the final output.

It is worth noting that the high-dimensional market conditions $\mathbf{x}_0^{(k)}, \mathbf{M}_{\text{short}}^{(k)}, \mathbf{M}_{\text{long}}^{(k)}$ are incorporated into the architecture differently than the scalar conditioning vector $\mathbf{c}^{(k)}$. This asymmetry follows from the difference in their dimensions, which determines how each can enter the U-Net. The conditions $\mathbf{x}_0^{(k)}, \mathbf{M}_{\text{short}}^{(k)}, \mathbf{M}_{\text{long}}^{(k)}$ share the same dimensions as the noisy target state $\mathbf{x}_t^{(k+1)}$ and are therefore channel-stacked with it at the input, allowing them to be processed jointly through the initial convolutional layer consistently with the standard U-Net design. By contrast, the scalar conditioning vector $\mathbf{c}^{(k)}$ does not match the dimensionality and cannot be channel-stacked without broadcasting it across the grid. We instead treat $\mathbf{c}^{(k)}$ in the same way as the diffusion timestep $t$, which is embedded and injected into the intermediate convolutional blocks rather than concatenated at the input. This allows $\mathbf{c}^{(k)}$ to adjust the convolutional blocks without requiring a dimensional match to the high-dimensional market conditions. We opted against broadcasting $\mathbf{c}^{(k)}$ across the grid and channel-stacking it alongside $\mathbf{x}_0^{(k)}, \mathbf{M}_{\text{short}}^{(k)}, \mathbf{M}_{\text{long}}^{(k)}$, since a broadcast scalar produces a spatially constant channel that convolutional kernels are not well-suited to exploit. This leads to spending spatial-processing capacity on information that carries no spatial variation, which would consume the limited channel budget without a corresponding gain in information content. FiLM avoids this by injecting $\mathbf{c}^{(k)}$ directly into the feature-map statistics at each convolutional block, giving it a more direct and parameter-efficient influence throughout the network. In our experiments, we did try broadcasting and channel-stacking of $\mathbf{c}^{(k)}$ versus the FiLM method. However, the accuracy and confidence intervals of the resulting model were worse.

Each convolutional layer in the network uses a $3 \times 3$ kernel with padding of size $1$ and stride $1$, preserving the spatial dimensions of the input throughout the encoder and decoder, followed by a SiLU activation \citep{Elfwing_Uchibe_Doya_2018}. The final output layer applies a $1 \times 1$ convolution with no subsequent activation, since the network is tasked to predict the unconstrained target noise $\hat{\boldsymbol{\epsilon}}_\theta$, which is not restricted to a bounded range. The overall network is deliberately shallow, comprising a single downsampling stage with narrow channel width, and the time and scalar-conditioning embedding dimensions are kept modest.  These choices reflect the limited size of the training dataset available for this application, as increasing the model capacity overfitted the model to the data. We manually tuned both smaller and larger configurations of channel width and bottleneck width. The configuration with half the channel widths $C_l$ of Figure~\ref{fig:arch} produced vol-surfaces with the correct shape (smile/skew structure) but incorrect overall volatility levels, indicating insufficient capacity to calibrate the absolute magnitude, while doubling the channel widths $C_l$ did not improve the validation performance. We therefore selected the reported configuration as the smallest capacity sufficient to reproduce realistic surface dynamics, consistent with the capacity constraints imposed by the limited size of the training data.

After $3 \times 3$ max-pooling with stride 3 the feature map is $3 \times 3$, so each $3 \times 3$ convolution in the bottleneck sees the entire pooled surface and acts essentially as a fully connected mixing layer. This step is intended as it gives the network a global view of the surface, which is needed to set its overall level. It is consistent with our observation that lower-capacity variants reproduce the shape of the surface but not its level. The decoder upsamples with a $3 \times 3$ transposed convolution with stride 3, and the MLP producing $\mathbf{e}_{\text{ctx}}^{(k)}$ has 10 hidden units and 2 hidden layers. The sinusoidal time embedding has dimension $d_{\text{emb}} = 24$ (smaller than is common in image diffusion models). With $T = 500$ steps, its 12 frequency pairs (periods from about 6 to about 29{,}000 steps) already give every timestep a distinct code, which the MLP then combines with $\mathbf{c}^{(k)}$. Varying $d_{\text{emb}}$ between 12 and 48 did not change the validation performance. All architecture choices were made on the validation set only with hyperparameter values reported in Table \ref{tab:hyperparameters}.

\subsection{Training and Sampling Algorithm}
\label{subsec:training_alg}
We adopt the training-validation-test framework to ensure that the training contributes to true out-of-sample performance. The training epochs are denoted by  $N_{\text{epochs}}$ and the batch size by $B.$ We use the AdamW optimizer \citep{Loshchilov_Hutter_2019}, which is the Adam optimizer with decoupled weight decay. We fix $\mu_t$ and $v_t$ in Section \ref{subsec:score_matching} with the cosine schedule introduced by \citet{Nichol_Dhariwal_2021}, which we discretize into $T$ timesteps for training. To ensure stability, the learning rate is dynamically managed via a plateau-based scheduler, which decays the learning rate by a factor of $\gamma_{\text{lr}}$ if the validation loss fails to improve for $P_{\text{lr}}$ consecutive epochs, strictly bounded by a minimum rate of $\eta_{\text{min}}$. Due to the fact that diffusion models are highly sensitive to exploding gradients, we enforce gradient clipping with a maximum $L_2$-norm of $c_{\text{grad}}$. 
The value of $c_{\text{grad}}$ in our experiment is found by observing the gradient norm oscillation of the neural network with EWMA parameters $\theta_{\text{EWMA}}$ (referred to as the EWMA diffusion model) of the training epochs. 
The gradient norm oscillation exhibits a clear lower bound over the training epochs and the value of $c_{\text{grad}}$ is set to the level slightly above the observed lower bound to enhance the training stability. Gradient norms are small in our setting because the network has few parameters. The median norm of $\nabla_\theta \mathcal{L}_{\text{total}}$ over training is 0.32 and its 95th percentile is 0.94. 
Also, we address the overfitting issue through early stopping with early stopping patience $P_{es}$ and threshold $T_{es}$. In our numerical experiments, we found that maintaining an EWMA of the network parameters $\theta$ with a decay rate of $\beta_{\text{EWMA}}$, denoted as $\theta_{\text{EWMA}}$, and referred to as the EWMA diffusion model, smooths the optimization trajectory and prevents parameter oscillation. Therefore, the vol-surfaces generated by the EWMA diffusion model are more accurate than generated vol-surfaces from the neural network with parameters $\theta$. The complete optimization subroutine is provided in Algorithm \ref{alg:training_ema}. The conditional forecast of the volatility surface for the next trading day (sampling) is generated by simulating the reverse diffusion process using the neural network with parameter $\theta_{\text{EWMA}}$. We initialize the reverse process for the target forecast day $k+1$ by drawing a sample $\mathbf{x}_T^{(k+1)} \sim \mathcal{N}(\mathbf{0}, \mathbf{I})$. We then iteratively denoise it across the discrete timesteps $t = T, \dots, 1,$ to create the final forecast of the volatility surface. The sampling subroutine is provided in Algorithm \ref{alg:sampling_ema}.

\begin{algorithm}[htbp]
\caption{Training the Conditional Diffusion Model with Arbitrage Penalty}
\label{alg:training_ema}
\begin{algorithmic}[1]
\Require Historical trading days in training set $k \in \{1, \dots, K\}$, denoising steps $T$, penalty weight $\lambda$, Min-SNR threshold $\gamma$, EWMA decay $\beta_{\text{EWMA}}$, learning rate $\eta$, max gradient norm $c_{\text{grad}}$.
\Require Schedule parameters: epochs $N_{\text{epochs}}$, batch size $B$, plateau patience $P_{\text{lr}}$, learning rate decay factor $\gamma_{\text{lr}}$, minimum learning rate $\eta_{\text{min}}$, early-stopping patience $P_{es}$ and threshold $T_{es}$.
\State Initialize neural network parameters $\theta$ randomly
\State Initialize learning rate $\eta \leftarrow \eta_0$
\State Initialize EWMA parameters $\theta_{\text{EWMA}} \leftarrow \theta$

\For{$\text{epoch} = 1, \dots, N_{\text{epochs}}$}
    \For{each batch of $B$ trading days from the training set}
        \State For each trading day included in the batch, retrieve $\mathcal{Y}^{(k)}$ and $\mathbf{x}_0^{(k+1)}$
        \State Sample discrete diffusion timestep $t \sim \mathrm{Uniform}(\{1, \dots, T\})$
        \State Sample Gaussian noise $\boldsymbol{\epsilon} \sim \mathcal{N}(\mathbf{0}, \mathbf{I})$
        \State Compute VP schedule parameters $\mu_t$, $v_t$ \eqref{eq: mean and variance}, and $\text{SNR}(t) = \frac{\mu_t^2}{v_t}$
        \State Construct the noisy target state: $\mathbf{x}_t^{(k+1)} = \mu_t \mathbf{x}_0^{(k+1)} + \sqrt{v_t} \boldsymbol{\epsilon}$
        \State Predict the added noise: $\hat{\boldsymbol{\epsilon}} = \boldsymbol{\epsilon}_\theta(\mathbf{x}_t^{(k+1)}, t, \mathcal{Y}^{(k)})$
        \State Compute MSE loss: $\mathcal{L}_{\text{simple}} = \| \boldsymbol{\epsilon} - \hat{\boldsymbol{\epsilon}} \|_2^2$
        \State Estimate the standardized vol-surface: $\hat{\mathbf{x}}_0^{(k+1)} = \frac{1}{\mu_t} \left( \mathbf{x}_t^{(k+1)} - \sqrt{v_t} \hat{\boldsymbol{\epsilon}} \right)$
        \State Project back to the vol-surface: $\hat{\boldsymbol{\sigma}}^{(k+1)} = \mathcal{T}^{-1}(\hat{\mathbf{x}}_0^{(k+1)})$
        \State Compute the arbitrage penalty: $\Phi(\hat{\boldsymbol{\sigma}}^{(k+1)}) = \Phi_{\tau}(\hat{\boldsymbol{\sigma}}^{(k+1)}) + \Phi_{b}(\hat{\boldsymbol{\sigma}}^{(k+1)})$
        \State Compute the total loss: $\mathcal{L}_{\text{total}} = \mathcal{L}_{\text{simple}} + \lambda \cdot \min(\text{SNR}(t), \gamma) \cdot \Phi(\hat{\boldsymbol{\sigma}}^{(k+1)})$
        \State Compute gradients: $\mathbf{g} = \nabla_\theta \mathcal{L}_{\text{total}}$
        \State Apply gradient clipping: $\mathbf{g} \leftarrow \mathbf{g} \cdot \min\left(1, \frac{c_{\text{grad}}}{\|\mathbf{g}\|_2}\right)$
        \State Update network parameters $\theta$ via AdamW using $\mathbf{g}$ and current learning rate $\eta$
        \State Update EWMA weights: $\theta_{\text{EWMA}} \leftarrow \beta_{\text{EWMA}} \theta_{\text{EWMA}} + (1 - \beta_{\text{EWMA}}) \theta$
    \EndFor
    
    \State Compute validation loss on the out-of-sample validation set using parameters $\theta$
    \If{the validation loss fails to improve for $P_{\text{lr}}$ consecutive epochs}
        \State Decay learning rate: $\eta \leftarrow \max(\eta \cdot \gamma_{\text{lr}}, \, \eta_{\text{min}})$
    \EndIf
    \If{validation early stopping criterion is met}
        \State \textbf{break}
    \EndIf
\EndFor
\State \Return $\theta_{\text{EWMA}}$
\end{algorithmic}
\end{algorithm}

\begin{algorithm}[htbp]
\caption{Sampling Conditional IV Surfaces using EWMA Diffusion Model}
\label{alg:sampling_ema}
\begin{algorithmic}[1]
\Require Trained EWMA network $\boldsymbol{\epsilon}_{\theta_{\text{EWMA}}}$.
\Require Market conditioning set $\mathcal{Y}^{(k)}$ for the current trading day $k$.
\Require VP schedule parameters $\mu_t$ and $v_t$.
\Require Training set mean and standard deviation $\mu_{ij}$ and $s_{ij}$.
\State Sample initial pure noise for the target day $k+1$: $\mathbf{x}_T^{(k+1)} \sim \mathcal{N}(\mathbf{0}, \mathbf{I})$
\For{$t = T, T-1, \dots, 1$}
    \State Sample standard Gaussian $\mathbf{z} \sim \mathcal{N}(\mathbf{0}, \mathbf{I})$ if $t > 1$, else $\mathbf{z} = \mathbf{0}$
    \State Discretize the continuous schedule for the current step:
    $$ \alpha_t = \frac{\mu_t^2}{\mu_{t-1}^2}, \quad \beta_t = 1 - \alpha_t $$
    \State Predict the added noise using the EWMA network: 
    $$ \hat{\boldsymbol{\epsilon}} = \boldsymbol{\epsilon}_{\theta_{\text{EWMA}}}\left(\mathbf{x}_t^{(k+1)}, t, \mathcal{Y}^{(k)}\right) $$
    \State Compute the mean of the reverse conditional transition:
    $$ \boldsymbol{\mu}_{\theta_{\text{EWMA}}}\left(\mathbf{x}_t^{(k+1)}, t, \mathcal{Y}^{(k)}\right) = \frac{1}{\sqrt{\alpha_t}} \left( \mathbf{x}_t^{(k+1)} - \frac{\beta_t}{\sqrt{v_t}} \hat{\boldsymbol{\epsilon}} \right) $$
    \State Compute the variance of the reverse step: $\sigma_t^2 = \frac{v_{t-1}}{v_t} \beta_t$
    \State Perform the reverse step to compute the denoised state:
    $$ \mathbf{x}_{t-1}^{(k+1)} = \boldsymbol{\mu}_{\theta_{\text{EWMA}}}\left(\mathbf{x}_t^{(k+1)}, t, \mathcal{Y}^{(k)}\right) + \sigma_t \mathbf{z} $$
\EndFor
\State Obtain the final generated standardized vol-surface sample $\mathbf{x}_0^{(k+1)}$
\State Project back to vol-surface using the inverse transformation $\mathcal{T}^{-1}$:
$$ \hat{\sigma}_{ij}^{(k+1)} = \mathcal{T}^{-1}\left(\mathbf{x}_{0, ij}^{(k+1)}\right) = \exp\left( \mathbf{x}_{0, ij}^{(k+1)} \cdot s_{ij} + \mu_{ij} \right) $$
\State \Return Generated day-ahead vol-surface forecast $\hat{\boldsymbol{\sigma}}^{(k+1)}$
\end{algorithmic}
\end{algorithm}




\section{Experiments}
\label{sec:experiments}
\subsection{Market Data Representation and Pre-Processing}
\label{subsec:data_preprocessing}
We use market data from the Option Prices file provided by OptionMetrics \citep{OptionMetrics_2023} spanning the period from January 4, 1996 to August 31, 2023. This data contains vol-surfaces derived from traded options on the S\&P 500 index. The dataset is split chronologically: the initial 80\% serves as the training set, the subsequent 10\% constitutes the validation set, and the final 10\% is reserved as the test set. As is standard practice, we use OTM option prices to construct vol-surfaces. Specifically, for $K \leq S$, we utilize put options data and for $K > S$, we use call options data. To ensure data quality, options with zero trading volume are excluded. Under this procedure, ITM and ATM ($K \leq S$) implied volatilities are derived from OTM puts, while OTM implied volatilities ($K > S$) are derived from OTM calls. Since equity market investors are predominantly concerned with downside risks, they frequently trade OTM puts as a protective hedge to limit losses from market downturns. 



\begin{figure}[htbp]
    \centering
    \includegraphics[width=0.7\textwidth]{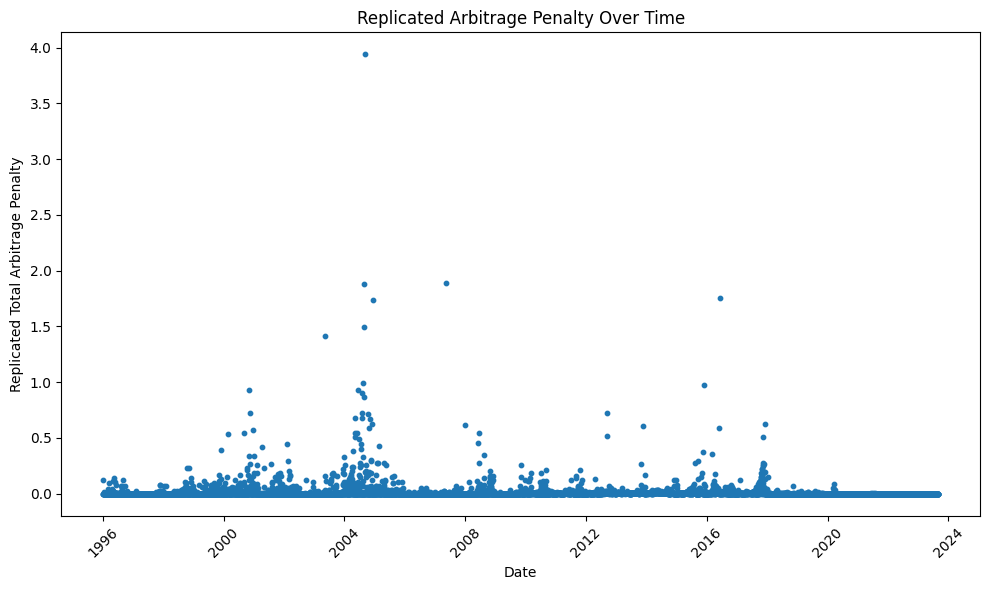}
    \caption{Arbitrage level of Vega-smoothed dataset from 1996--2023.}
    \label{fig:arb}
\end{figure}

\begin{figure}[htbp]
    \centering
    \begin{subfigure}[b]{\textwidth}
        \centering
        \includegraphics[width=0.95\textwidth]{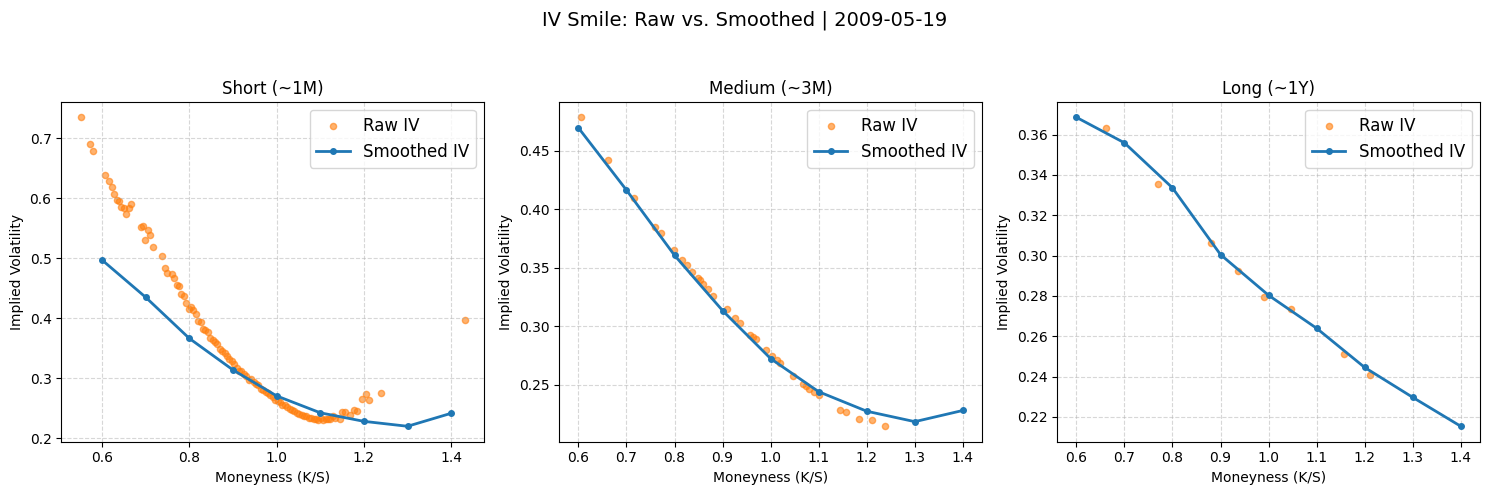}
        \caption{Implied volatility vs. smoothed surface on May 19, 2009.}
        \label{fig:arb_1}
    \end{subfigure}
    
    \vspace{1em} 
    
    \begin{subfigure}[b]{\textwidth}
        \centering
        \includegraphics[width=0.95\textwidth]{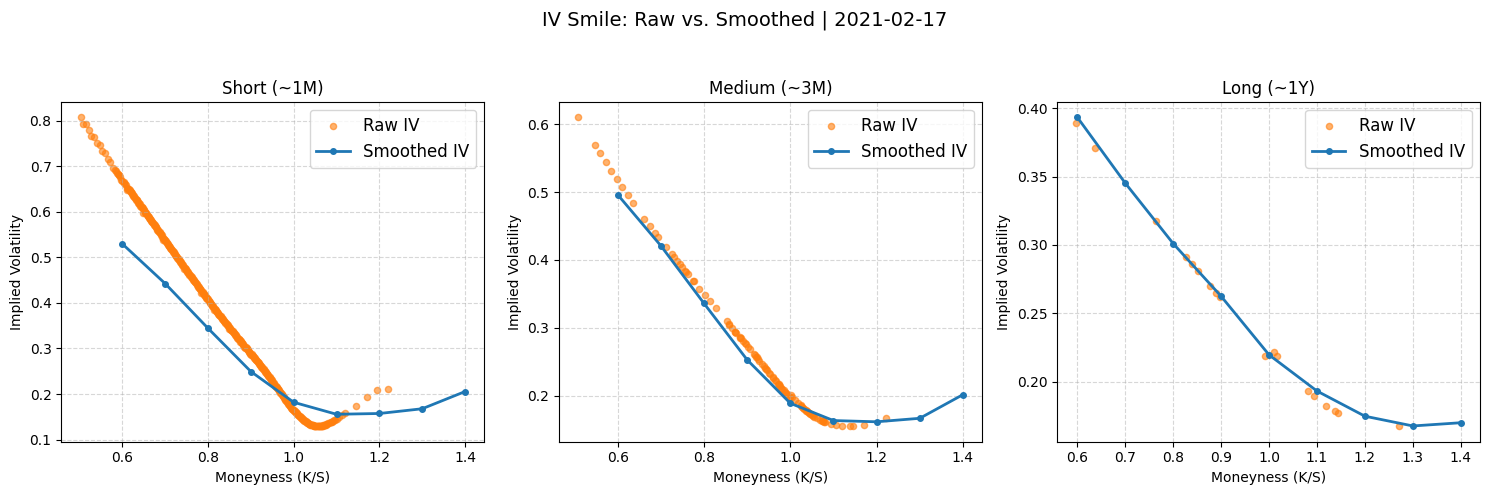}
        \caption{Implied volatility vs. smoothed surface on February 17, 2021.}
        \label{fig:arb_2}
    \end{subfigure}
    \caption{Implied volatility and smoothed implied volatility data comparison of two arbitrarily chosen dates from the OptionMetrics dataset. The figures demonstrate the extrapolation required in the OTM regions due to the sparsity of market data.}
    \label{fig:raw_smooth}
\end{figure}

In our dataset, we consider the vol-surfaces over a discrete grid $(\mathbf{m}, \boldsymbol{\tau})$ consisting of moneyness levels 
\eqstar{
    \mathbf{m} \in \{0.6, 0.7, 0.8, 0.9, 1.0, 1.1, 1.2, 1.3, 1.4\},
} 
and times-to-maturity (in years) 
\eqstar{
    \boldsymbol{\tau} \in \{1/252, 1/52, 2/52, 1/12, 1/6, 1/4, 1/2, 3/4, 1\}.
}
Since market option prices are scattered across moneyness and time-to-maturity rather than quoted on a fixed grid as defined above, directly constructing vol-surfaces from market option prices may result in irregular and non-smooth surfaces. Therefore, we apply a non-parametric smoothing technique based on the Vega-weighted Nadaraya-Watson kernel estimator, proposed by \citet{Cont_Da_Fonseca_2002}. Let $\Omega^{(k)} = \{(m_n, \tau_n)\}_{n=1}^{N_k}$ denote the unstructured set of coordinate pairs for all $N_k$ available market option quotes observed on trading day $k$. For each observed coordinate $n$, let $\sigma_n^{(k)}$ represent the market implied volatility and $\mathcal{V}_n^{(k)}$ represent the corresponding option Vega. Let $\sigma_{ij}^{(k)}$ denote the implied volatility at grid point $(m_i, \tau_j)$ on trading day $k$ where $i$ and $j$ index the moneyness and time-to-maturity dimensions respectively. The estimator is defined as:
\eq{*}{
    \sigma_{ij}^{(k)} = \frac{\sum_{n=1}^{N_k} \mathcal{V}_n^{(k)} K(m_n - m_i, \tau_n - \tau_j) \sigma_n^{(k)}}{\sum_{n=1}^{N_k} \mathcal{V}_n^{(k)} K(m_n - m_i, \tau_n - \tau_j)},
}
where $K(u, v)$ is a two-dimensional Gaussian kernel:
\eq{*}{
    K(u, v) = \frac{1}{2\pi} \exp\left(-\frac{u^2}{2h_m} - \frac{v^2}{2h_\tau}\right).
}
The bandwidth hyperparameters $h_m$ and $h_\tau$ control the spatial degree of smoothing across the moneyness and time-to-maturity dimensions, respectively. 
 We adopt the bandwidth hyperparameter values reported by \citet{Vuletic_Cont_2024} rather than calibrating them. Figure \ref{fig:arb} illustrates that the Vega-smoothed dataset displays significant arbitrage, concentrated mainly in the earlier part of the sample.  Another observation worth noting is that the Vega-smoothing and interpolation technique above at times extrapolates the OTM region of the implied volatility smile where quotes are missing at required points of the implied volatility grid as shown in Figure \ref{fig:raw_smooth}. Given that we penalize arbitrage, learning from extrapolated points adversely affects the calibration of the predictive distribution in the OTM region, as we discuss in Section~\ref{subsec:real_results}. Figure~\ref{fig:raw_smooth} also shows that the smoothing flattens the low-moneyness wing of the smile at short maturities.

\subsection{Choice of models for performance comparison}
To evaluate the forecasting performance of our proposed diffusion model, we benchmark it against a diverse set of models including traditional time-series and stochastic volatility models, as well as deep learning generative architectures. Specifically, we compare our model against a Vector Autoregression (VAR(1)) model, a principal-component factor model with autoregressive factor dynamics (PCA-AR(1)), the SABR model \citep{Hagan_2002}, an LSTM network  augmented with an attention layer \citep{LSTMIV}, and the VolGAN model \citep{Vuletic_Cont_2024}.
\begin{itemize}
\item The VAR(1) model treats the vol-surface as a single vector and predicts the one-day-ahead vol-surface as a linear function of the current vol-surface. Mathematically, the model is:
\eqstar{
\mathbf{x}^{(k+1)} = \boldsymbol{\beta}_{0}^{(k)} + \boldsymbol{\beta}_{1}^{(k)} \mathbf{x}^{(k)} + \bm{\epsilon}^{(k+1)},
}
where $\mathbf{x}^{(k)}$ represents the vol-surface on trading day $k$, $\boldsymbol{\beta}_{0}^{(k)}$ is the intercept vector, $\boldsymbol{\beta}_{1}^{(k)}$ is the coefficient matrix and $\bm{\epsilon}^{(k+1)} \sim \mathcal{N}(\mathbf{0}, \boldsymbol{\Sigma})$ with residual covariance matrix $\boldsymbol{\Sigma}$. The parameters $\boldsymbol{\beta}_{0}^{(k)}$ and $\boldsymbol{\beta}_{1}^{(k)}$ are estimated via ordinary least squares using an expanding window where on each trading day $k$, $\boldsymbol{\beta}_{0}^{(k)}$ and $\boldsymbol{\beta}_{1}^{(k)}$ are recomputed using all available historical data up to trading day $k$, and the updated model is then used to predict the one-day-ahead vol-surface.

\item The PCA-AR(1) model follows the factor approach of \citet{Cont_Da_Fonseca_2002}. On each trading day $k$, we compute the principal components of the log vol-surfaces observed up to day $k$ (expanding window), fit an AR(1) model to each of the first 10 factor scores, forecast the scores one day ahead and reconstruct the log vol-surface.

\item The SABR model is calibrated separately for each time to maturity on the vol-surface using Hagan's closed-form approximation, which maps the underlying model parameters $\alpha$, $\rho$, and $\nu$ to implied volatility as a function of strike. For each trading day $k$ and specific time to maturity, the parameter vector $\boldsymbol{\theta}^{(k)} = [\alpha^{(k)}, \rho^{(k)}, \nu^{(k)}]^{\top}$ is calibrated by minimizing the squared error between the model-predicted and the observed market volatilities. To produce a forecast, these calibrated parameters are projected forward using an EWMA of their historical values. The predicted parameter vector for the next trading day $k+1$, denoted as $\hat{\boldsymbol{\theta}}^{(k+1)}_{\text{SABR}}$, is formulated as:
\eqstar{
\hat{\boldsymbol{\theta}}^{(k+1)}_{\text{SABR}} = (1 - \kappa_{\text{S}}) \boldsymbol{\theta}^{(k)}_\text{SABR} + \kappa_{\text{S}} \hat{\boldsymbol{\theta}}^{(k)}_\text{SABR}
}
where $\kappa_{\text{S}} \in (0,1)$ is the decay factor (we use $\kappa_{\text{S}} = 0.5$ and $\beta =1.0$), assigning greater weight to more recent market calibrations. The projected parameters are then mapped back through the SABR formula to reconstruct the predicted implied volatilities for that time to maturity. This process is repeated across all maturities to construct the full predicted vol-surface, which is subsequently compared against the observed market surface to evaluate forecasting accuracy.

\item In the attention-based LSTM model considered, we treat the vol-surface as a single vector and predict its one-day-ahead value from a historical window of daily, weekly, and monthly averaged implied volatilities at each surface point. These inputs are passed through a first LSTM layer, whose hidden states are reweighted by an attention layer that learns to emphasize the more informative points in the input window before passing them to a second LSTM layer. The output of the second LSTM layer is mapped through a fully connected layer to produce the predicted vol-surface for the next trading day. We apply a dropout mechanism within the network during training to reduce overfitting. Following the same evaluation protocol as our other benchmarks, the model is re-estimated over rolling windows of historical data to produce a sequence of one-day-ahead forecasts, which are compared against the observed vol-surfaces over the test period.

\item VolGAN is structured as a conditional GAN architecture constructed with dense neural network layers to model the dynamics of the underlying return and the vol-surface. The model input consists of one-day and two-day underlying returns, the one-month realized volatility, and the current log vol-surface, alongside a Gaussian noise vector. These inputs are processed through the dense layers to output a simulated underlying return and the predicted log vol-surface increment for the next trading day. To regularize the shape of the predicted vol-surface, the model is trained using an augmented loss function that incorporates a smoothness penalty. This additional term mitigates the jaggedness often associated with predictions from the model without this additional penalty term. 

\item Additionally, we train the VolGAN model using the conditioning variables $\mathcal{Y}^{(k)}$ introduced in Section \ref{subsec:conditions}. By aligning the conditioning sets between our proposed diffusion model and VolGAN, we can isolate and evaluate the specific contribution of our proposed conditioning framework to the overall prediction accuracy versus the GAN approach.
\end{itemize}
The more recent generative models discussed in Section~1 \citep{Bergeron_2021, ning2023arbitrage, choudhary2024funvol, buehler2026dysanos} are designed to simulate realistic surfaces rather than to forecast the day-ahead vol-surface conditional on the current market state. Thus, VolGAN remains our main generative benchmark. Note that  VAR(1) and PCA-AR(1) are re-estimated on an expanding window, and the LSTM on rolling windows during the test period, whereas the diffusion model is trained only once on the training set; this favours the benchmarks. The hyperparameters of all benchmarks are listed in Table~\ref{tab:benchmark_hyperparameters} in Appendix~\ref{sec:hyperp}.

\subsection{Forecasting Performance Comparison}
\label{subsec:real_results}
\begin{figure}[p]
    \centering
    \begin{subfigure}[b]{0.32\textwidth}
        \centering
        \includegraphics[width=\textwidth]{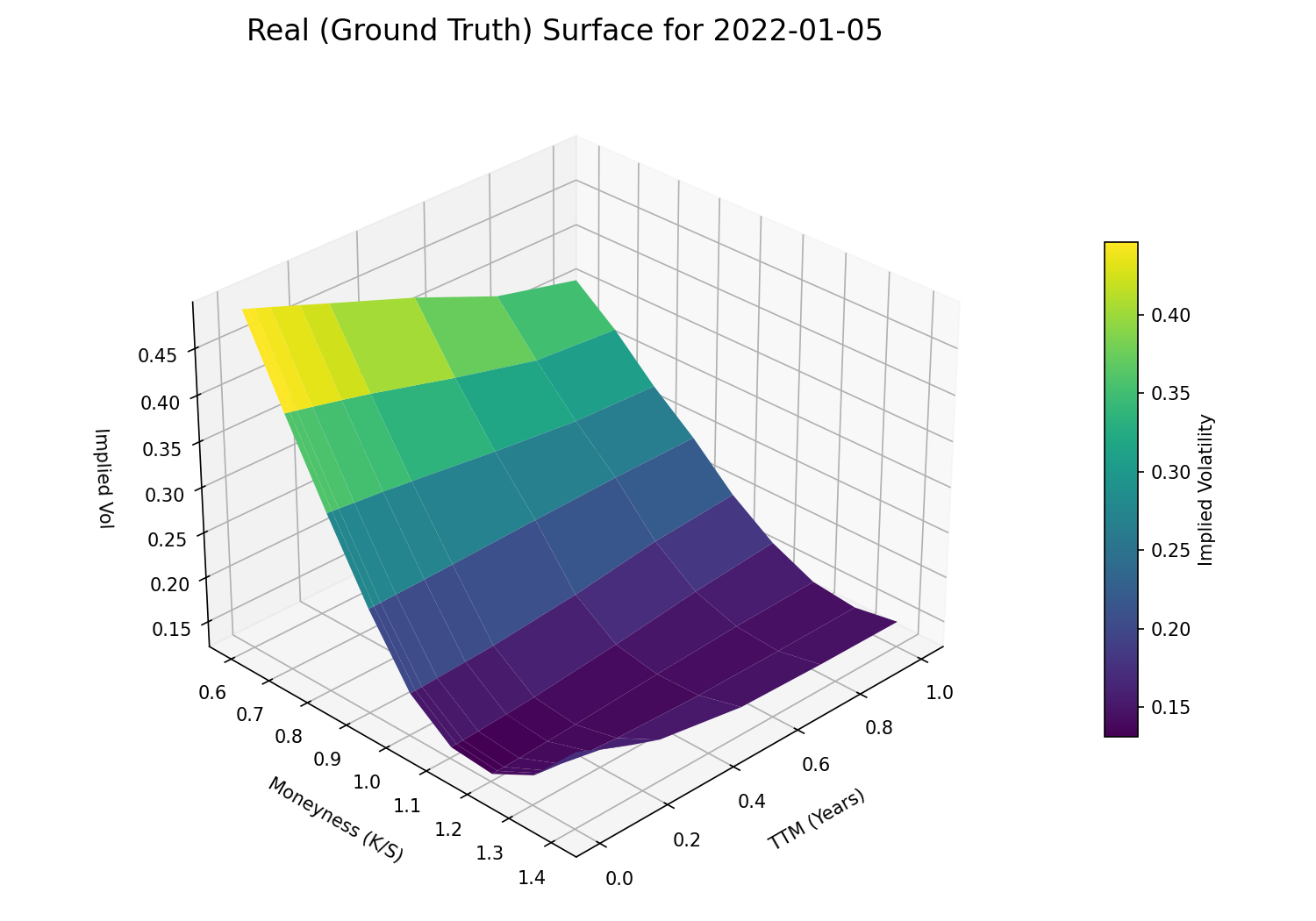}
        \caption{Real.}
        \label{fig:real_surf}
    \end{subfigure}
    \hfill
    \begin{subfigure}[b]{0.32\textwidth}
        \centering
        \includegraphics[width=\textwidth]{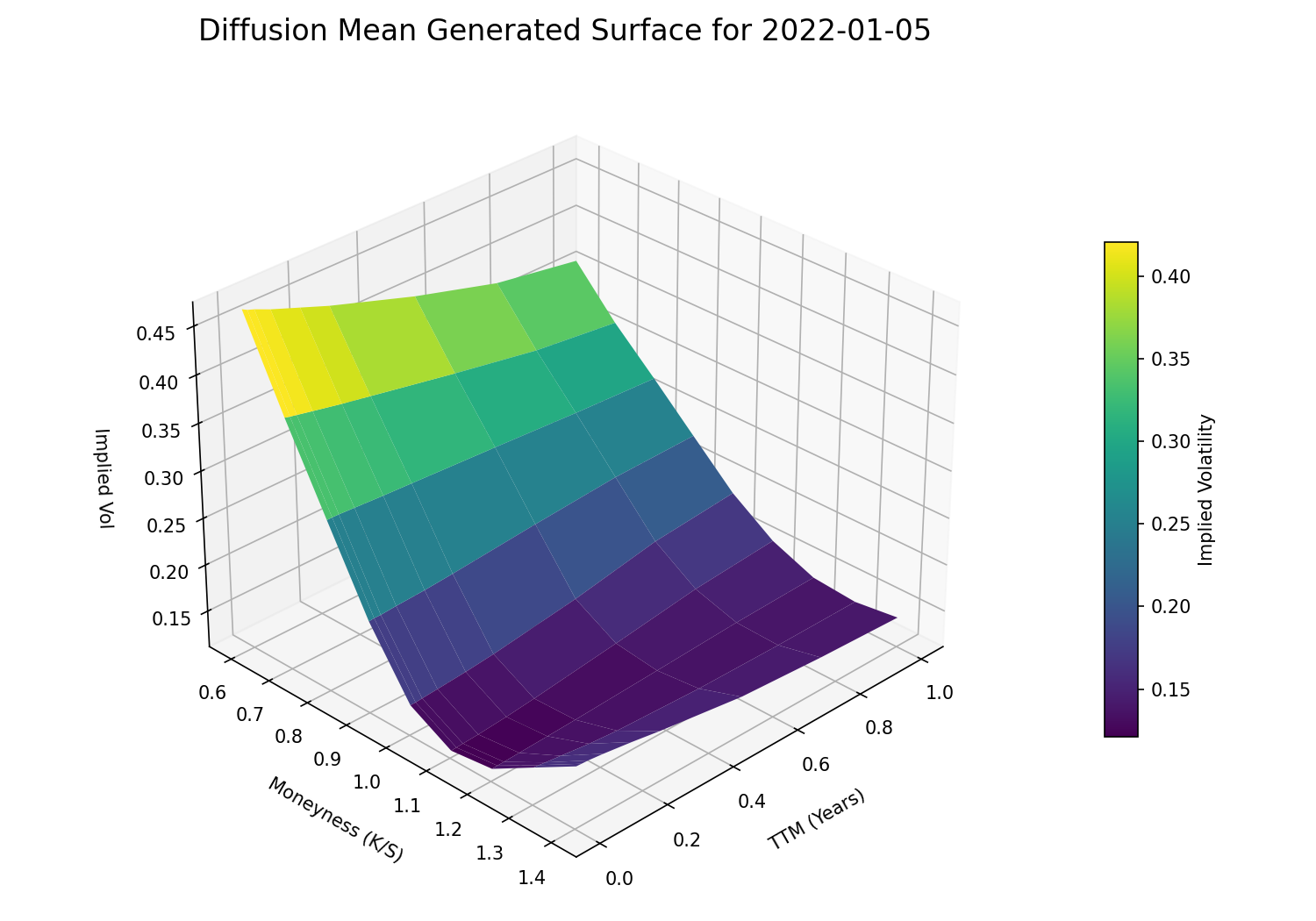}
        \caption{Diffusion Model (Mean).}
        \label{fig:diff_surf}
    \end{subfigure}
    \hfill
    \begin{subfigure}[b]{0.32\textwidth}
        \centering
        \includegraphics[width=\textwidth]{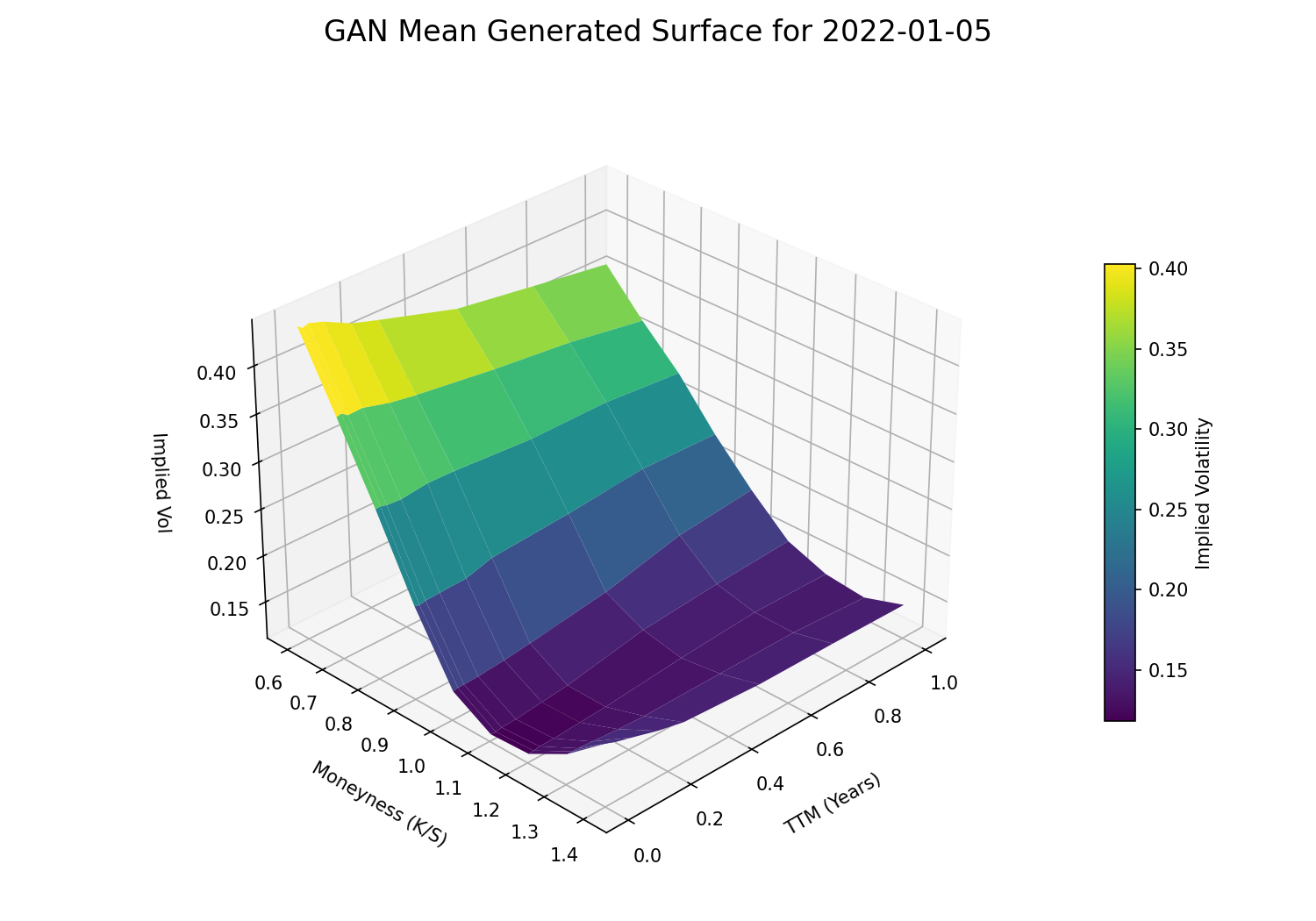}
        \caption{VolGAN (Mean).}
        \label{fig:gan_surf}
    \end{subfigure}

    \vspace{0.5em}

    \begin{subfigure}[b]{0.32\textwidth}
        \centering
        \includegraphics[width=\textwidth]{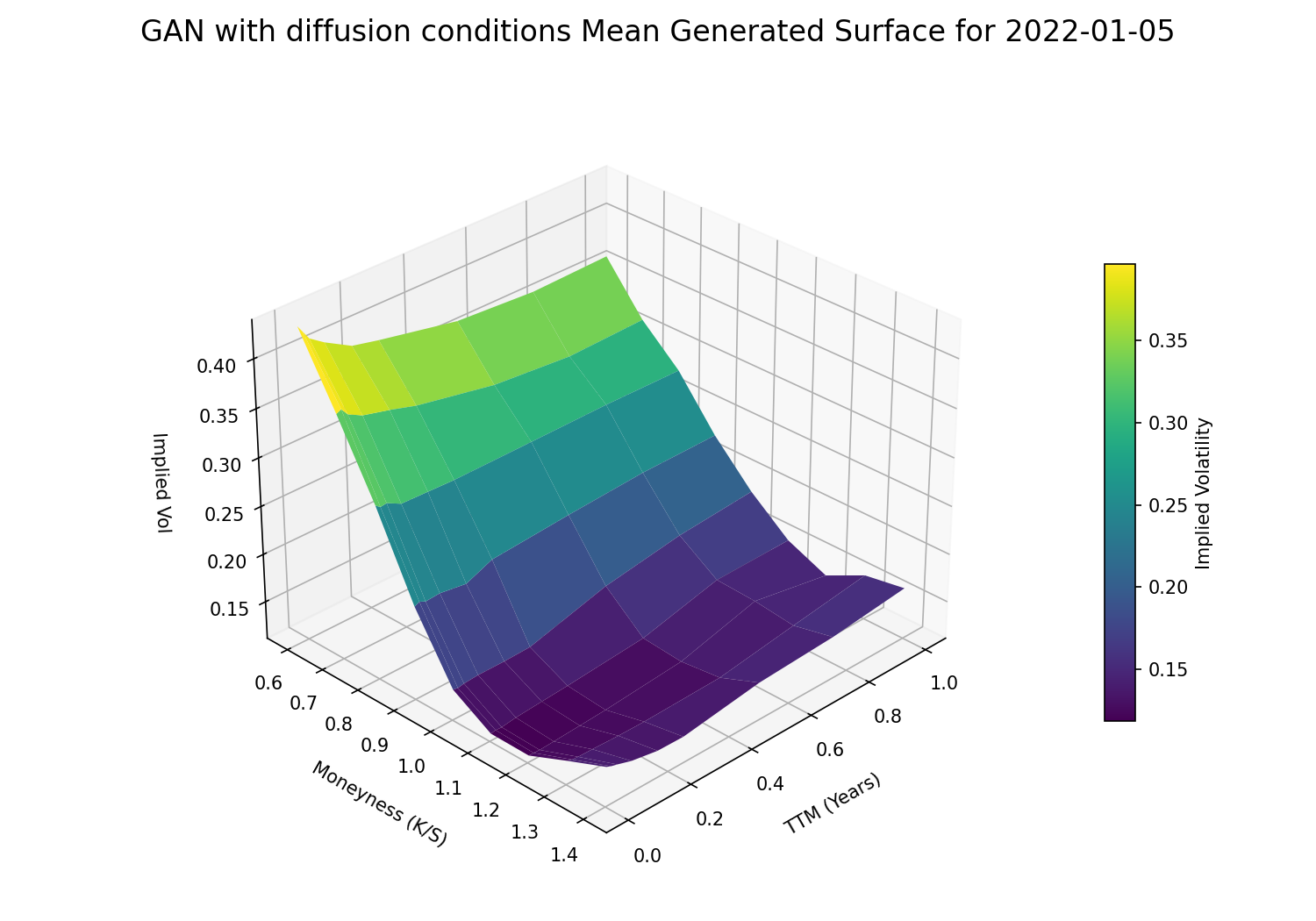}
        \caption{VolGAN with $\mathcal{Y}^{(k)}$ (Mean).}
        \label{fig:ganyk_surf}
    \end{subfigure}
    \hfill
    \begin{subfigure}[b]{0.32\textwidth}
        \centering
        \includegraphics[width=\textwidth]{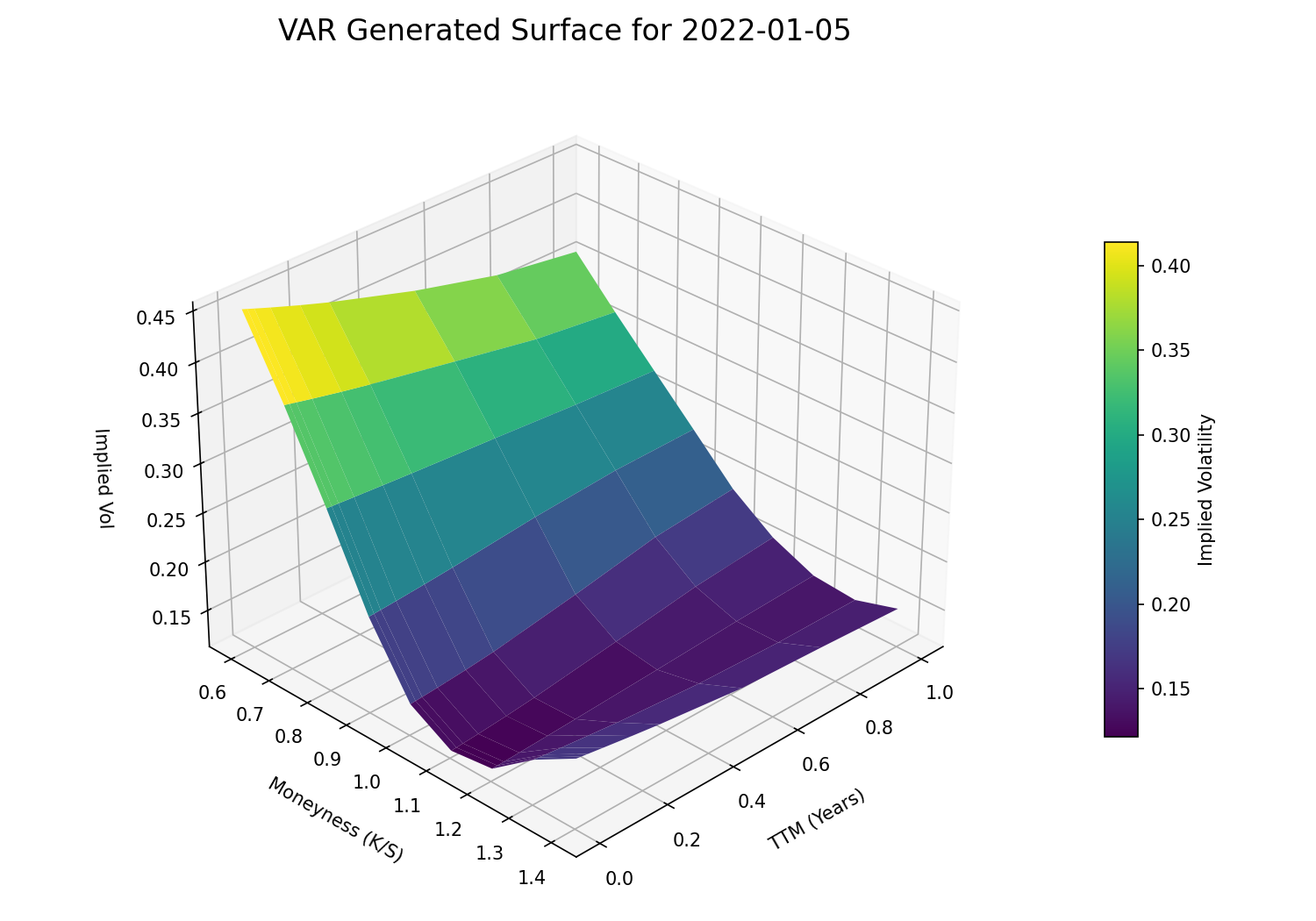}
        \caption{VAR(1).}
        \label{fig:var_surf}
    \end{subfigure}
    \hfill
    \begin{subfigure}[b]{0.32\textwidth}
        \centering
        \includegraphics[width=\textwidth]{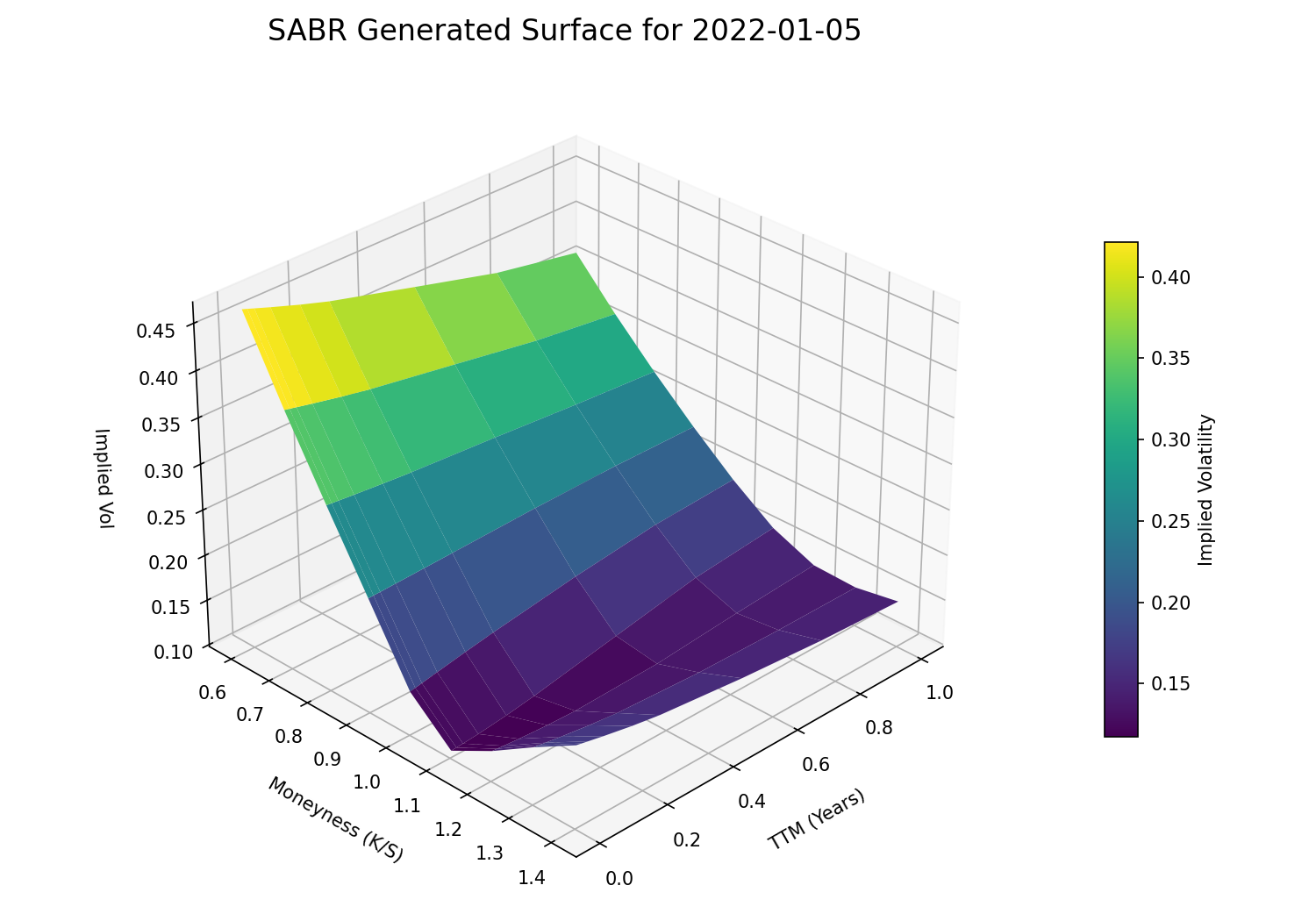}
        \caption{SABR.}
        \label{fig:sabr_surf}
    \end{subfigure}

    \vspace{0.5em}

    \begin{subfigure}[b]{0.32\textwidth}
        \centering
        \includegraphics[width=\textwidth]{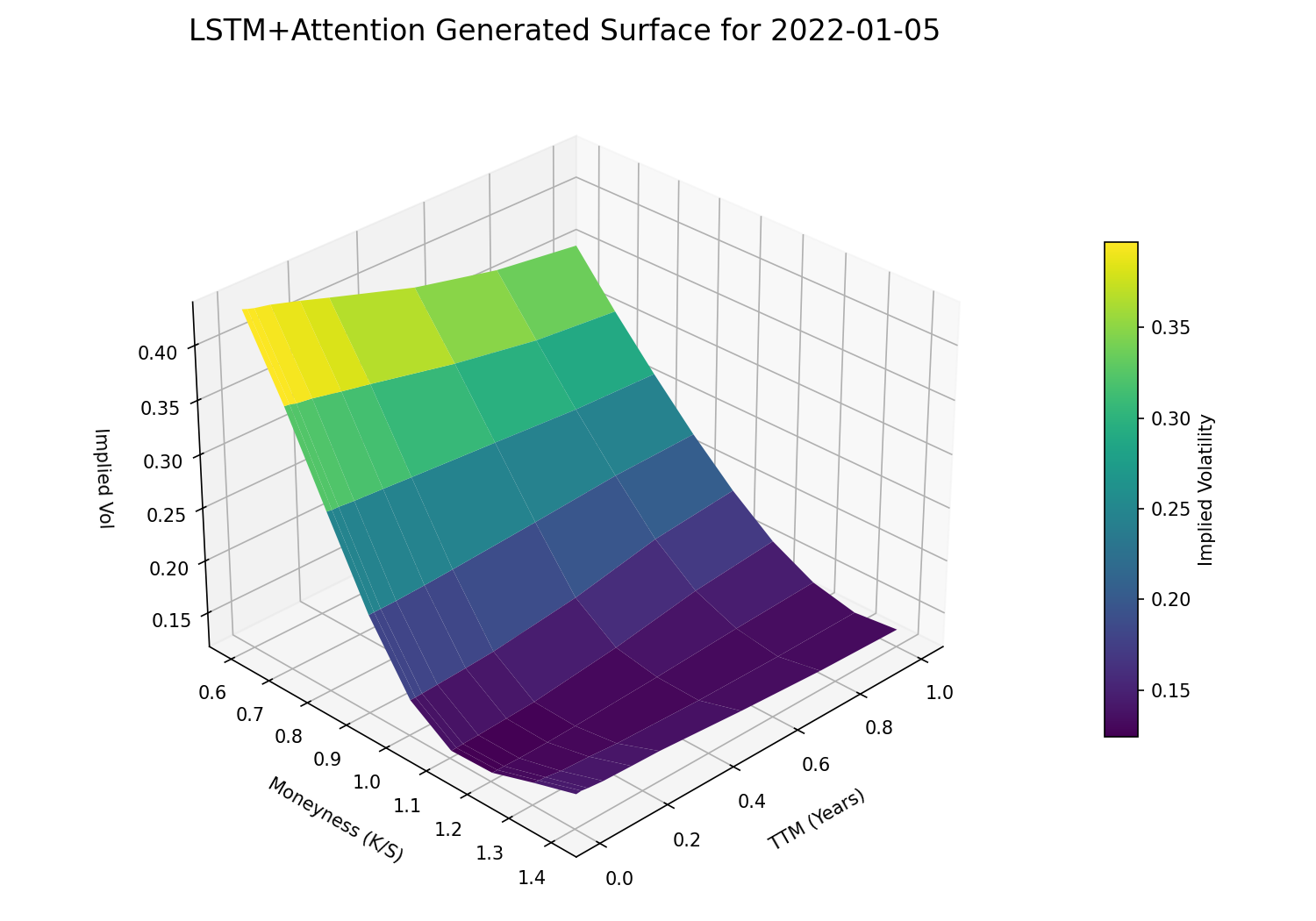}
        \caption{LSTM+Attention.}
        \label{fig:lstm_surf}
    \end{subfigure}
    \begin{subfigure}[b]{0.32\textwidth}
        \centering
        \includegraphics[width=\textwidth]{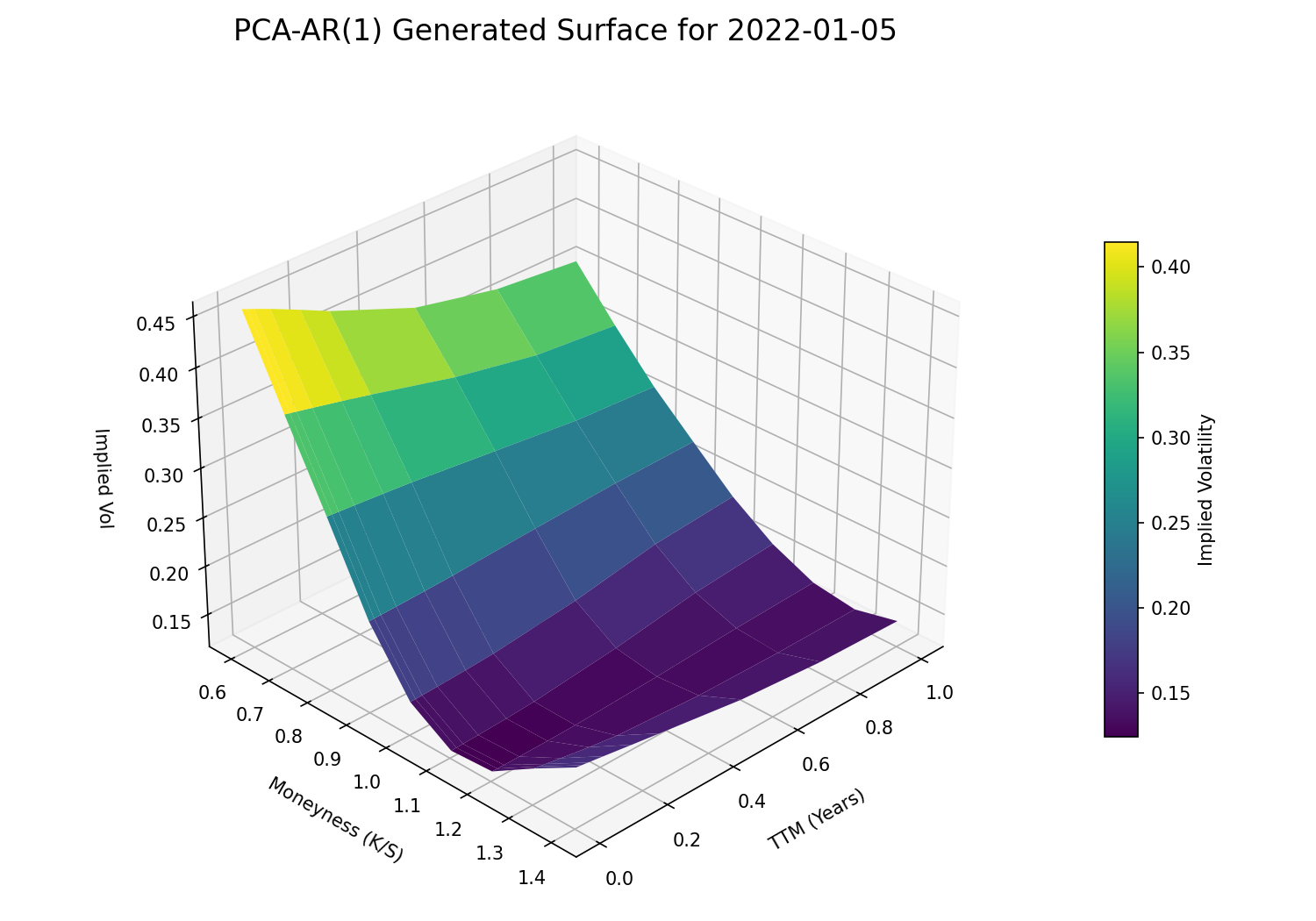}
        \caption{PCA-AR(1).}
        \label{fig:pcaar_surf}
    \end{subfigure}
    \hfill
    \begin{minipage}[b]{0.32\textwidth}\end{minipage} 

    \caption{Qualitative comparison of generated vol-surfaces across all benchmark models for a single test-set date (2022-01-05). The (a) real vol-surface shows a smooth smirk, with implied volatility highest at low moneyness and short maturity. This shape and magnitude are closely replicated by our (b) diffusion model. The (c) VolGAN surface captures a comparable overall shape and peak level but exhibits a visible ``crease'' at the short-term, low-moneyness corner. The (d) VolGAN with $\mathcal{Y}^{(k)}$ surface preserves the smirk shape but noticeably underestimates the peak implied volatility relative to the real surface. The (e) VAR(1) and (f) SABR surfaces both closely track the shape and range of the real surface, while (g) LSTM+Attention, like (d), underestimates the peak implied volatility level. The (h) PCA-AR(1) surface closely tracks the shape and level of the real surface.}

\label{fig:qualitative_comparison}
\end{figure}

For evaluating model performance based on the market data, we generate 100 vol-surfaces for each trading day and compare the quality of the generated vol-surfaces. Point accuracy is measured by the mean absolute percentage error of the mean forecast; for trading day $k$, $\text{MAPE}^{(k)} := \frac{1}{N_m N_\tau} \sum_{i=1}^{N_m} \sum_{j=1}^{N_\tau} \big| (\hat{\sigma}_{i,j}^{(k)} - \sigma_{i,j}^{(k)}) / \sigma_{i,j}^{(k)} \big|$, and the overall MAPE is its average over the test set. Our main comparison benchmark is the original VolGAN model and we first report those comparison results in detail. Since GAN-based methodologies suffer from well-known instability issues, we select the most accurate VolGAN model among ten different trained models based on the test MAPE of the vol-surface and use it to compare with the conditional diffusion model. This choice can only favour the benchmark since we report the best of ten VolGAN runs as if the best run had been known in advance, so it cannot inflate the advantage of the diffusion model. Note that we also trained ten different diffusion models and randomly picked one for performance comparison. With the best trained model based on validation MAPE, the diffusion model performance is even stronger than reported in the tables below. If both models are instead selected by validation MAPE, VolGAN run 6 and diffusion seed 9 are chosen, with test MAPEs of 4.2254\% and 2.9748\%, respectively. Table~\ref{tab:seed_summary} reports means and standard deviations over all ten runs of each model: the overall MAPE of the diffusion model is $2.996\% \pm 0.055\%$, against $4.790\% \pm 1.541\%$ for VolGAN, and every diffusion seed outperforms the best VolGAN run. The model reported in the main text (3.0026\%) is close to the ten-seed mean and therefore representative. Each VolGAN is trained until a pre-determined hard-stop of 10,000 epochs as done originally by its authors and the model is included in the comparison if no exploding or vanishing gradients are observed and the gradient norm oscillation is stable at the end of the training. VolGAN 9 did not meet this criterion and Table~\ref{tab:seed_summary} reports statistics with and without it. Figure \ref{fig:qualitative_comparison} presents a side-by-side comparison for an arbitrarily chosen date (January 5, 2022) from the test set only for illustration purposes. It displays the real vol-surface and the mean vol-surface generated by our conditional diffusion model, together with the forecasts of all six benchmark models.

\begin{figure}[htbp]
\centering
\includegraphics[width=0.75\textwidth]{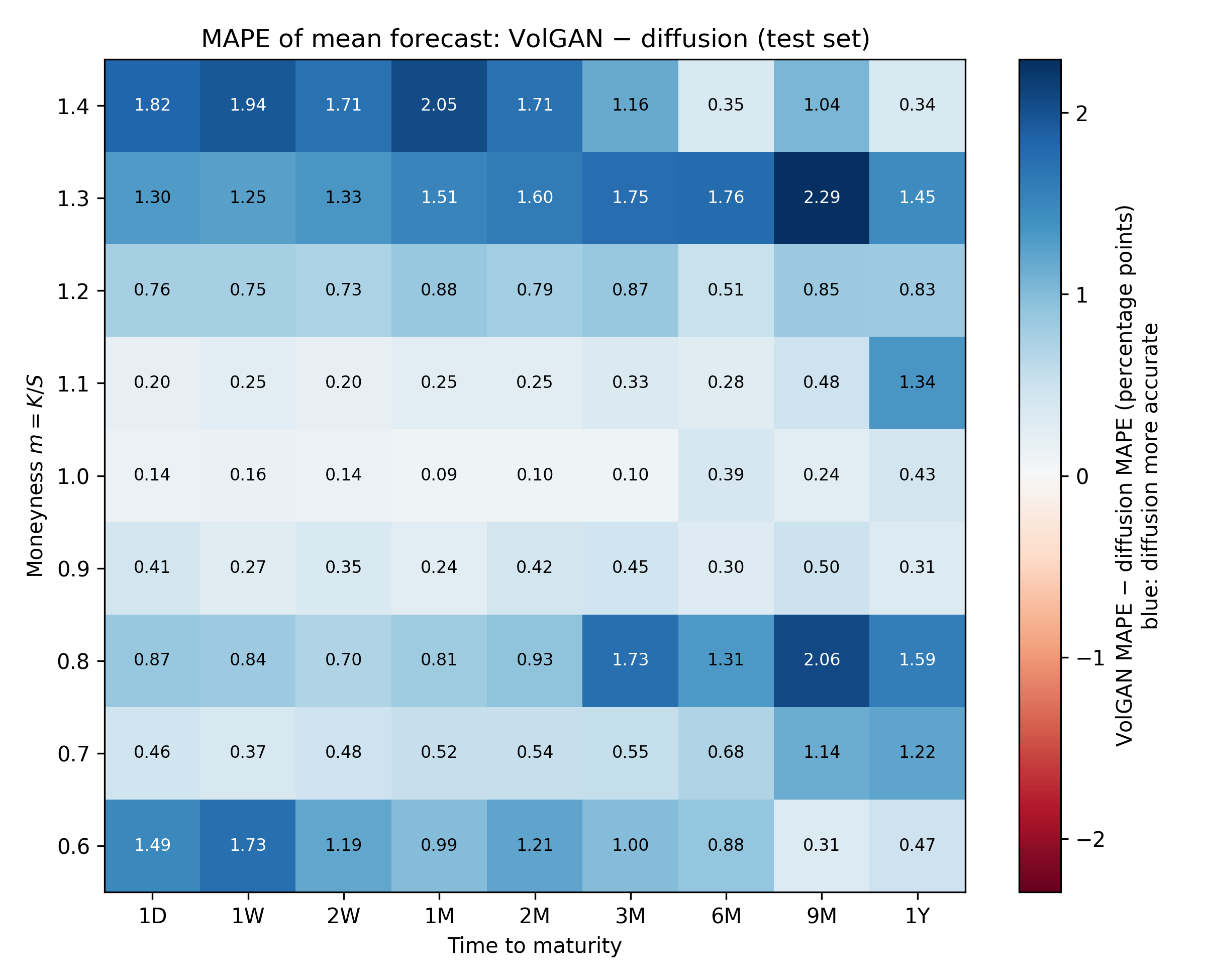}
\caption{Difference in mean absolute percentage error of the mean forecast between VolGAN and the diffusion model at each of the $9 \times 9$ grid points over the test period (VolGAN MAPE minus diffusion MAPE, in percentage points). Rows are moneyness levels and columns are maturities; blue cells indicate grid points where the diffusion model is more accurate and red cells where VolGAN is more accurate.}
\label{fig:error_heatmaps}
\end{figure}

The diffusion model's generated surface closely reproduces both the overall smirk shape and the magnitude of the real surface, with no visually apparent distortion at any region of the surface. By contrast, several benchmarks exhibit visible discrepancies. VolGAN reproduces the general smirk shape and peak magnitude but shows a localized ``crease'' at the short-term, low-moneyness corner of the surface, suggesting difficulty modelling the joint curvature in this region. VolGAN conditioned on $\mathcal{Y}^{(k)}$ and LSTM-Attention both preserve the qualitative smirk shape but visibly underestimate the peak implied volatility relative to the real surface. VAR(1) and SABR both track the shape and magnitude of the real surface closely. To show that these features are systematic rather than specific to this date, Figure~\ref{fig:error_heatmaps} compares the mean absolute percentage error of VolGAN and the diffusion model at every grid point over the whole test period. The diffusion model is more accurate than VolGAN at all 81 grid points. The difference is small near the money ($m = 0.9$--$1.1$), mostly below 0.5 percentage points, where both models forecast the liquid part of the surface well, and is largest in the wings of the smile: up to 2.3 percentage points in the OTM region ($m = 1.3$--$1.4$), between 1.0 and 1.7 percentage points at low moneyness and short maturities ($m = 0.6$, up to two months).

\begin{figure}[h!]
    \centering
    \caption{Time series comparison for ATM points. Left: Diffusion Model vs. Real. Right: VolGAN vs. Real.}
    \label{fig:atm_slices}

    \begin{subfigure}[b]{0.49\textwidth}
        \centering
        \includegraphics[width=\textwidth]{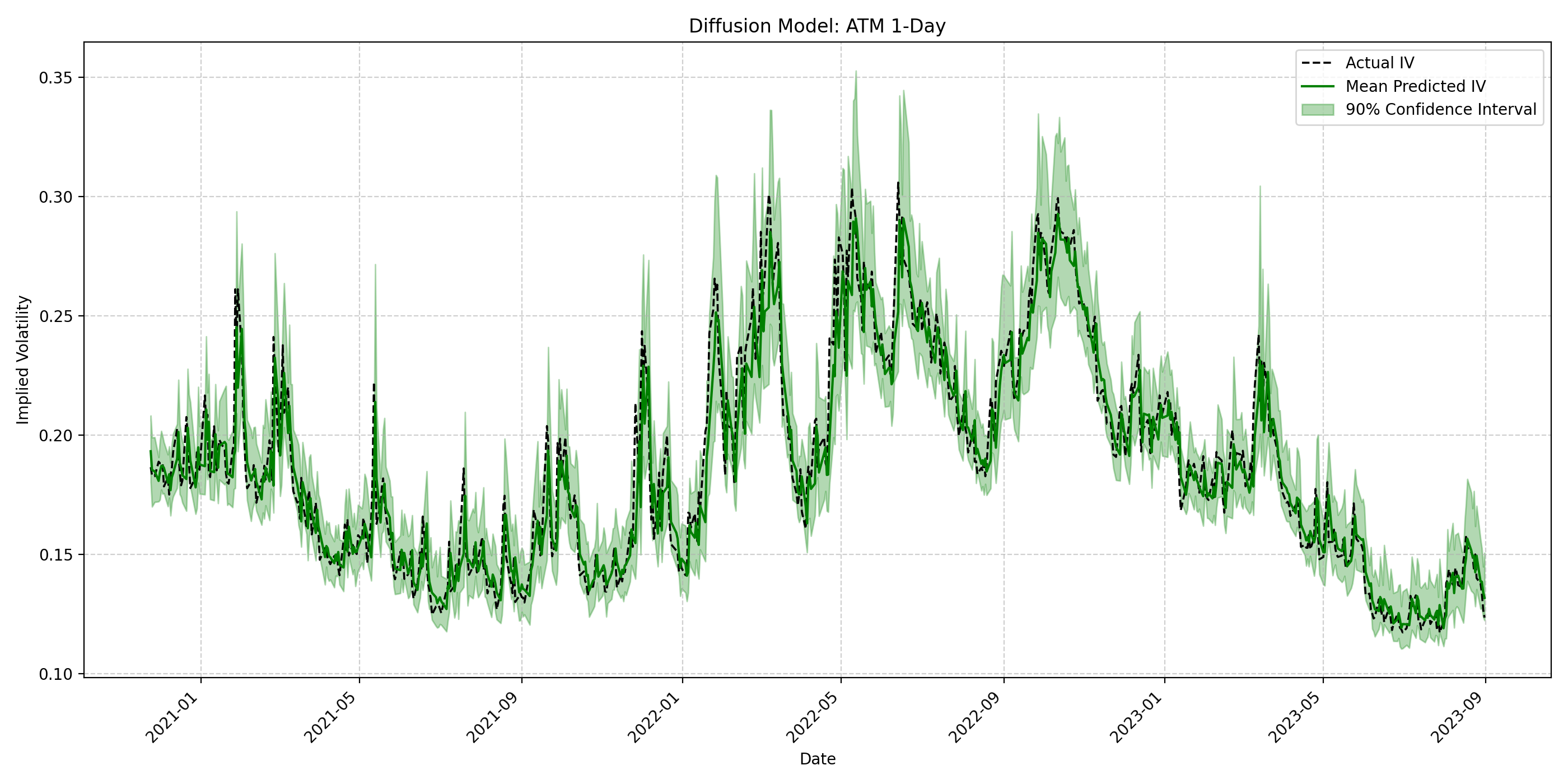}
        \caption{Diffusion Model (ATM 1-Day).}
    \end{subfigure}
    \hfill
    \begin{subfigure}[b]{0.49\textwidth}
        \centering
        \includegraphics[width=\textwidth]{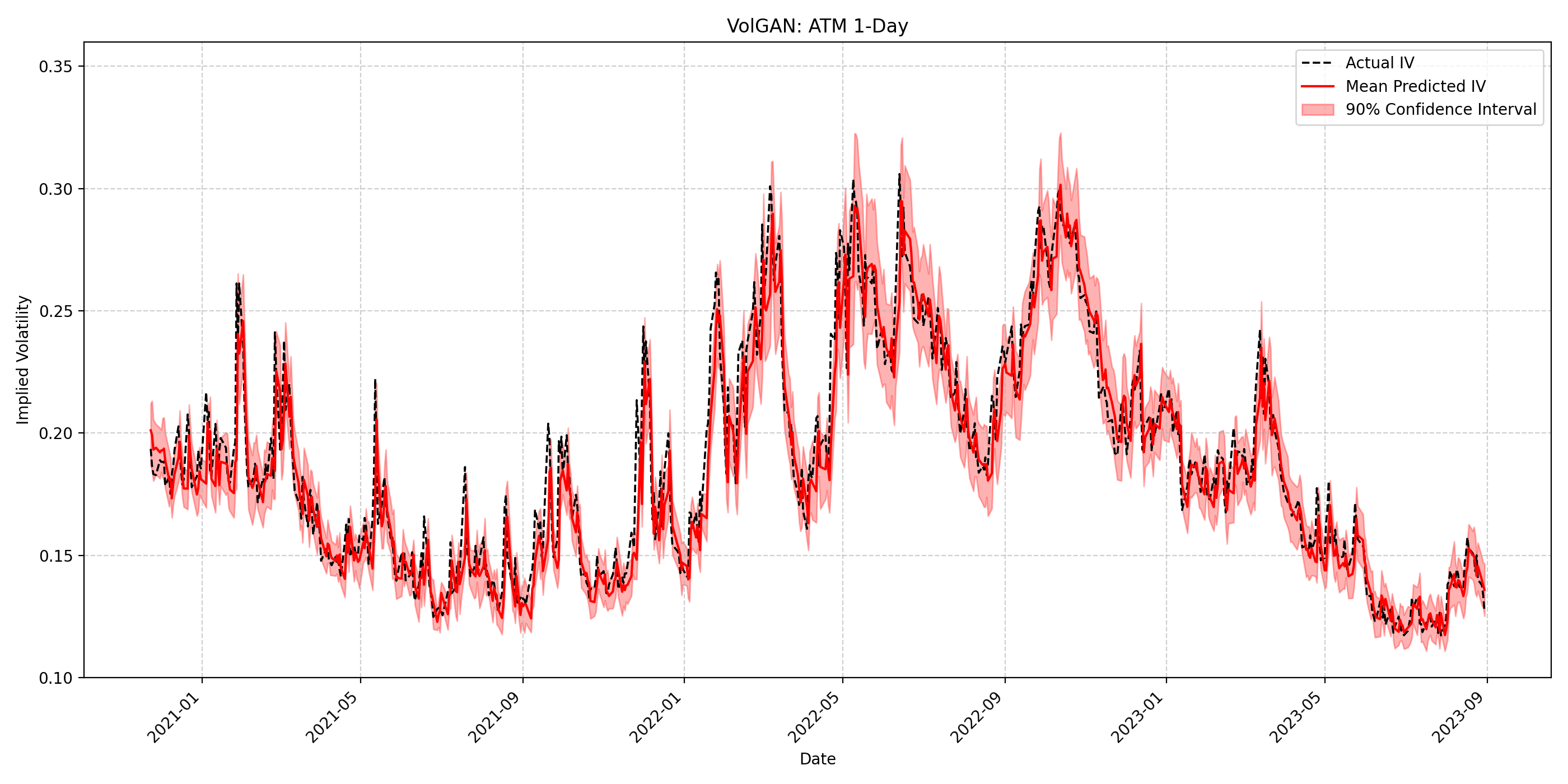}
        \caption{VolGAN Model (ATM 1-Day).}
    \end{subfigure}

    \vspace{0.5cm}

    \begin{subfigure}[b]{0.49\textwidth}
        \centering
        \includegraphics[width=\textwidth]{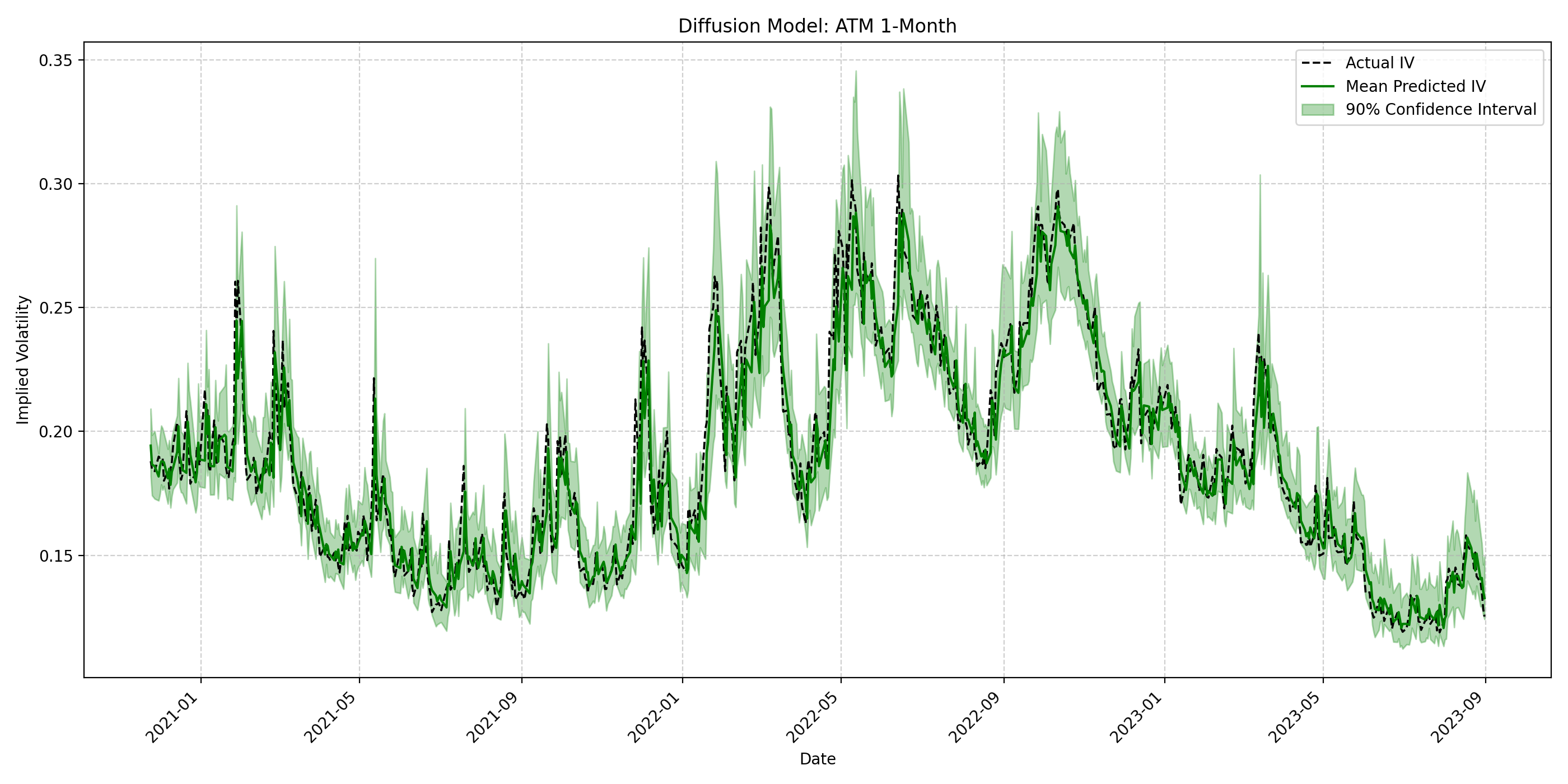}
        \caption{Diffusion Model (ATM 1-Month).}
    \end{subfigure}
    \hfill
    \begin{subfigure}[b]{0.49\textwidth}
        \centering
        \includegraphics[width=\textwidth]{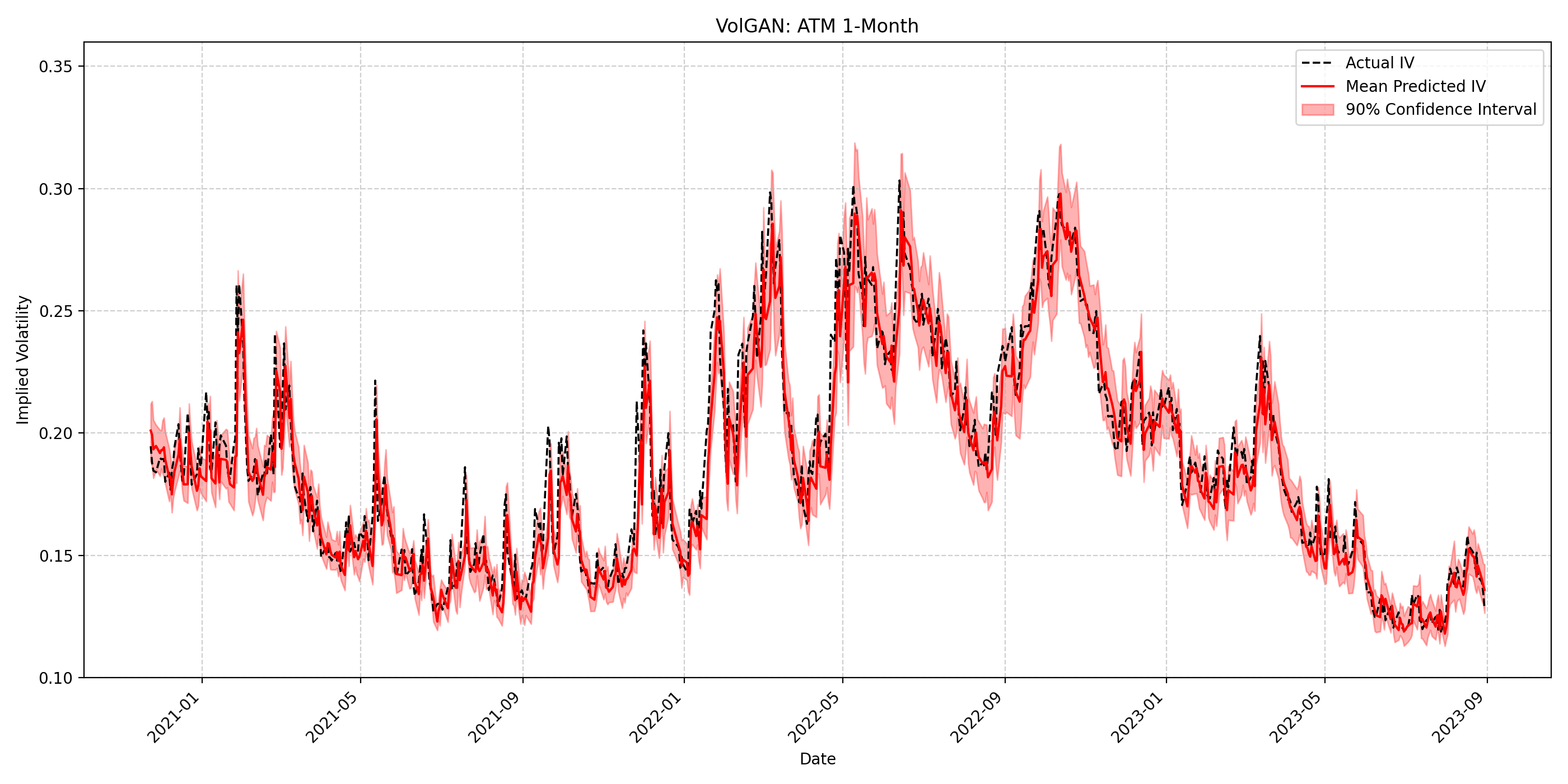}
        \caption{VolGAN Model (ATM 1-Month).}
    \end{subfigure}
    
    \vspace{0.5cm}

    \begin{subfigure}[b]{0.49\textwidth}
        \centering
        \includegraphics[width=\textwidth]{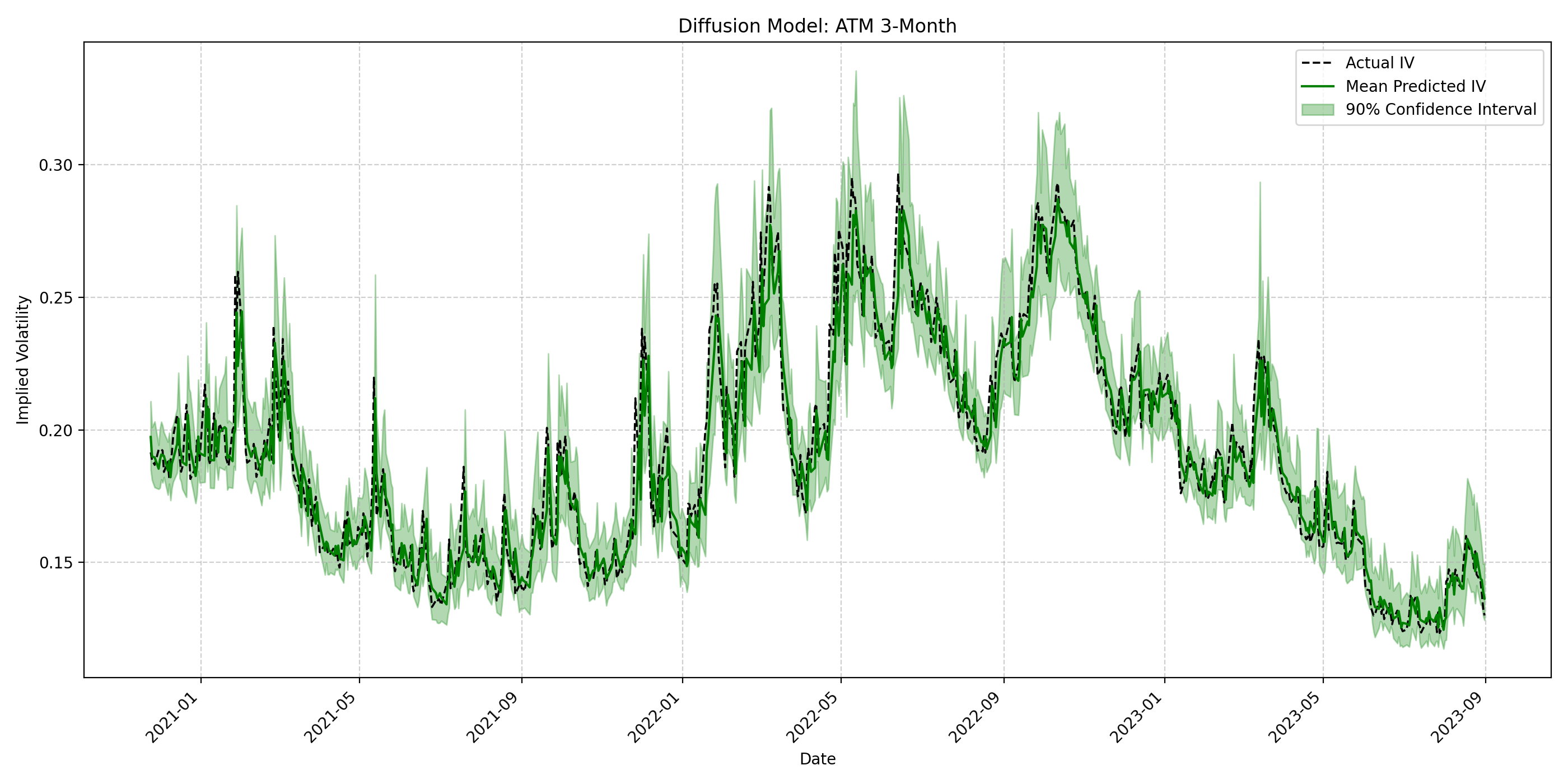}
        \caption{Diffusion Model (ATM 3-Month).}
    \end{subfigure}
    \hfill
    \begin{subfigure}[b]{0.49\textwidth}
        \centering
        \includegraphics[width=\textwidth]{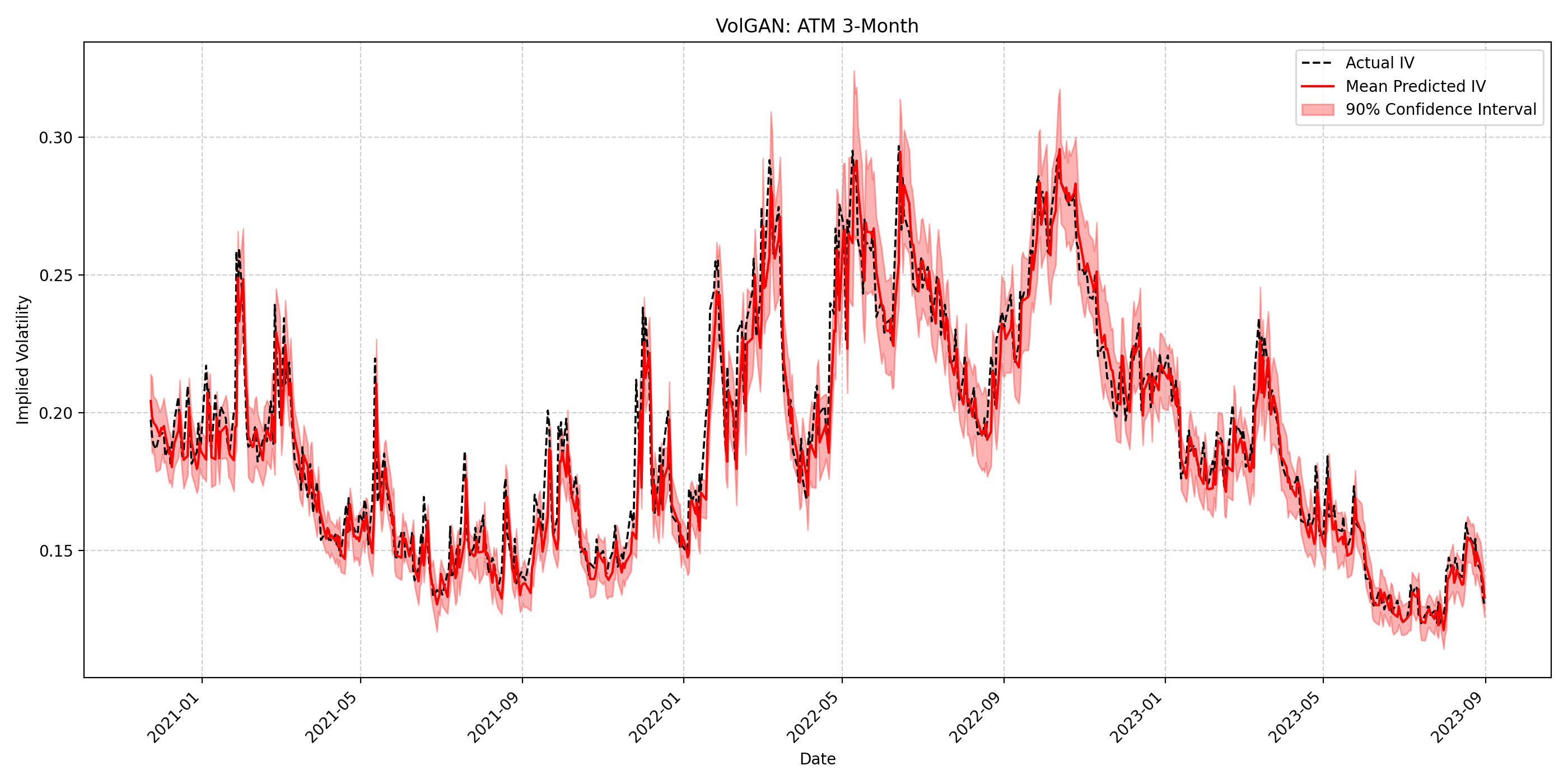}
        \caption{VolGAN Model (ATM 3-Month).}
    \end{subfigure}
\end{figure}

\begin{figure}[h!]
    \centering
    \caption{Time series comparison for ITM points. Left: Diffusion Model vs. Real. Right: VolGAN vs. Real.}
    \label{fig:itm_slices}

    \begin{subfigure}[b]{0.49\textwidth}
        \centering
        \includegraphics[width=\textwidth]{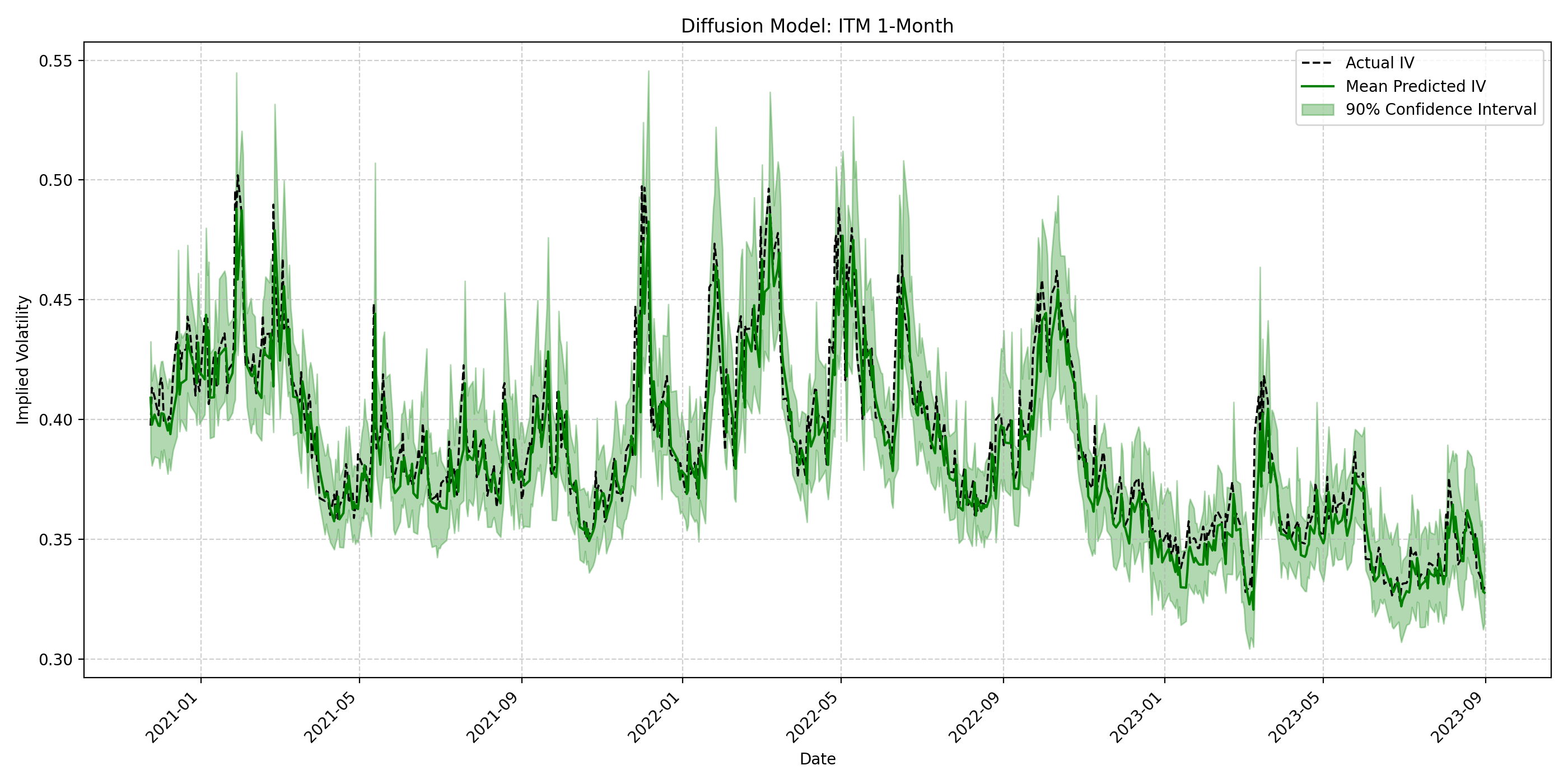}
        \caption{Diffusion Model (ITM 1-Month).}
    \end{subfigure}
    \hfill
    \begin{subfigure}[b]{0.49\textwidth}
        \centering
        \includegraphics[width=\textwidth]{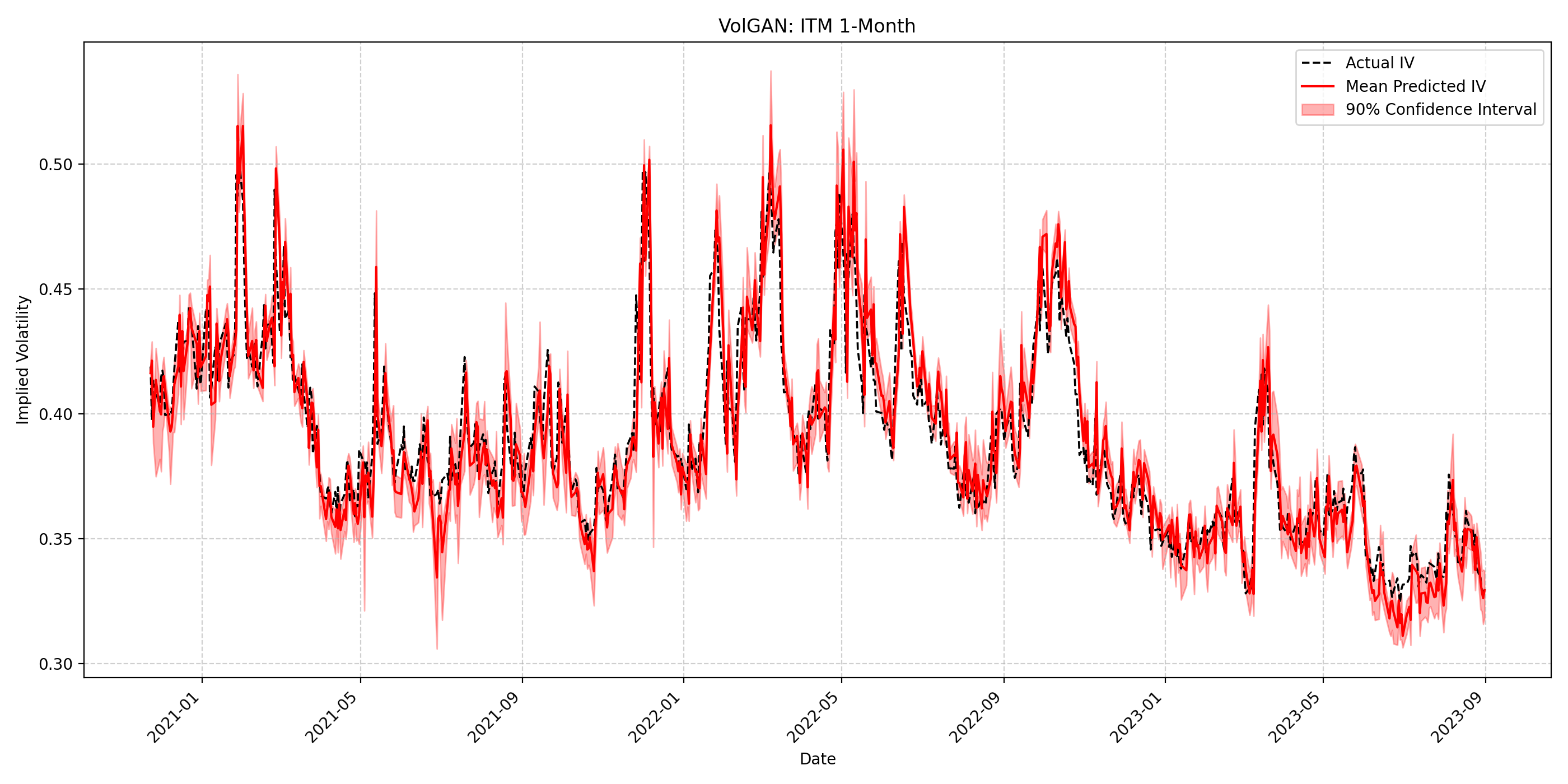}
        \caption{VolGAN Model (ITM 1-Month).}
    \end{subfigure}
    
    \vspace{0.5cm}

    \begin{subfigure}[b]{0.49\textwidth}
        \centering
        \includegraphics[width=\textwidth]{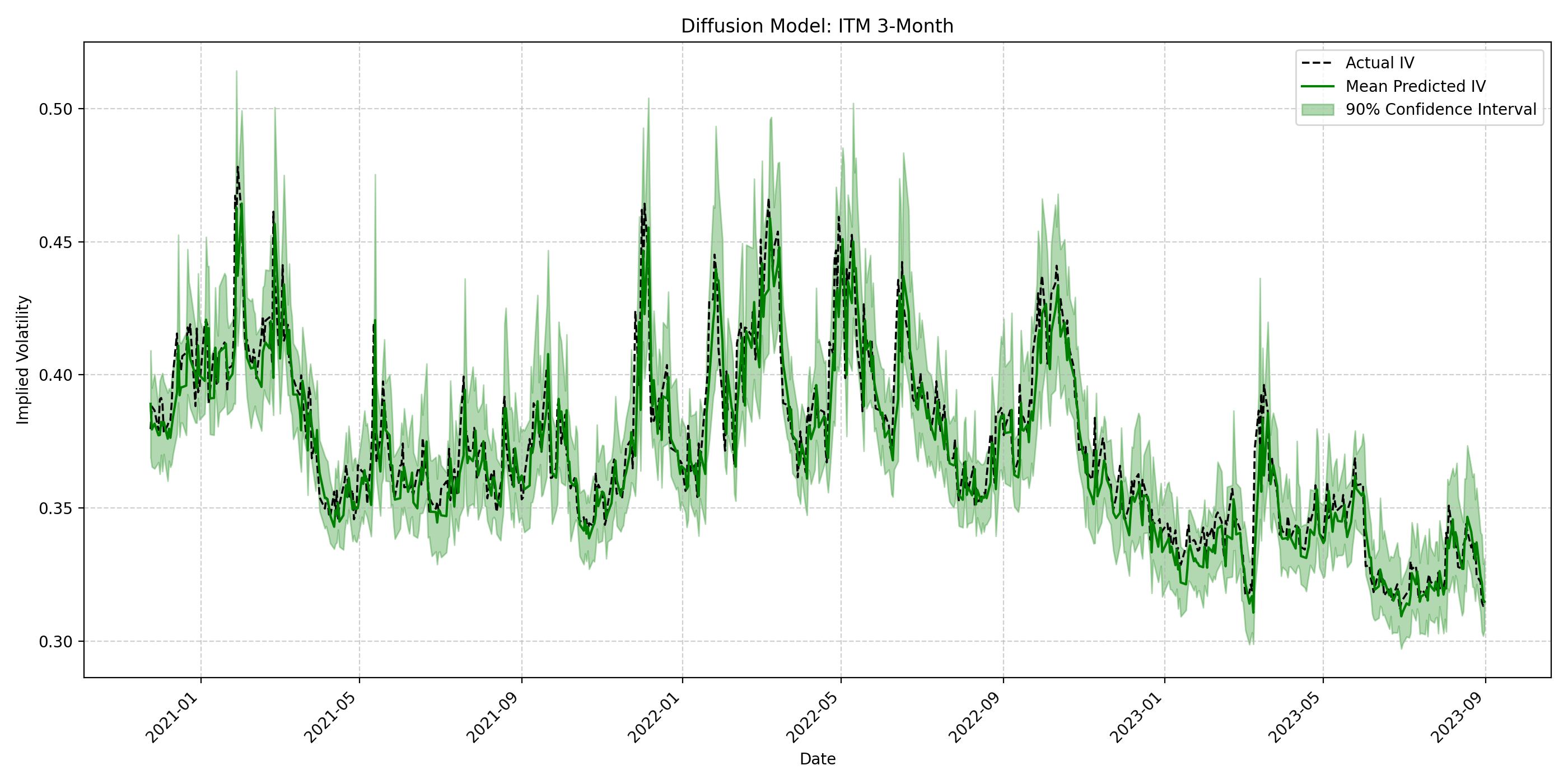}
        \caption{Diffusion Model (ITM 3-Month).}
    \end{subfigure}
    \hfill
    \begin{subfigure}[b]{0.49\textwidth}
        \centering
        \includegraphics[width=\textwidth]{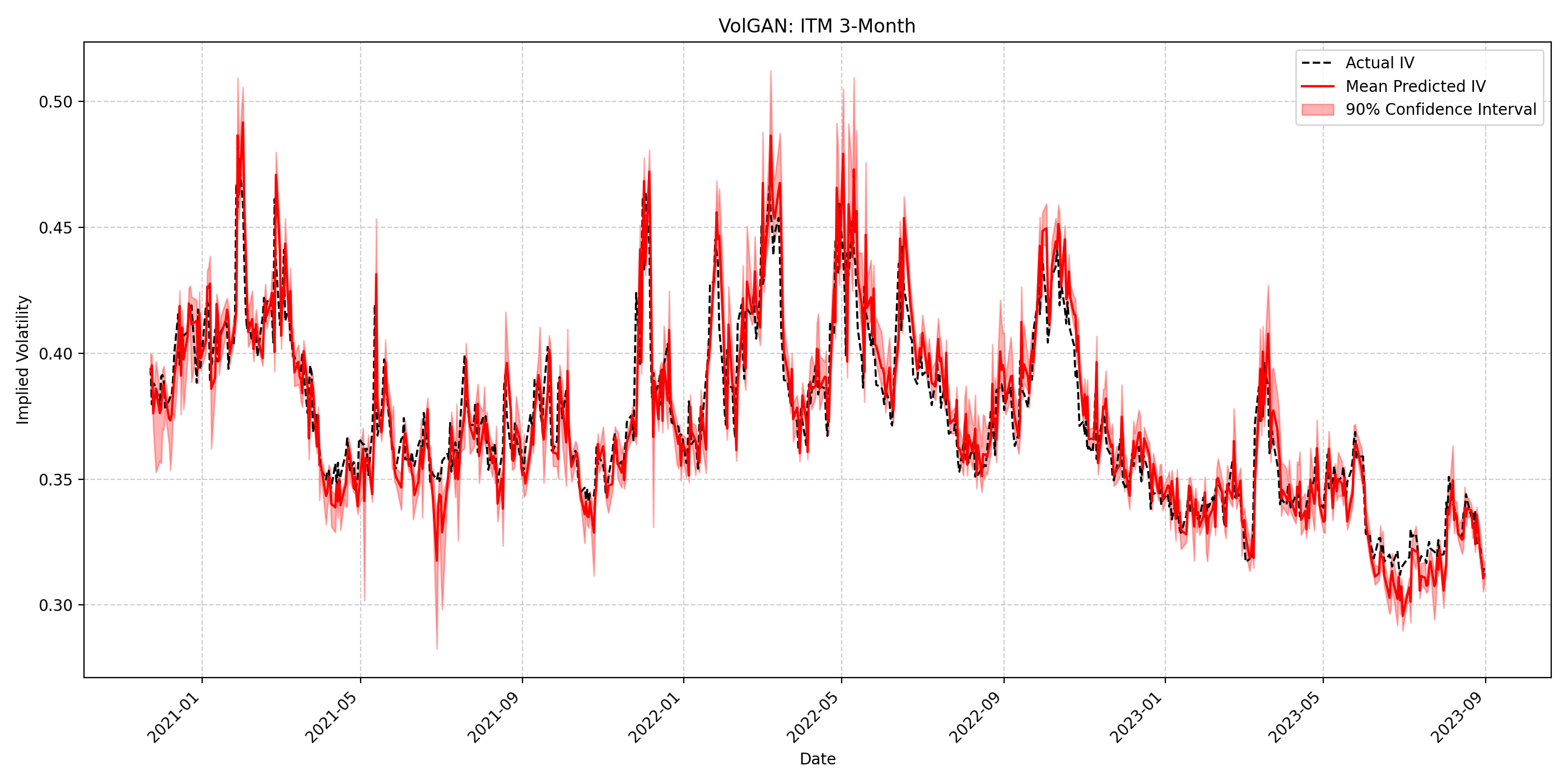}
        \caption{VolGAN Model (ITM 3-Month).}
    \end{subfigure}
\end{figure}

\begin{figure}[h!]
    \centering
    \caption{Time series comparison for OTM points. Left: Diffusion Model vs. Real. Right: VolGAN vs. Real.}
    \label{fig:otm_slices}

    \begin{subfigure}[b]{0.49\textwidth}
        \centering
        \includegraphics[width=\textwidth]{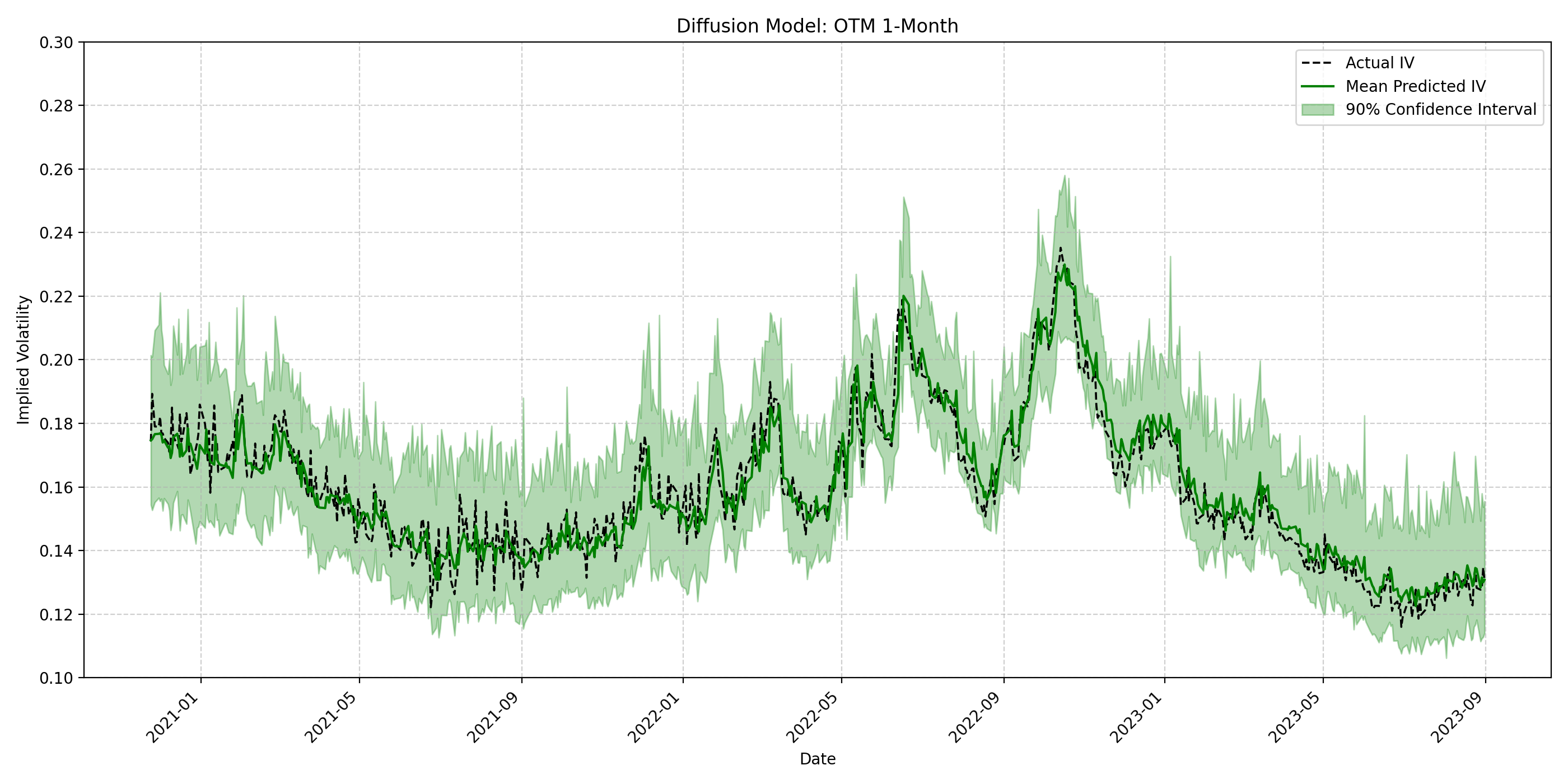}
        \caption{Diffusion Model (OTM 1-Month).}
    \end{subfigure}
    \hfill
    \begin{subfigure}[b]{0.49\textwidth}
        \centering
        \includegraphics[width=\textwidth]{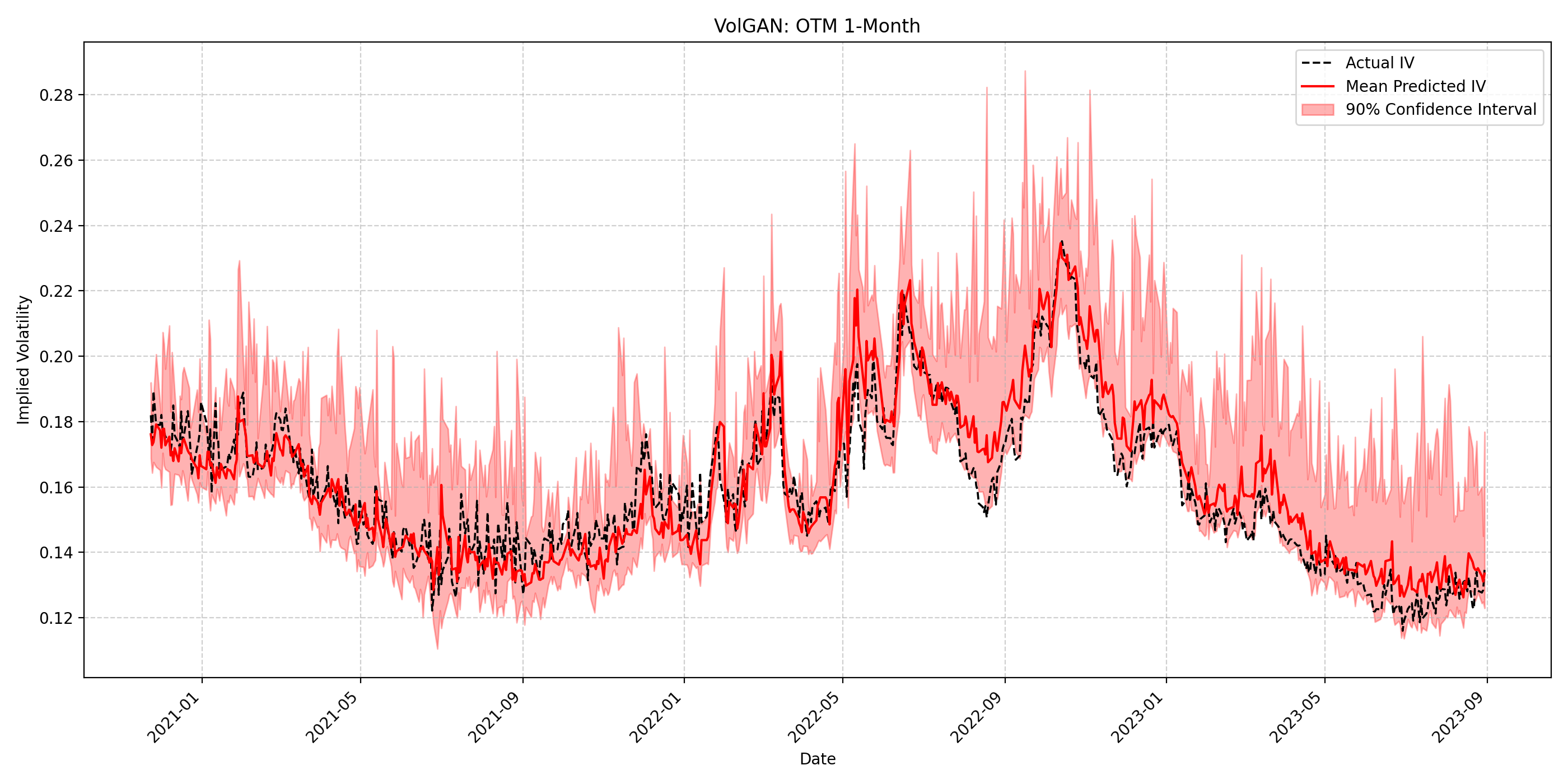}
        \caption{VolGAN Model (OTM 1-Month).}
    \end{subfigure}
    
    \vspace{0.5cm}

    \begin{subfigure}[b]{0.49\textwidth}
        \centering
        \includegraphics[width=\textwidth]{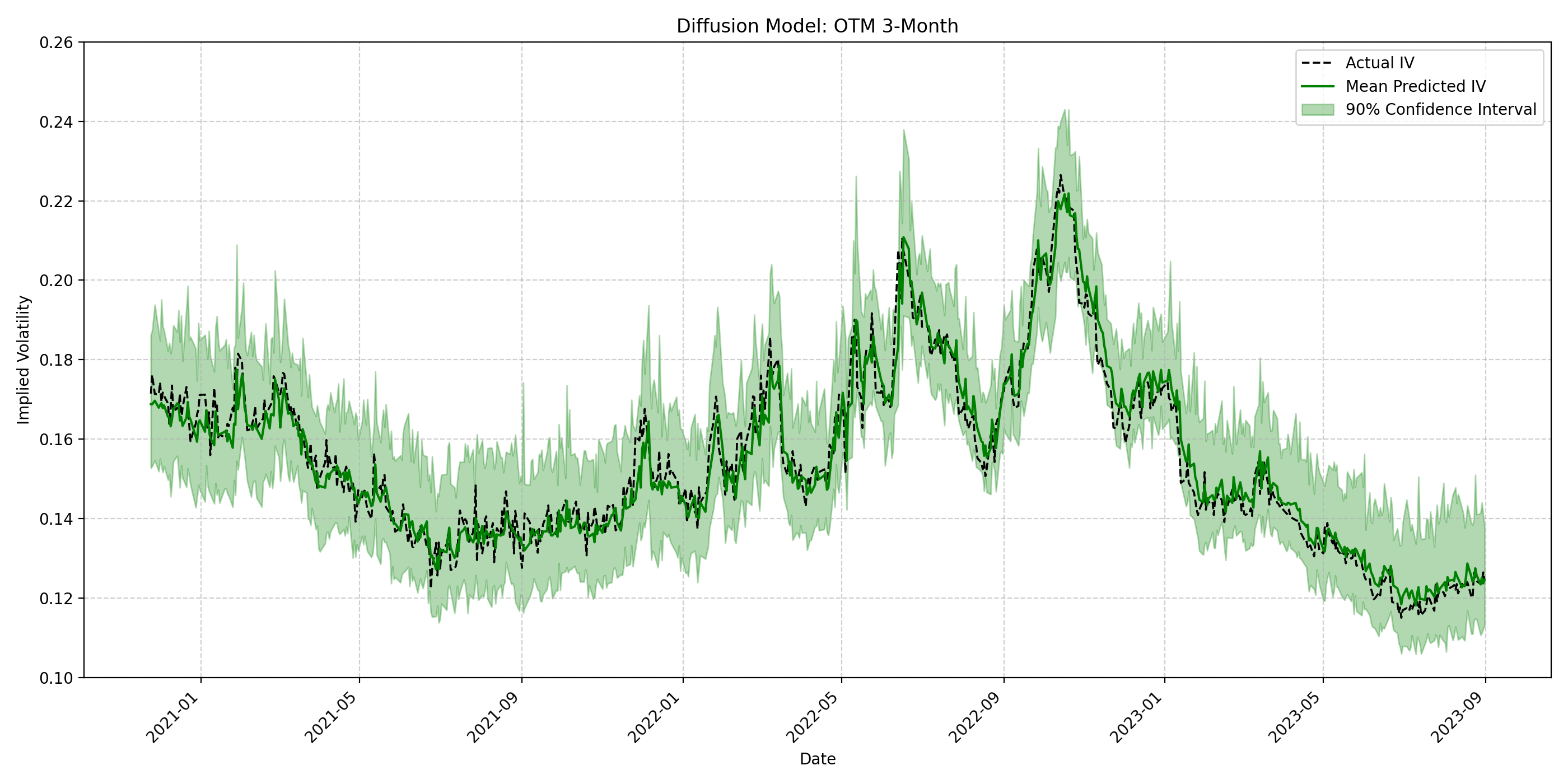}
        \caption{Diffusion Model (OTM 3-Month).}
    \end{subfigure}
    \hfill
    \begin{subfigure}[b]{0.49\textwidth}
        \centering
        \includegraphics[width=\textwidth]{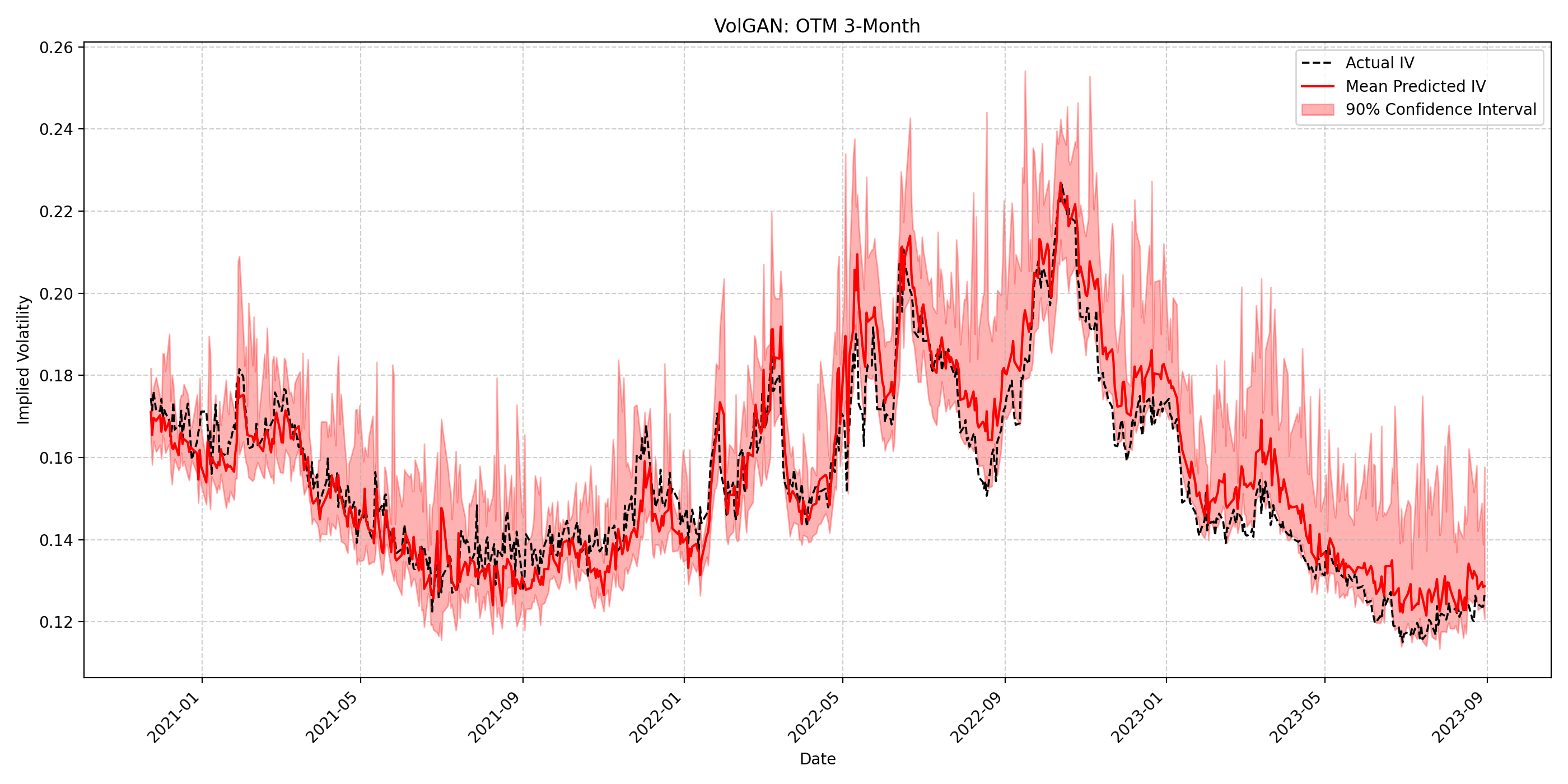}
        \caption{VolGAN Model (OTM 3-Month).}
    \end{subfigure}
\end{figure}

We perform further tests to evaluate the forecasting performance of the diffusion model. For each trading day in the test set, we compute a mean forecast and a 90\% confidence interval (CI). The lower and upper bounds are defined by the 5th and 95th percentiles of the samples, respectively. Figures \ref{fig:atm_slices}, \ref{fig:itm_slices}, and \ref{fig:otm_slices} plot these time series results for seven representative points on the moneyness/time-to-maturity grid: ATM 1-Day, 1-Month, and 3-Month options, ITM 1-Month and 3-Month options, and OTM 1-Month and 3-Month options. For each point, we compare the forecast between the diffusion model (left panel) and the VolGAN model (right panel). Both plots overlay the mean prediction and 90\% CI against the same real data. 

\begin{table}[h!]
\centering
\begin{threeparttable}
\caption{Comparison of Model Performance Metrics}
\label{tab:performance_metrics}
\footnotesize\setlength{\tabcolsep}{2.5pt}
\begin{tabular}{l rrrrrrr}
\toprule
Metric & Diffusion & VolGAN & SABR & VAR(1) & \shortstack{LSTM+\\Attention} & \shortstack{VolGAN (with\\$\mathcal{Y}^{(k)}$)}  & PCA-AR(1) \\
\midrule
Overall RMSE & \textbf{1.0814} & 1.2983 & 1.3278 & \underline{1.1710} & 1.3310 & 1.3658 & 1.2739  \\
Overall MAPE & \textbf{3.0026} & 3.8425 & 4.3108 & \underline{3.3231} & 4.3582 & 4.1152 & 3.9048  \\
ATM 1M MAPE & \textbf{4.5146} & 4.6083 & 6.0174 & 4.5400 & 4.9945 & \underline{4.5247} & 5.0731  \\
ITM 1M MAPE & \underline{2.4184} & 2.9407 & 2.7894 & \textbf{2.3819} & 3.1470 & 3.8690 & 2.4870 \\
OTM 1M MAPE & \textbf{3.1542} & \underline{4.6687} & 4.8501 & 5.0063 & 5.2559 & 6.9765 & 6.1273 \\
ATM 1M CI Breach \% & \textbf{11.0632} & \underline{26.7241} & N/A & N/A & N/A & 29.4540 & N/A \\
ITM 1M CI Breach \% & \textbf{8.4770} & 55.0287 & N/A & N/A & N/A & \underline{49.4253} & N/A  \\
OTM 1M CI Breach \% & \underline{0.5747} & \textbf{14.2241} & N/A & N/A & N/A & 28.5920 & N/A \\
\bottomrule
\end{tabular}
\begin{tablenotes}
\footnotesize
\item \textit{Note:} This table compares the forecasting errors across seven models. The best performing model for each metric across the row is highlighted in \textbf{bold}, and the second-best is \underline{underscored}. For the CI Breach \%, performance is ranked by absolute proximity to the theoretical 10\% target. Deterministic models (SABR, VAR(1), PCA-AR(1) and LSTM+Attention) do not produce probabilistic confidence intervals and are therefore marked as N/A. MAPE and breach rates are in \%; RMSE is in volatility points (\%); 1M denotes the one-month maturity.
\end{tablenotes}
\end{threeparttable}
\end{table}

Table \ref{tab:performance_metrics} summarizes the forecasting performance of our proposed diffusion model compared to the six benchmark models: PCA-AR(1), VolGAN, SABR, VAR(1), LSTM+Attention, and VolGAN with our conditioning set $\mathcal{Y}^{(k)}$. Our diffusion model demonstrates strong predictive accuracy, achieving the best overall MAPE of 3.0026\% and an overall RMSE of 1.0814, outperforming the second-best VAR(1) benchmark, which achieved a MAPE of 3.3231\% and an RMSE of 1.1710. This indicates a consistent ability to capture the dynamics of the vol-surface. When evaluating the 1-month predictions across different moneyness levels, the diffusion model remains highly competitive. It achieves the lowest MAPE for ATM points at 4.5146\% and outperforms all benchmarks in OTM forecasting with a MAPE of 3.1542\%. For in-the-money (ITM) points, only VAR(1) achieves a slightly lower error (2.3819\% versus 2.4184\%), and our model maintains strong and stable performance across the surface.

To evaluate the reliability of our uncertainty estimates, we measure the CI breach rate (CI Breach \%) for the generative models. This metric calculates the percentage of days in the test set where the realized vol-surface value fell outside the 90\% CI generated by the models. A perfectly calibrated model should target a 10\% breach rate.
Our diffusion model achieves stable and well-calibrated breach rates in the most liquid regions of the surface with 11.0632\% for 1-month ATM points and 8.4770\% for 1-month ITM points. The VolGAN and VolGAN with $\mathcal{Y}^{(k)}$ exhibit much higher breach rates in those regions, with ITM breach rates of 55.0287\% and 49.4253\%, respectively. This suggests that their generated confidence intervals are structurally biased, failing to capture the true market distribution.
For 1-month OTM points, the diffusion model produces a highly conservative breach rate of 0.5747\%. The mean forecast in the OTM region is most accurate in absolute terms (RMSE of 0.62 volatility points, against 1.25 at ATM and 1.41 at ITM; Table~\ref{tab:dispersion}), but the predictive distribution is too wide. The mean interval width is 6.6 times the RMSE at OTM, against 2.9 at ATM and 3.1 at ITM, whereas a calibrated Gaussian forecast would give about 3.3. The moments in Table~\ref{tab:moments} also show that the tails are heavier in the generated OTM distribution than in the data. The day-to-day variability of OTM implied volatilities and the share of days with 1-month OTM grid point being extrapolated are 5.64 and 1.77 times higher, respectively, in the training period than in the test period (Table~\ref{tab:shift} in Appendix~\ref{sec:otm_diagnostics}). Moreover, almost 80\% of the arbitrage penalty in the training set comes from the OTM points (Table~\ref{tab:shift}). OTM breach rates are 0.86\% on days when the grid point is interpolated and 0.29\% when it is extrapolated (Table~\ref{tab:otm_split}). The model therefore carries the dispersion of the noisier, sparsely quoted and ``arbitrage-heavy'' training period into a test period in which OTM quotes are more abundant. 
Coverage at other nominal levels is also reported in Table~\ref{tab:coverage_levels}. In the Heston model experiment of Appendix~\ref{subsec:heston_results}, where no quotes are missing and no smoothing is needed, breach rates are similar across ATM, ITM and OTM regions, which indicates that the over-dispersion stems from the data rather than from the model.

\begin{table}[htbp]
\centering
\caption{Dispersion diagnostic for the 1-month grid points of the diffusion model (test set). For a calibrated Gaussian predictive distribution, the 90\% interval width is about $2 \times 1.645 = 3.29$ times the RMSE.}
\label{tab:dispersion}
\small
\begin{tabular}{lcccc}
\toprule
1-month point & \shortstack{RMSE\\(vol.\ points)} & \shortstack{Mean 90\% CI width\\(vol.\ points)} & Width / RMSE & \shortstack{Breach\\rate (\%)} \\
\midrule
ATM & 1.246 & 3.573 & 2.87 & 11.06 \\
ITM & 1.407 & 4.378 & 3.11 & 8.48 \\
OTM & 0.623 & 4.123 & 6.62 & 0.57 \\
\bottomrule
\end{tabular}
\end{table}

\begin{figure}[h!]
    \centering
    \caption{Daily MAPE of the 100 generated surfaces against the smoothed real surface on the test set: median (dashed line) and 90\% confidence interval (shaded area).}
    \label{fig:mape_timeseries}
    \includegraphics[width=0.9\textwidth]{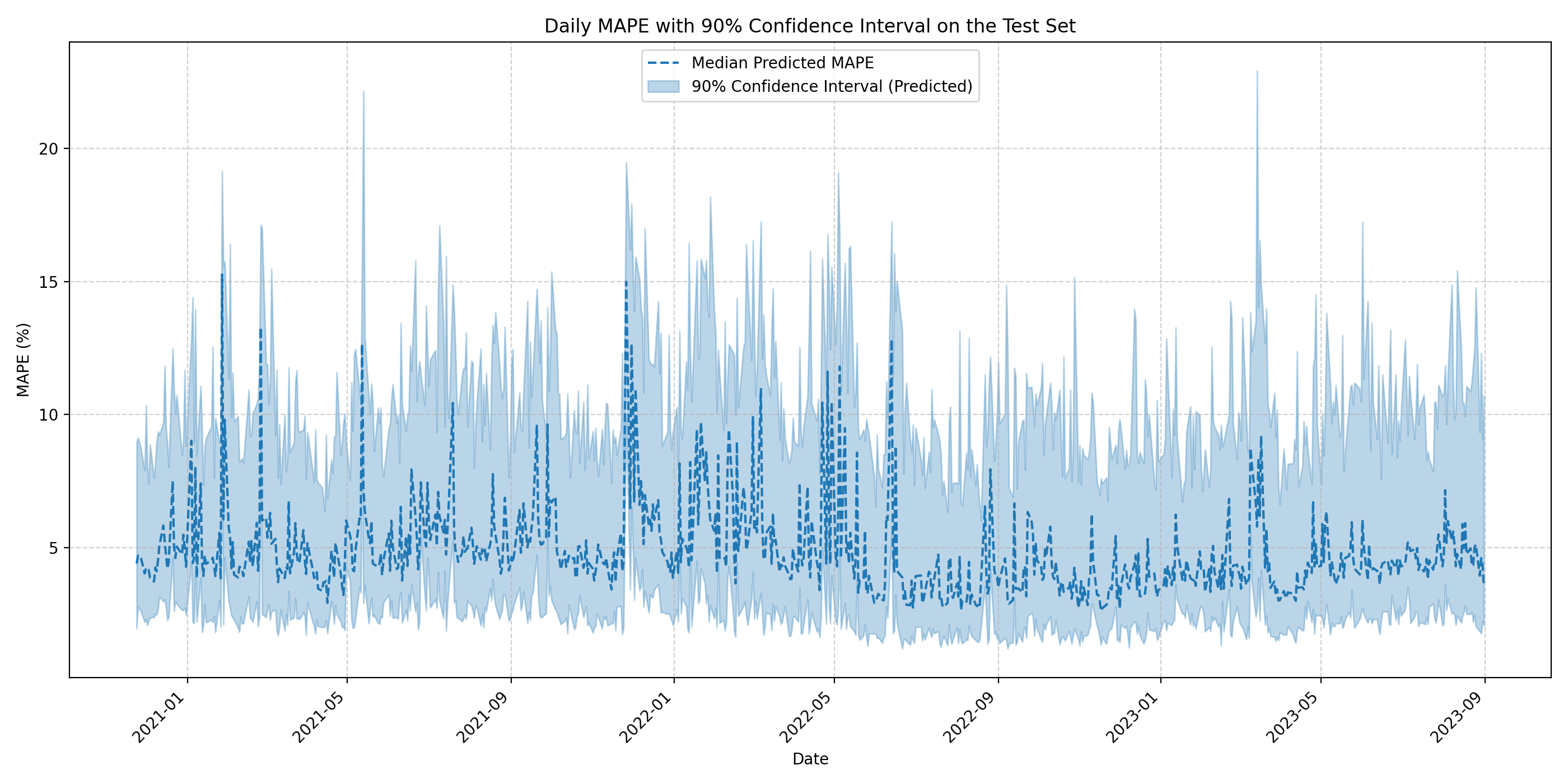}
\end{figure}

Beyond analyzing individual grid points, we evaluate the accuracy across the entire vol-surface. A robust model should not only be accurate at certain points but also maintain a low error across the vol-surface structure. To measure this, we calculate the daily MAPE for the predicted surface against the smoothed real surface for each trading day in the test set. Figure \ref{fig:mape_timeseries} plots this daily surface MAPE over the entire test period for our conditional diffusion model. The plot displays the median MAPE of the predicted surface (dashed line), calculated from the 100 generated samples for each trading day, as well as the 90\% CI (shaded area) of this prediction.
The error is consistently low for most of the time, with the median MAPE fluctuating in the 3\% to 7\% range for a long period. The error is not static with obvious fluctuations where it spikes to 10--15\% during periods of market turmoil.
The upper bound of the CI can be wide during these turmoil periods, indicating that some generated samples in the distribution have a higher error; however, the median forecast remains robust and tracks well within the low single-digit percentages.

\begin{figure}[h!]
    \centering
    \caption{Distributional Comparison for Selected vol-surface grid-points.}
    \label{fig:histograms}

    \begin{subfigure}[b]{0.49\textwidth}
        \centering
        \includegraphics[width=\textwidth]{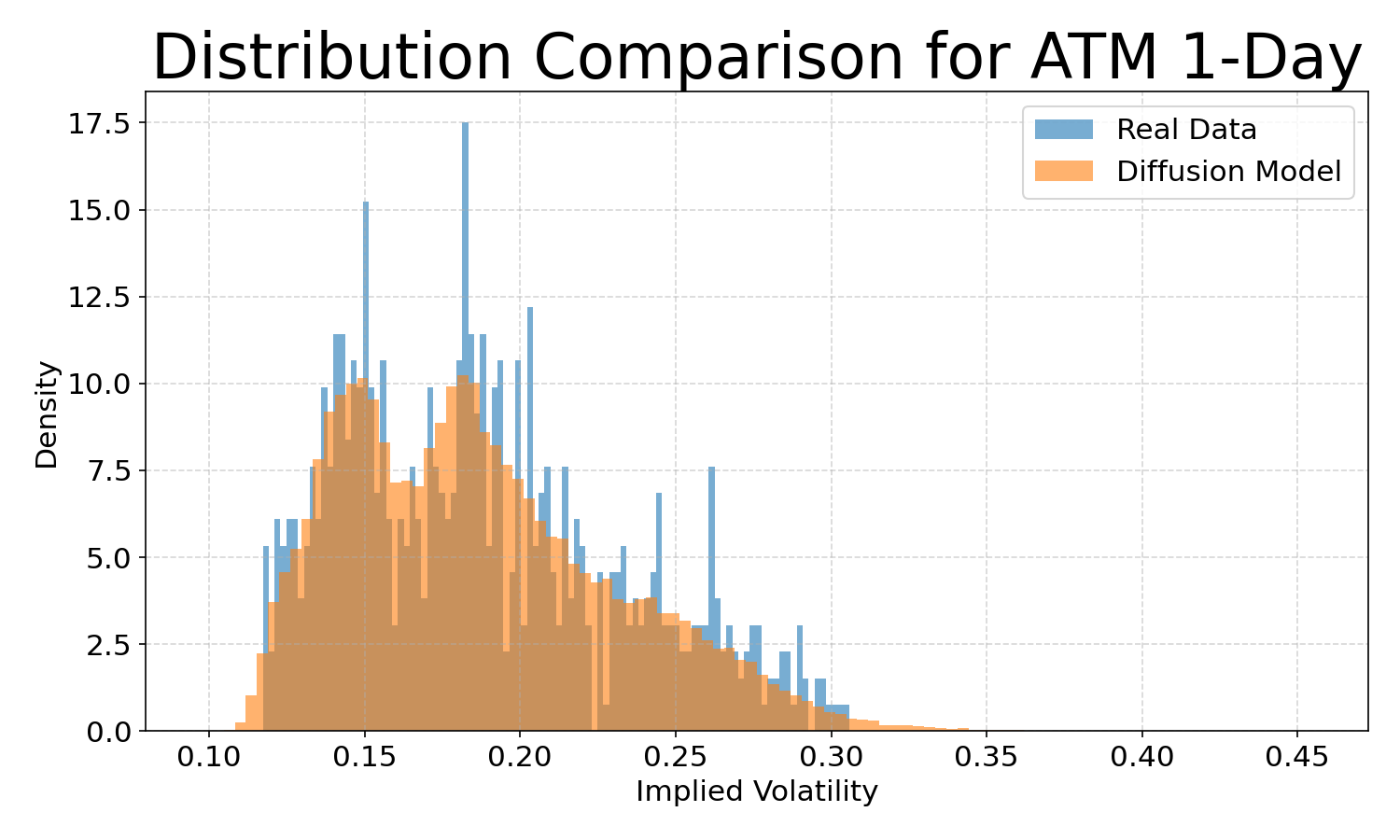}
        \caption{ATM 1-Day distribution histogram plot of real data and generated data.}
    \end{subfigure}
    \hfill
    \begin{subfigure}[b]{0.49\textwidth}
        \centering
        \includegraphics[width=\textwidth]{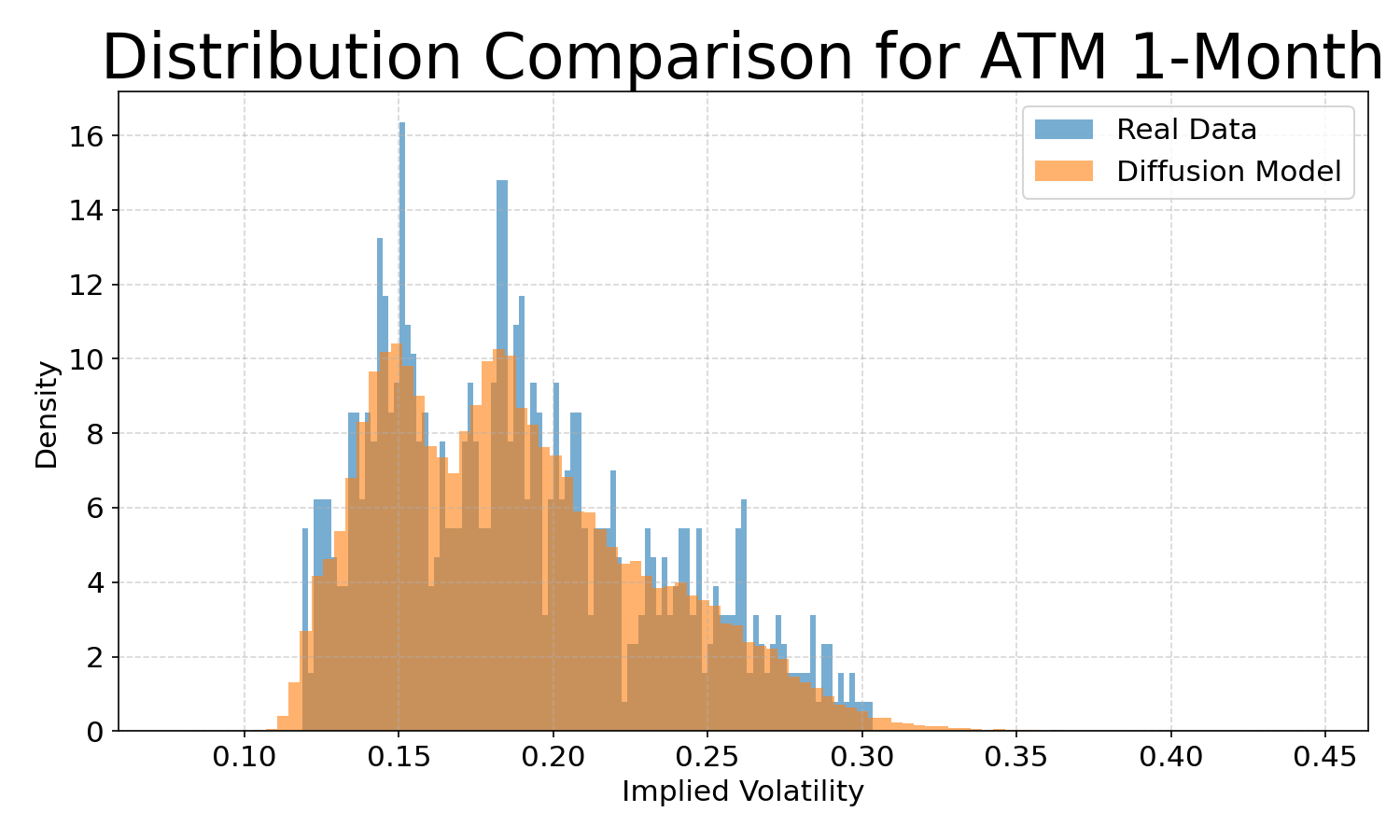}
        \caption{ATM 1-Month distribution histogram plot of real data and generated data.}
    \end{subfigure}

    \vspace{0.5cm}

    \begin{subfigure}[b]{0.49\textwidth}
        \centering
        \includegraphics[width=\textwidth]{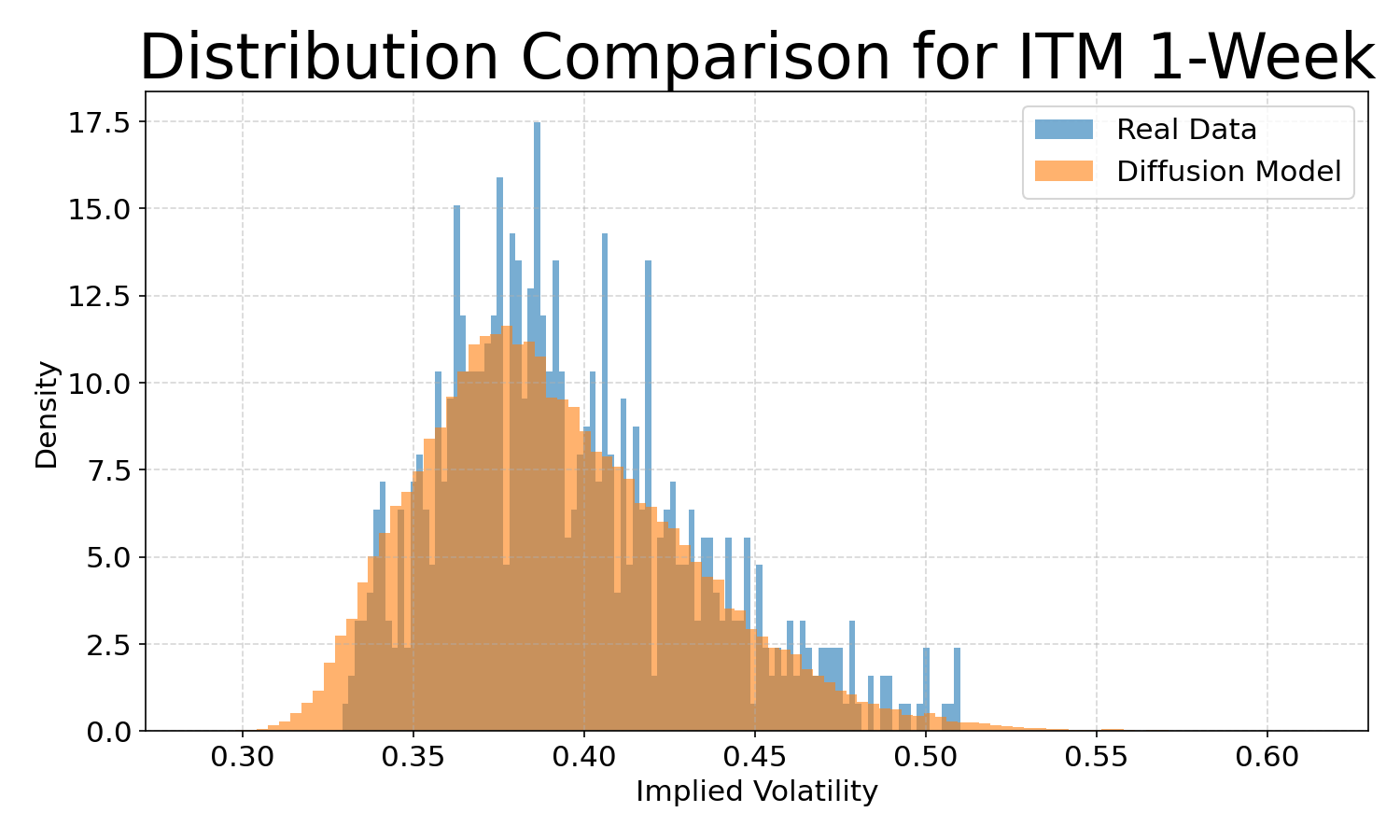}
        \caption{ITM 1-Week distribution histogram plot of real data and generated data.}
    \end{subfigure}
    \hfill
    \begin{subfigure}[b]{0.49\textwidth}
        \centering
        \includegraphics[width=\textwidth]{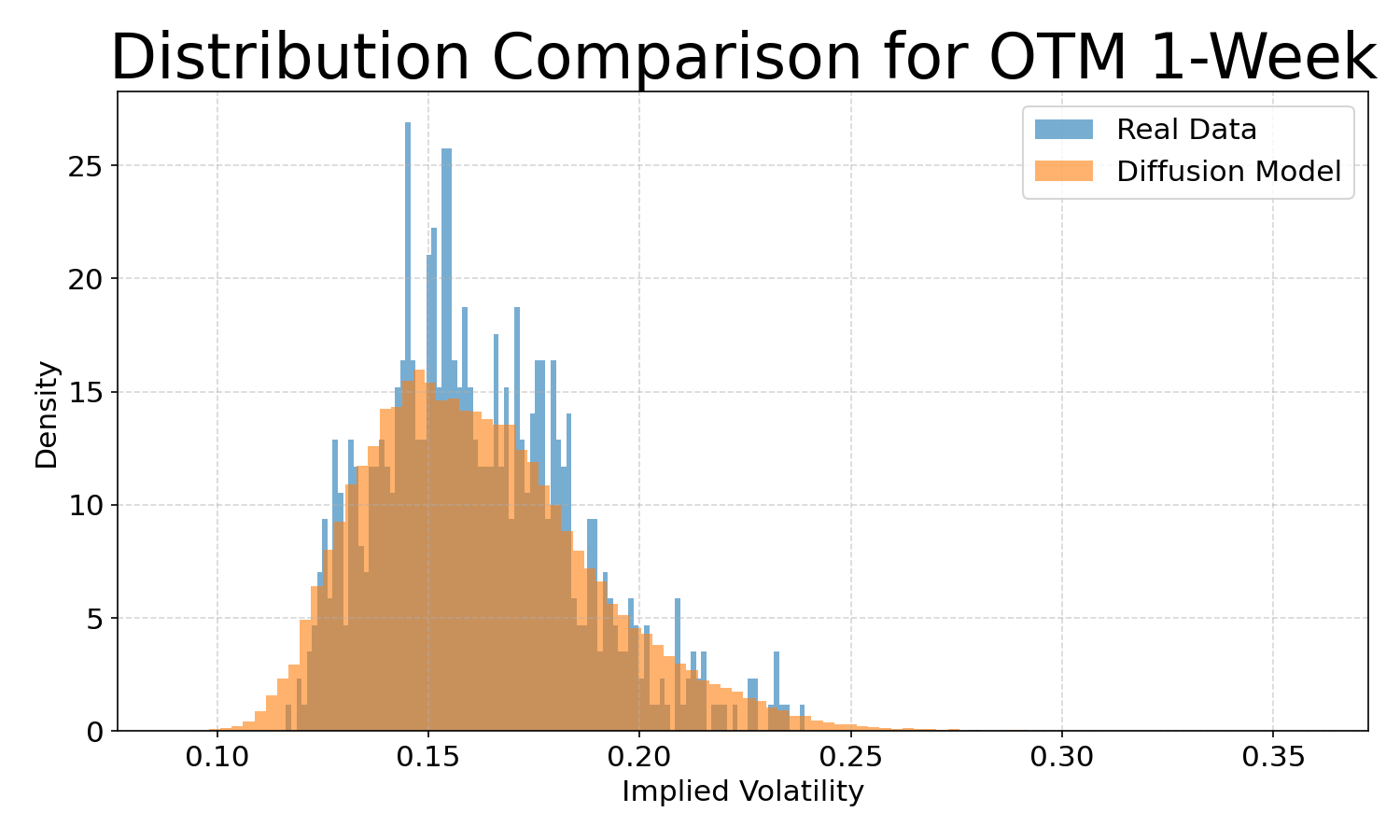}
        \caption{OTM 1-Week distribution histogram plot of real data and generated data.}
    \end{subfigure}

\end{figure}

Furthermore, a robust generative model should replicate the statistical distribution. To assess distributional fidelity, we compare the statistical moments of the generated data against the real data. The distribution of volatility is non-Gaussian and often exhibits significant skewness and fat tails. A successful model must capture these features. We analyze the distributions for several grid points by aggregating all 100 generated samples for each trading day in the test set. Table \ref{tab:moments} compares the first four moments (Mean, Standard Deviation, Skewness, and Kurtosis) for five selected grid points.
\begin{table}[h!]
    \centering
    \caption{Comparison of Statistical Moments for Selected Grid Points (Test Set).}
    \label{tab:moments}
    \resizebox{\textwidth}{!}{%
    \begin{tabular}{lrrrrrrrrrr}
        \toprule
        Metric & \multicolumn{2}{c}{ATM 1-Day} & \multicolumn{2}{c}{ATM 1-Month} & \multicolumn{2}{c}{OTM 1-Week} & \multicolumn{2}{c}{OTM 3-Month} & \multicolumn{2}{c}{ITM 1-Week} \\
        \cmidrule(lr){2-3} \cmidrule(lr){4-5} \cmidrule(lr){6-7} \cmidrule(lr){8-9} \cmidrule(lr){10-11}
        & Real Data & Diffusion & Real Data & Diffusion & Real Data & Diffusion & Real Data & Diffusion & Real Data & Diffusion \\
        \midrule
        Mean & 0.1884 & 0.1868 & 0.1891 & 0.1876 & 0.1616 & 0.1622 & 0.1546 & 0.1548 & 0.3960 & 0.3914 \\
        Std Dev & 0.0448 & 0.0441 & 0.0439 & 0.0431 & 0.0230 & 0.0272 & 0.0221 & 0.0241 & 0.0371 & 0.0385 \\
        Skewness & 0.5487 & 0.6993 & 0.5339 & 0.6651 & 0.6318 & 0.7804 & 0.6201 & 0.6723 & 0.6887 & 0.7034 \\
        Kurtosis (Fisher) & -0.5404 & 0.1520 & -0.5562 & 0.0679 & 0.2771 & 1.0518 & 0.1709 & 0.4273 & 0.0930 & 0.6775 \\
        \bottomrule
    \end{tabular}
    }

    \caption*{Comparison of the first four statistical moments between the real data distribution and the distribution of all generated samples from the diffusion model over the test set. Kurtosis is reported as Fisher's kurtosis (normal=0).}
\end{table}
The results show the first two moments are very closely matched, confirming the point-wise accuracy observed in the previous subsection. Furthermore, the model correctly captures the positive skewness present in all the real data points. The most notable discrepancy, however, appears in the kurtosis. The real data consistently exhibits smaller kurtosis, indicating a distribution with thinner tails and a more stable central peak. This is visually confirmed in Figure \ref{fig:histograms}, where the real data histograms in blue appear wider and more rectangular. In contrast, our diffusion model generates data with a larger kurtosis. This shows the model is producing fatter tails and a spikier central peak than that of the real data. The orange distribution is more sharply peaked and has more mass in the extreme tails. This implies that the model, while accurate on average, tends to overestimate the frequency of extreme volatility events. For tail-risk applications such as stress-testing portfolios by computing VaR-type quantities from the simulated surfaces, or pricing far-out-of-the-money exotics, our model will thus result in conservative estimates. A natural direction for future work would be to modify the VP-SDE \eqref{eq: forward diffusion} with a non-Gaussian forward process and investigate if it better matches the empirical tail behaviour.

We further compare the predictive accuracy of the diffusion model and the benchmark models using the Diebold--Mariano (DM) test \citep{diebold1995comparing}, which evaluates whether two competing forecasts have equal expected forecasting error. The null hypothesis states that both models have the same expected predictive error. Let $d^{(k)}$ denote the difference between the MAPEs of the diffusion model and a benchmark on trading day $k$ in the test set. A negative value of $d^{(k)}$ indicates that the diffusion model achieves lower forecasting error on trading day $k$, while a positive value indicates the opposite. 

A larger magnitude of the DM test statistic implies the sample mean of $d^{(k)}$ is more statistically significant, whereas a small value indicates the opposite.

\begin{table}[htbp]
\centering
\caption{Diebold--Mariano test results comparing the diffusion model against benchmark models}
\label{tab:dm_results}
\begin{tabular}{lccccc}
\toprule
Benchmark & Mean $d^{(k)}$ (\%) & DM Statistic & p-value & Diffusion Win Days & Benchmark Win Days \\
\midrule
PCA-AR(1) & $-0.902$ & $-24.76$ & $<0.001$ & 593 & 103 \\
VAR(1)                          & $-0.320$ & $-11.49$ & $<0.001$ & 600 & 96  \\
SABR                            & $-1.308$ & $-23.63$ & $<0.001$ & 581 & 115 \\
LSTM-Attention                  & $-1.356$ & $-27.34$ & $<0.001$ & 478 & 218 \\
VolGAN                          & $-0.840$ & $-9.95$  & $<0.001$ & 484 & 212 \\
VolGAN with $\mathcal{Y}^{(k)}$ & $-1.113$ & $-13.03$ & $<0.001$ & 525 & 171 \\
\bottomrule
\end{tabular}
\end{table}

\begin{figure}[h!]
    \centering
    \caption{Distribution of $d^{(k)}$ in the test set across all benchmark models.}
    \label{fig:dm_hist}
    \begin{subfigure}[b]{0.32\textwidth}
        \centering
        \includegraphics[width=\textwidth]{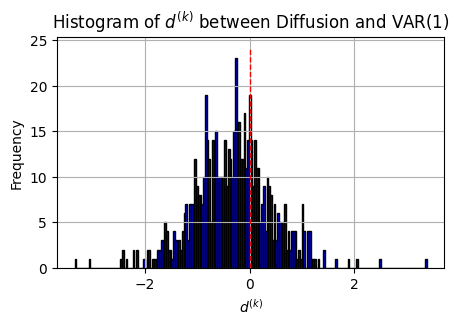}
        \caption{VAR(1)}
        \label{fig:dm_hist_var}
    \end{subfigure}
    \hfill
    \begin{subfigure}[b]{0.32\textwidth}
        \centering
        \includegraphics[width=\textwidth]{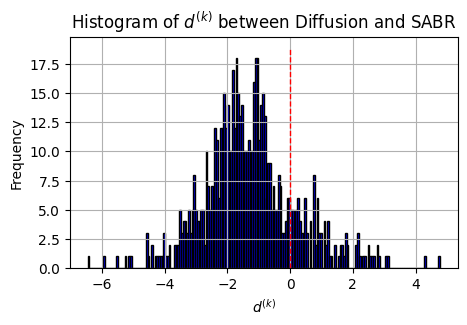}
        \caption{SABR}
        \label{fig:dm_hist_sabr}
    \end{subfigure}
    \hfill
    \begin{subfigure}[b]{0.32\textwidth}
        \centering
        \includegraphics[width=\textwidth]{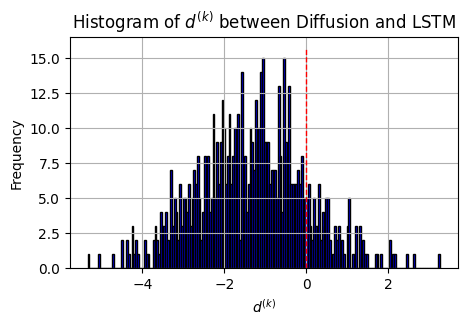}
        \caption{LSTM-Attention}
        \label{fig:dm_hist_lstm}
    \end{subfigure}

    \vspace{0.5em}

    \begin{subfigure}[b]{0.32\textwidth}
        \centering
        \includegraphics[width=\textwidth]{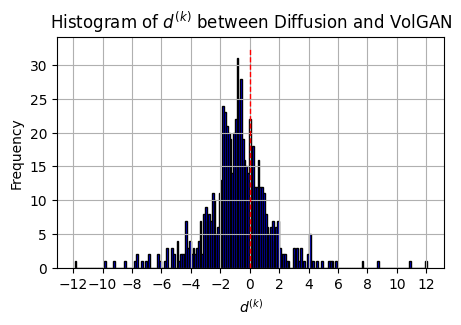}
        \caption{VolGAN}
        \label{fig:dm_hist_volgan}
    \end{subfigure}
    \hfill
    \begin{subfigure}[b]{0.32\textwidth}
        \centering
        \includegraphics[width=\textwidth]{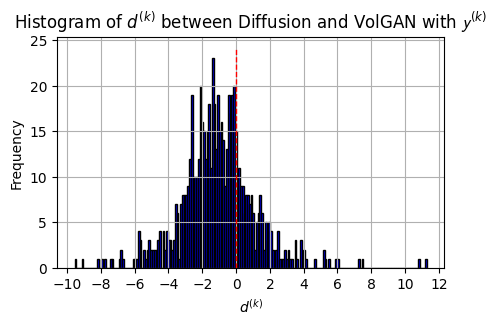}
        \caption{VolGAN with $\mathcal{Y}^{(k)}$}
        \label{fig:dm_hist_volgan_yk}
    \end{subfigure}
    \hfill
    \begin{subfigure}[b]{0.32\textwidth}
        \centering
        \includegraphics[width=\textwidth]{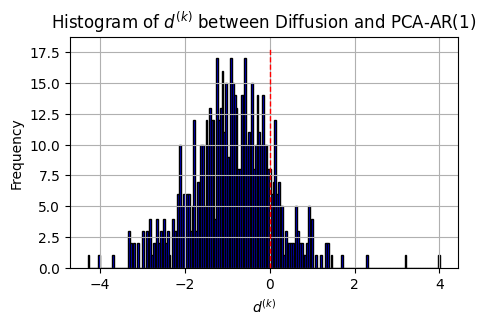}
        \caption{PCA-AR(1)}
        \label{fig:dm_hist_pcaar}
    \end{subfigure}
\end{figure}

Table \ref{tab:dm_results} reports the DM test results across all six benchmarks. In every case, the sample mean of $d^{(k)}$ is negative, indicating that the diffusion model achieves lower average forecasting error than each benchmark. The magnitude of this reduction varies considerably across benchmark types: the gap is smallest against VAR(1) ($-0.32$ percentage points), the purely linear, statistical baseline, and largest against LSTM-Attention ($-1.36$ percentage points) and SABR ($-1.31$ percentage points). All DM statistics are significant at the $0.1\%$ level, indicating that these average differences in forecasting error are highly consistent across the test set and statistically robust, rather than driven by a small number of outlier trading days. This conclusion is further supported by the day-count comparison and Figure \ref{fig:dm_hist} which illustrates the distribution of $d^{(k)}$ for all six benchmarks.

Figure \ref{fig:dm_hist_volgan} shows the distribution of $d^{(k)}$ between the diffusion model and VolGAN. The other histograms can be interpreted similarly across all six benchmarks. We compare the interval $d^{(k)} \in [-2, 0]$ with $d^{(k)} \in [0, 2]$. These regions correspond to cases where the diffusion model outperforms VolGAN by up to 2\% and vice versa, respectively. The density in the $[-2, 0]$ region is significantly higher, implying that the diffusion model more frequently outperforms VolGAN by a modest margin. The tail regions of $d^{(k)} \in [-12, -2]$ and $d^{(k)} \in [2, 12]$ show the density in the positive interval is significantly lower. This implies that cases where VolGAN significantly outperforms the diffusion model are much rarer than the opposite scenario.

 We also evaluate the computational requirements of the diffusion model and VolGAN. All experiments are implemented in PyTorch and executed on a high-performance computing cluster of the Digital Research Alliance of Canada, using an NVIDIA H100 GPU (10 GB of allocated memory). The proposed conditional diffusion model is trained for 2000 epochs with a mini-batch size of 64, requiring approximately two hours. This training phase remains computationally competitive with the VolGAN baseline, which requires three hours to complete its 10,000 training epochs. During sampling, generating 69,600 vol-surfaces takes 520 seconds for the diffusion model. Although this iterative reverse sampling process is more computationally demanding than the forward pass of VolGAN, which completes the same generation in under 10 seconds, it is highly efficient in absolute terms. Once the model is trained, the sampling process poses no operational bottleneck for daily end-of-day risk reporting or batch processing applications. We also tried deterministic DDIM sampling \citep{song2021denoising} with 50 steps, but the resulting samples were noisier, so we retain the original sampling with $T = 500$ steps. Faster solvers could reduce the cost further, but at these timings acceleration is not needed for end-of-day applications.

\subsection{Arbitrage Penalty Effectiveness Comparison}
\label{subsec:arb_comparison}
\begin{figure}[tbp]
    \centering
    \includegraphics[width=0.9\textwidth]{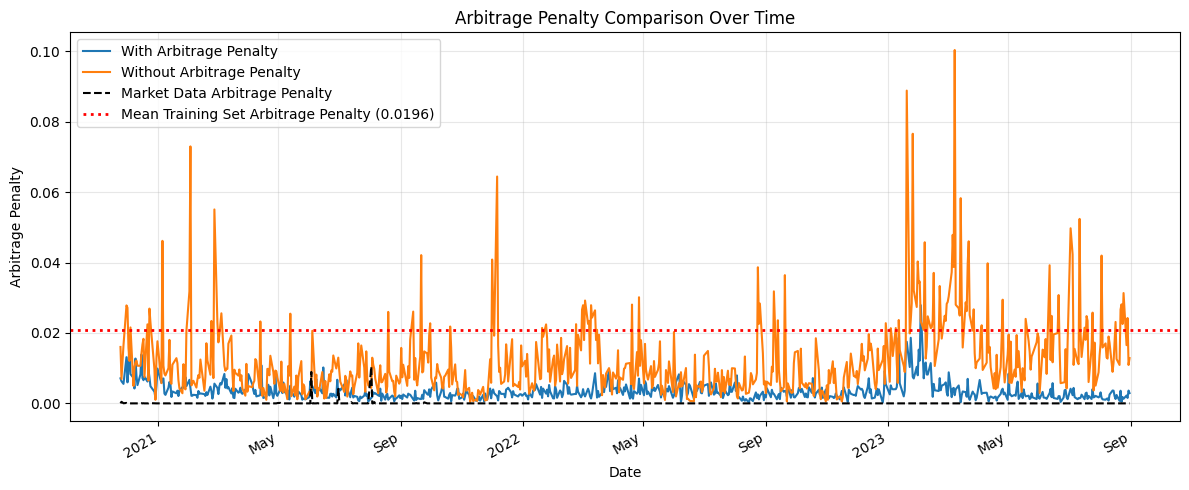}
    \caption{Time-series comparison of arbitrage penalties of the generated vol-surfaces. The model trained without the regularization term ($\mathcal{L}_{\text{simple}}$, orange) generated vol-surfaces with higher arbitrage penalty than the vol-surfaces from the model with SNR-weighted arbitrage penalty ($\mathcal{L}_{\text{total}}$, blue).}
    \label{fig:arbitrage_comparison}
\end{figure}

\begin{figure}[htbp]
    \centering
    \includegraphics[width=0.9\textwidth]{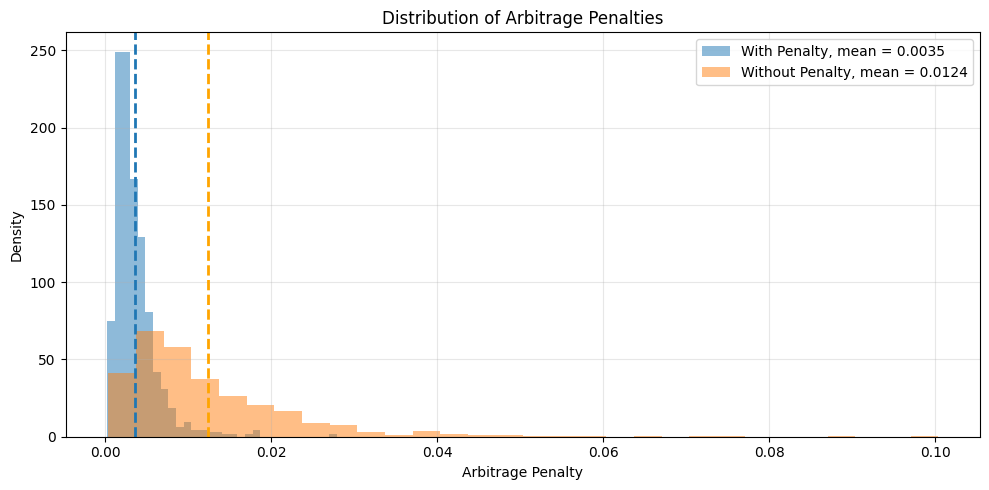}
    \caption{The distribution comparison of arbitrage penalties for generated vol-surfaces. The baseline model trained with $\mathcal{L}_{\text{simple}}$, shown in orange, exhibits a right-skewed penalty distribution, indicating more frequent arbitrage violations. The conditional diffusion model trained with $\mathcal{L}_{\text{total}}$, shown in blue, shifted the distribution toward zero.}
    \label{fig:arbitrage_dist}
\end{figure}

In Section \ref{subsec:score_matching}, we introduced an SNR-weighted arbitrage regularization term in the loss function to enforce the arbitrage constraints. To evaluate the efficacy of this regularization, we compare a model trained with the complete loss $\mathcal{L}_{\text{total}}$ \eqref{eq:final loss function} against a model trained on $\mathcal{L}_{\text{simple}}$ \eqref{eq:noiseloss}, which does not include the arbitrage penalty term.
As reported in Table \ref{tab:penalty_stats}, including the regularization term reduces all four statistics of the arbitrage penalties without compromising predictive accuracy, as measured by MAPE. By lowering both the mean and standard deviation, the regularization term shifts the arbitrage penalty distribution to the left and concentrates it more tightly around zero. This effect is more clearly illustrated in Figure \ref{fig:arbitrage_dist}, where the distribution of the regularized model (blue) exhibits a sharper, taller mode at zero. In contrast, the orange distribution shows a wider distribution with the mean shifted to the right compared to the blue distribution mean. Furthermore, Figure \ref{fig:arbitrage_comparison} demonstrates that the model trained without the regularization term generates vol-surfaces with consistently higher arbitrage penalties with a peak around $0.1$. The regularization term mitigates the arbitrage penalties across the majority of the time series and suppresses the peak to $0.028$. Together, these results confirm that the SNR-weighted regularization term successfully minimizes arbitrage violations in the generated vol-surfaces. 

\begin{table}[htbp]
    \centering
    \caption{Summary statistics of the arbitrage penalty $\Phi$ of the generated test-set daily mean vol-surfaces. The mean of MAPE column shows the mean MAPE of the test set.}
    \label{tab:penalty_stats}
    \small\setlength{\tabcolsep}{4pt}
    \begin{tabular}{l c c c c c}
        \hline
        \textbf{Model} & \textbf{Mean} & \textbf{Std} & \textbf{Min} & \textbf{Max} & \textbf{Mean of MAPE} \\
        \hline
        With Arbitrage Penalty    & 0.003524 & 0.002645 & 0.000225 & 0.027898 & 3.0026\\
        Without Arbitrage Penalty & 0.012378 & 0.010938 & 0.000380 & 0.100377 & 3.0941\\
        VolGAN                    & 0.004699   & 0.002067 & 0 & 0.0343  & 3.8425 \\
        \hline
    \end{tabular}
\end{table}

As discussed in Section \ref{sec:background}, the penalty term $\Phi_b$ in \eqref{eq:arb pen butterfly} regularizes the convexity of the implied volatility smile and is only a proxy for the exact no-butterfly-arbitrage condition. We now verify how closely this proxy agrees with the Durrleman condition \eqref{eq:durrleman-m-sigma} on the generated vol-surfaces. For a fixed maturity $\tau_j$, the function $g(m)$ defined in \eqref{eq:durrleman-m-sigma} being non-negative across moneyness ensures that the call price is a convex function of the strike. To compare $g(m)$ with the convexity penalty, we separate the second-order term of \eqref{eq:durrleman-m-sigma} from the terms involving only the level and slope of the smile as:
\eqlnostar{eq:durrleman-decomp}{
g(m) = \underbrace{\tau_j\, \sigma\, m^2\, \partial_{mm}\sigma}_{\text{curvature}}
+ \underbrace{\left( 1 - \frac{m \log m \, \partial_m \sigma}{\sigma} \right)^2
- \frac{\left( m\, \sigma\, \partial_m \sigma\, \tau_j \right)^2}{4}
+ \tau_j\, \sigma\, m\, \partial_m \sigma}_{=:\, R(m, \tau_j, \sigma, \partial_m \sigma)} \;\geq\; 0.
}
The curvature term in \eqref{eq:durrleman-decomp} measures the curvature of the smile, and since  $\tau_j > 0$, $\sigma > 0$ and $m^2 > 0$, the curvature term has the same sign as $\partial_{mm}\sigma$ which is measured by $\left[ \sigma_{i-1,j} - 2\sigma_{i,j} + \sigma_{i+1,j} \right]$ in the penalty term \eqref{eq:arb pen butterfly}. Therefore, a grid point violates the convexity condition penalized by $\Phi_b$ if and only if the curvature term is negative. The remaining term $R$ in \eqref{eq:durrleman-decomp} depends only on the level and the slope of the smile. When $R \geq 0$, a convex smile ($\partial_{mm}\sigma \geq 0$) implies $g \geq 0$, so that $\Phi_b$ is sufficient for the absence of butterfly arbitrage in this case. However, the convexity condition is not necessary, since a negative curvature term is admissible as long as it is offset by $R$. Around the ATM region where $m \approx 1$,  the first term of $R$ is close to one and $R$ is typically close to one as well, which allows  a negative curvature. Thus, the convexity condition only fails to detect a violation of \eqref{eq:durrleman-decomp} when $R < 0$, which requires a steep skew at longer maturities.

To quantify the difference between the Durrleman condition and our convexity-based penalty term, we evaluate $g$ on the vol-surfaces. Since during the training process, the total arbitrage penalty $\Phi$ is only applied on the predicted vol-surface $\hat{\boldsymbol{\sigma}}^{(k+1)}$, we use generated vol-surfaces of the test set for this evaluation. The derivatives in \eqref{eq:durrleman-decomp} are approximated on the moneyness grid as: $\partial_m \sigma \approx (\sigma_{i+1,j} - \sigma_{i-1,j})/(2h),$ and $\partial_{mm} \sigma \approx (\sigma_{i+1,j} - 2\sigma_{i,j} + \sigma_{i-1,j})/h^2,$ with grid spacing $h = 0.1$. Each grid point is then classified according to whether it satisfies the convexity condition ($\partial_{mm}\sigma \geq 0$) and whether it satisfies the Durrleman condition ($g \geq 0$). The results are reported in Table \ref{tab:durrleman_agreement}.

\begin{table}[htbp]
\centering
\begin{threeparttable}
\caption{Agreement between the convexity condition and the Durrleman condition on the generated vol-surfaces of the test set.}
\label{tab:durrleman_agreement}
\begin{tabular}{l cc}
\toprule
 & Durrleman satisfied ($g \geq 0$) & Durrleman violated ($g < 0$) \\
\midrule
Convexity satisfied ($\partial_{mm}\sigma \geq 0$) & 82.98\% & $<0.001$\% \\
Convexity violated ($\partial_{mm}\sigma < 0$)     & 16.93\% & 0.09\% \\
\bottomrule
\end{tabular}
\begin{tablenotes}
\footnotesize
\item \textit{Note:} Percentages are shares of all 4{,}384{,}800 interior grid points of the generated vol-surfaces (100 vol-surfaces per trading day over 696 trading days in the test set). The upper-right cell corresponds to butterfly arbitrage that is not detected by the convexity condition.
\end{tablenotes}
\end{threeparttable}
\end{table}

As reported in Table \ref{tab:durrleman_agreement}, grid points where the smile is convex but the Durrleman condition is violated account for fewer than 0.001\% of all grid points. Therefore, the difference between our convexity-based penalty $\Phi_b$ and the Durrleman condition, in terms of butterfly arbitrage that the penalty fails to detect, is negligible. The convexity condition is violated at 16.93\% of all grid points while satisfying the Durrleman condition, which implies that the convexity condition is more restrictive than necessary. The restrictiveness of the convexity condition may penalize arbitrage-free grid points, but it does not admit butterfly arbitrage beyond the negligible fraction. For a regularization term whose purpose is to suppress arbitrage, this conservative direction is preferable to the opposite. Moreover, this restrictiveness has limited practical consequences. As reported in Table \ref{tab:penalty_stats}, the model trained with the arbitrage penalty achieves a lower MAPE than the model trained without it (3.0026\% versus 3.0941\%), so the conservativeness of the penalty does not come at the expense of forecasting accuracy. An alternative is to penalize the Durrleman condition by replacing $\Phi_b$ in \eqref{eq: total_arb} with $\sum_{i,j} \max\left(0, -g(m_i, \tau_j)\right)^2$, where $g$ is evaluated with the finite difference approximations as discussed above. We also trained the model with this penalty in place of $\Phi_b$, keeping all other settings unchanged and tuning $\lambda$ in the same way. Compared with the model trained with $\Phi_b$, however, this version produced a much higher overall MAPE (19\% versus 3.0026\%) and wider confidence intervals (mean width of about 10 volatility points).
A possible explanation is that $g$ involves the ratio $\partial_m \sigma / \sigma$ and products of derivatives, which makes it more sensitive to the noise in the estimate $\hat{\boldsymbol{\sigma}}^{(k+1)}$ evaluated during training than the curvature term appearing in $\Phi_b$. Given that the convexity penalty misses only a negligible fraction of butterfly arbitrage while yielding better forecasting accuracy and calibration, our use of the penalty $\Phi_b$ is justified.



\subsection{Ablation Study}
\label{subsec:ablation}

To assess the contribution of each conditioning variable to the forecasting performance, we conduct a leave-one-out ablation study, in which each component of the conditioning set $\mathcal{Y}^{(k)}$ is removed in turn while holding the remaining architecture and training procedure fixed. All configurations are trained and evaluated using the same ten random seeds, which ensures that the configurations differ only in their conditioning. Each configuration is evaluated by point-forecast accuracy that is measured by MAPE, and predictive interval calibration, measured by the empirical breach rate of the 90\% confidence interval relative to the 10\% target. Since the test set contains $n = 696$ trading days, the standard error of an empirical breach rate under perfect calibration (true rate of 10\%) is approximately $\sqrt{0.10 \times 0.90 / 696} \approx 1.1\%$. We use this metric to judge whether an observed breach rate deviation from a comparison configuration is likely to reflect a genuine calibration effect rather than sampling variation over the test window. A compact summary of the results is reported in Table~\ref{tab:ablation_compact} while the full set of results is reported in Table~\ref{tab:ablation_full}.

Removing the current vol-surface condition produces a substantial degradation in point-forecast accuracy, with overall MAPE rising from 3.003\% (original model) to 3.349\%. This configuration also shows a pronounced shift in calibration where the ITM breach rate falls to roughly 4--5\%, more than 3 percentage points or equivalently approximately 3 standard errors below the original ITM breach rate of approximately 7.9--8.5\%. This indicates that the current-surface condition anchors both the point forecast and the calibration of predictive uncertainty in the ITM region specifically.

Removing the long-horizon surface condition produces a small degradation in point accuracy (MAPE rising to 3.036\%) but a comparably large shift in ATM calibration: the ATM breach rate falls to approximately 8.3--9.2\%, compared to approximately 10.1--11.1\% in the original model, a gap of approximately 2 percentage points, or close to 2 standard errors. This suggests the long-horizon surface condition contributes to ATM calibration beyond its modest effect on point accuracy.
Removing the return condition leaves point accuracy essentially unchanged with MAPE of 3.010\% versus 3.003\% for the original model but shifts ATM calibration substantially in the opposite direction where the ATM breach rate rises to approximately 12.6--13.7\%, roughly 2--3 percentage points above the original model and 2 to 3 standard errors from the nominal $10\%$ target. Unlike the current-surface and long-horizon conditions, whose removal tightens calibration below target, removing the return condition loosens calibration above target, indicating this condition specifically prevents ATM interval overconfidence.

Removing the short-horizon surface or VIX condition leaves overall MAPE essentially unchanged or marginally improved (2.999\% and 2.949\%, respectively, versus 3.003\% for the original model). This would suggest these two conditions are dispensable. However, removing the short-horizon surface raises the ATM breach rate to approximately 11--13\%, comparable in magnitude to the return-condition effect described above, whereas the ATM breach rate without VIX (9.3--11.1\%) is essentially unchanged from the original model. The short-horizon surface condition, like the return condition, therefore contributes to constraining ATM interval width even though its removal does not measurably harm point-forecast accuracy, while the effect of the VIX condition cannot be distinguished from seed-to-seed variation. These conditions therefore leave the point forecast essentially unchanged but affect the calibration of the predictive intervals, which is arguably more important for a probabilistic forecasting framework intended for risk management or option pricing.

The squared-return condition presents the most pronounced calibration effect observed in the study. Its removal leaves MAPE nearly unchanged (3.015\% versus 3.003\%) but pushes the ATM breach rate to 11.8--14.7\%, up to 3.6 percentage points (about 3 standard errors) above the original model and 4.7 points above the nominal 10\%. This is the largest calibration shift among all single-condition ablations and suggests the squared-return condition plays an outsized role in controlling ATM interval width despite its negligible contribution to point accuracy. Finally, the ``Only Current Surface'' configuration (MAPE 3.337\%) keeps only the current surface among the surface conditions, together with all scalar conditions, and ``Only VIX'' (MAPE 3.094\%) keeps only the VIX return among the scalar conditions, together with all surface conditions. ``Only Current Surface'' performs clearly worse than the original model on point accuracy, confirming that the current surface alone is not sufficient. Compared with ``No Scalar'', ``Only VIX'' improves point accuracy (overall MAPE of 3.094\% versus 3.215\%) but returns wider confidence intervals in the OTM and ITM regions. Thus, we need other scalar conditions to calibrate the model better, even if the VIX alone is informative for point-forecast accuracy.

Furthermore, we also performed studies labelled ``No Scalar'', in which all scalar conditions (return, squared-return, and VIX) are removed while all high-dimensional conditions are kept, and ``No Surface Cond'', in which all surface conditions (current surface, short-horizon EWMA, and long-horizon EWMA) are excluded while all scalar conditions are retained.
Removing all scalar conditions degrades overall MAPE to 3.215\%, worse than removing the return, squared-return, or VIX conditions individually (3.010\%, 3.015\%, and 2.949\%, respectively). While removing the VIX condition alone did not harm the point-forecast accuracy, removing all scalar conditions together produces a clear degradation relative to the full model, indicating that the scalar conditions carry overlapping information individually but are jointly necessary. The results show that no single scalar condition is indispensable on its own, but the scalar conditioning category as a whole is.

The effect of removing all surface conditions is more severe. Overall MAPE rises from 3.003\% in the full model to 11.448\%. Additionally, CI breach rates rise from approximately 8--11\% across ATM and ITM regions in the full model to about 37\% in the ITM region under ``No Surface Cond'', far above the nominal 10\%, while the OTM region remains overly conservative (1.3\%). Considered together with the ``No Current Surface'' (overall MAPE of 3.349\%) and the ``Only Current Surface'' (overall MAPE of 3.337\%) results, it is clear that the forecasting performance degrades substantially as surface conditioning is progressively removed from the full set. This confirms that the high-dimensional conditioning set is essential to the forecasting framework and cannot be substituted by scalar conditioning alone.

From Table \ref{tab:ablation_compact}, we conclude that the conditioning variables serve complementary roles.  The surface conditions, in particular the current surface, primarily drive point-forecast accuracy and ITM calibration, while the return, short-horizon surface and squared-return conditions contribute comparatively little to point accuracy but measurably shift ATM interval calibration. These findings are further supported by the fact that removing all surface conditions produces the most severe degradation in this study, while removing all scalar conditions shows that they are jointly necessary.

\begin{table}[h!]
\centering
\begin{threeparttable}
\caption{Ablation Study: Comparison of Model Performance Metrics}
\label{tab:ablation_compact}
\footnotesize\setlength{\tabcolsep}{3pt}
\begin{tabular}{l rrrrrrr}
\toprule
Configuration & MAPE & ATM & ITM & OTM & ATM Breach & ITM Breach & OTM Breach \\
\midrule
Original                & 3.0026 & 4.5146 & 2.4184 & 3.1542 & 11.0632 & 8.4770 & 0.5747 \\
No Current Surface      & 3.3488 & 5.4376 & 2.7042 & 3.4547 & 9.7701 & 4.5977 & 0.7184 \\
No Long Surface         & 3.0359 & 4.5202 & 2.3702 & 3.1683 & 9.0517 & 7.4713 & 0.1437 \\
No Return               & 3.0098 & 4.4543 & 2.3541 & 3.0443 & 13.5057 & 10.3448 & 1.2931 \\
No Short Surface        & \underline{2.9994} & 4.6040 & 2.3750 & 2.9895 & 12.9310 & 8.4770 & 0.8621 \\
No Square Return        & 3.0152 & 4.5367 & 2.4479 & 3.0579 & 14.6552 & 10.2011 & 2.2989 \\
No VIX                  & \textbf{2.9492} & 4.5047 & 2.3566 & 3.1039 & 10.7759 & 7.1839 & 0.5747 \\
Only Current Surface    & 3.3368 & 4.5431 & 2.5801 & 3.5182 & 10.9195 & 9.9138 & 0.5747 \\
Only VIX                & 3.0941 & 4.6080 & 2.3615 & 3.3755 & 10.9195 & 6.1782 & 0.2874 \\
No Scalar                & 3.2149 & 4.6032 & 2.3994 & 3.2999 & 11.4943 & 8.0460 & 1.0057 \\
No Surface Cond          & 11.4483 & 11.4592 & 13.8045 & 9.8315 & 7.9023 & 36.7816 & 1.2931 \\
\bottomrule
\end{tabular}
\begin{tablenotes}
\footnotesize
\item \textit{Note:} This table compares forecasting errors across the Original (full) conditioning framework and each ablation configuration. The ATM, ITM, OTM columns show the 1-month MAPE at those moneyness levels. ``No Scalar'' removes all scalar conditions (return, squared-return, VIX) while retaining all surface conditions; ``No Surface Cond'' removes all surface conditions (current, short-horizon EWMA, long-horizon EWMA) while retaining all scalar conditions. ``Only Current Surface'' and ``Only VIX'' retain only the named condition within their respective conditioning category, with the other category entirely retained. The best-performing configuration for the overall MAPE column is highlighted in \textbf{bold}, and the second-best is \underline{underscored}. The original model is the representative diffusion model and other configurations correspond to the average of the results obtained across ten different seeds. 
\end{tablenotes}
\end{threeparttable}
\end{table}

\subsection{Two-Day-Ahead Volatility Surface Forecasting}
While our primary empirical analysis focuses on one-day-ahead predictions, the proposed conditional diffusion methodology can be extended to forecast the vol-surface multiple days ahead. As is well-known in the broader regression and time-series forecasting literature, predictive accuracy decays as the forecasting horizon increases due to the accumulated uncertainty over the longer period and market noise. Here, we demonstrate the performance of the two-day-ahead forecasting exercise. The model is trained directly to forecast $\mathbf{x}_0^{(k+2)}$ from $\mathcal{Y}^{(k)}$, with all other settings unchanged. Table \ref{tab:2_day_ahead} summarizes the performance metrics for the two-day-ahead predictions across all moneyness and time to maturity. As we increase the forecast horizon to 2 days, the overall MAPE increases to 3.8412\%, compared to the 3.0026\% achieved in the one-day-ahead framework. However, the breach rates of the 90\% confidence intervals at ATM points range from 10.5036\% to 11.5108\%, close to the nominal 10\%. The OTM points remain overly conservative, which is consistent with the data patterns previously observed in the day-ahead analysis. These results show that the conditional diffusion framework extends naturally to a two-day horizon.

\begin{table}[h!]
\centering
\caption{Performance Metrics for Two-Day-Ahead Forecasting}
\label{tab:2_day_ahead}
\resizebox{\textwidth}{!}{%
\begin{tabular}{l rrrr rrrr rrrr r}
\toprule
& \multicolumn{4}{c}{ATM} & \multicolumn{4}{c}{OTM} & \multicolumn{4}{c}{ITM} & \\
\cmidrule(lr){2-5} \cmidrule(lr){6-9} \cmidrule(lr){10-13}
Metric & 1-Day & 1-Week & 1-Month & 3-Month & 1-Day & 1-Week & 1-Month & 3-Month & 1-Day & 1-Week & 1-Month & 3-Month & Overall \\
\midrule
MAPE (\%) & 6.4826 & 6.4380 & 6.1946 & 5.5021 & 4.0755 & 4.0030 & 3.7141 & 3.0127 & 3.3824 & 3.3558 & 3.2384 & 2.9070 & 3.8412 \\
Std. of APE (\%) & 5.2086 & 5.1714 & 5.0248 & 4.5017 & 3.0942 & 3.0458 & 2.8141 & 2.3325 & 2.9790 & 2.9614 & 2.8274 & 2.4715 & \\
Mean CI Width & 0.0484 & 0.0480 & 0.0468 & 0.0436 & 0.0604 & 0.0591 & 0.0542 & 0.0417 & 0.0628 & 0.0624 & 0.0598 & 0.0533 & \\
Std. CI Width & 0.0143 & 0.0142 & 0.0138 & 0.0124 & 0.0111 & 0.0110 & 0.0096 & 0.0073 & 0.0134 & 0.0133 & 0.0132 & 0.0119 & \\
CI Breach \% & 11.0791 & 11.5108 & 10.7914 & 10.5036 & 0.1439 & 0.4317 & 0.4317 & 0.5755 & 7.9137 & 8.4892 & 7.3381 & 5.4676 & \\
\bottomrule
\end{tabular}
}
\end{table}

\section{Conclusion}
\label{sec:conclusion}

In this paper, we introduced a conditional Denoising Diffusion Probabilistic Model (DDPM) for forecasting one-day-ahead vol-surfaces. We demonstrated that the diffusion model provides an improved alternative to existing generative models when combined with the market conditioning set. We incorporated an explicit arbitrage penalty into the loss function that is dynamically weighted by the SNR, which suppresses arbitrage violations without destabilizing the training process. Our model not only achieved a lower overall MAPE than all six benchmarks, consistently across ten random seeds, but also was shown to be better calibrated.

We also verified that the convexity-based penalty misses only a negligible fraction of butterfly arbitrage relative to the exact Durrleman condition, and the ablation study shows that the surface conditions drive forecast accuracy while the return-based conditions mainly affect the calibration of ATM intervals. The predictive distribution is over-dispersed in the OTM region, which we trace to the sparsely quoted, extrapolated OTM data of the training period. A natural direction for future work is to first project the vol-surface onto an arbitrage-free space and use diffusion modelling in that space for forecasting day-ahead surfaces.



\section*{Acknowledgements}
The authors would like to thank two anonymous referees whose valuable feedback helped significantly improve the quality and accuracy of this manuscript. This research was partially supported by the Natural Sciences and Engineering Research Council of Canada (NSERC) Discovery Grant RGPIN-2026-05644.
\newpage
\bibliographystyle{elsarticle-harv}
\bibliography{references}

\appendix
\numberwithin{table}{section}
\numberwithin{figure}{section}

\section{Simulation Study with the Heston Model}
\label{subsec:heston_results}
\begin{figure}[h!]
    \centering
    \caption{Time series comparison for 1-Month points. Left: Diffusion Model vs. Real. Right: VolGAN vs. Real. Note: Dates are arbitrary.}
    \label{fig:heston_ts}

    \begin{subfigure}[b]{0.49\textwidth}
        \centering
        \includegraphics[width=\textwidth]{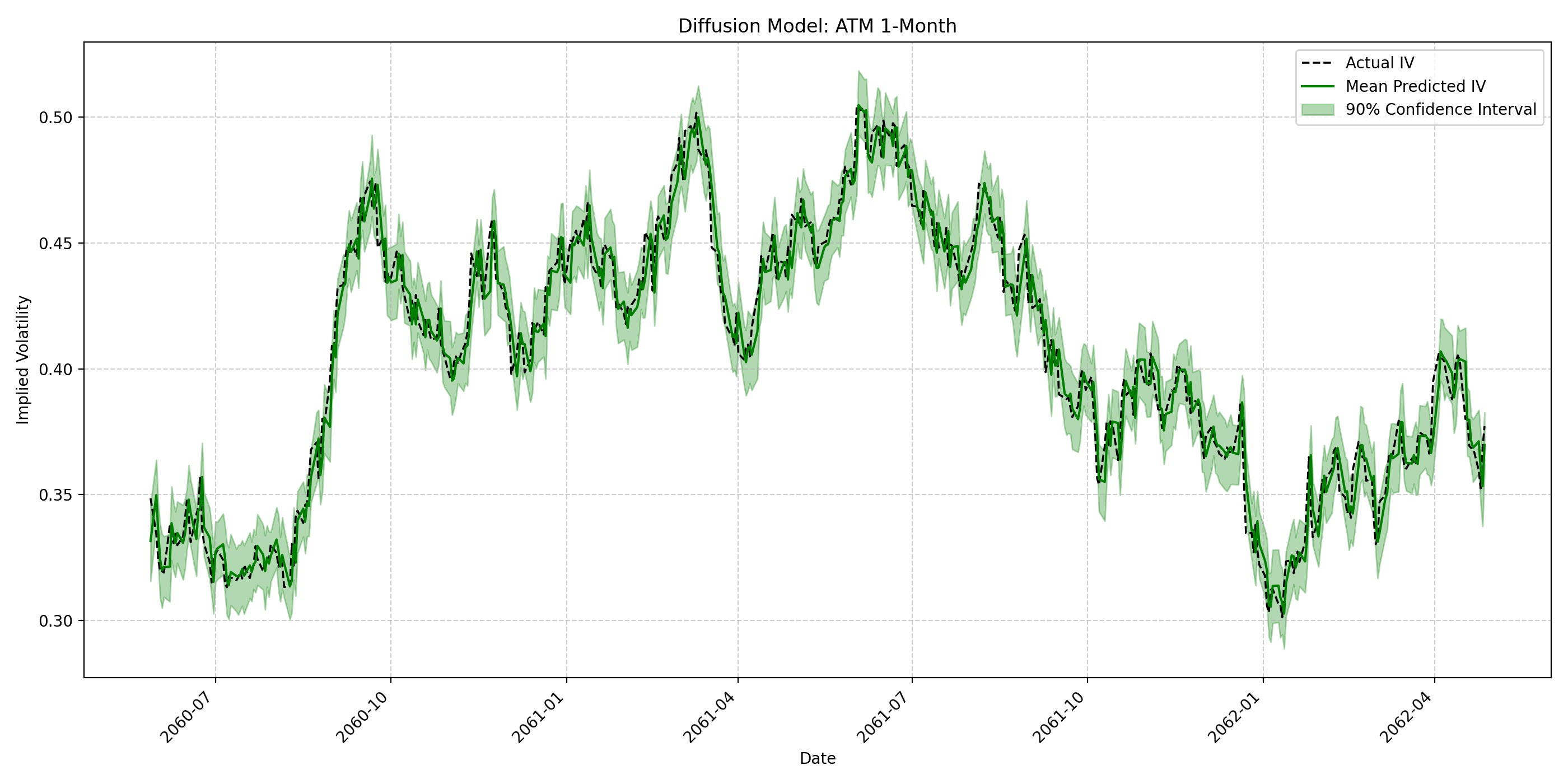}
        \caption{Diffusion Model with Heston Data (ATM 1-Month).}
    \end{subfigure}
    \hfill
    \begin{subfigure}[b]{0.49\textwidth}
        \centering
        \includegraphics[width=\textwidth]{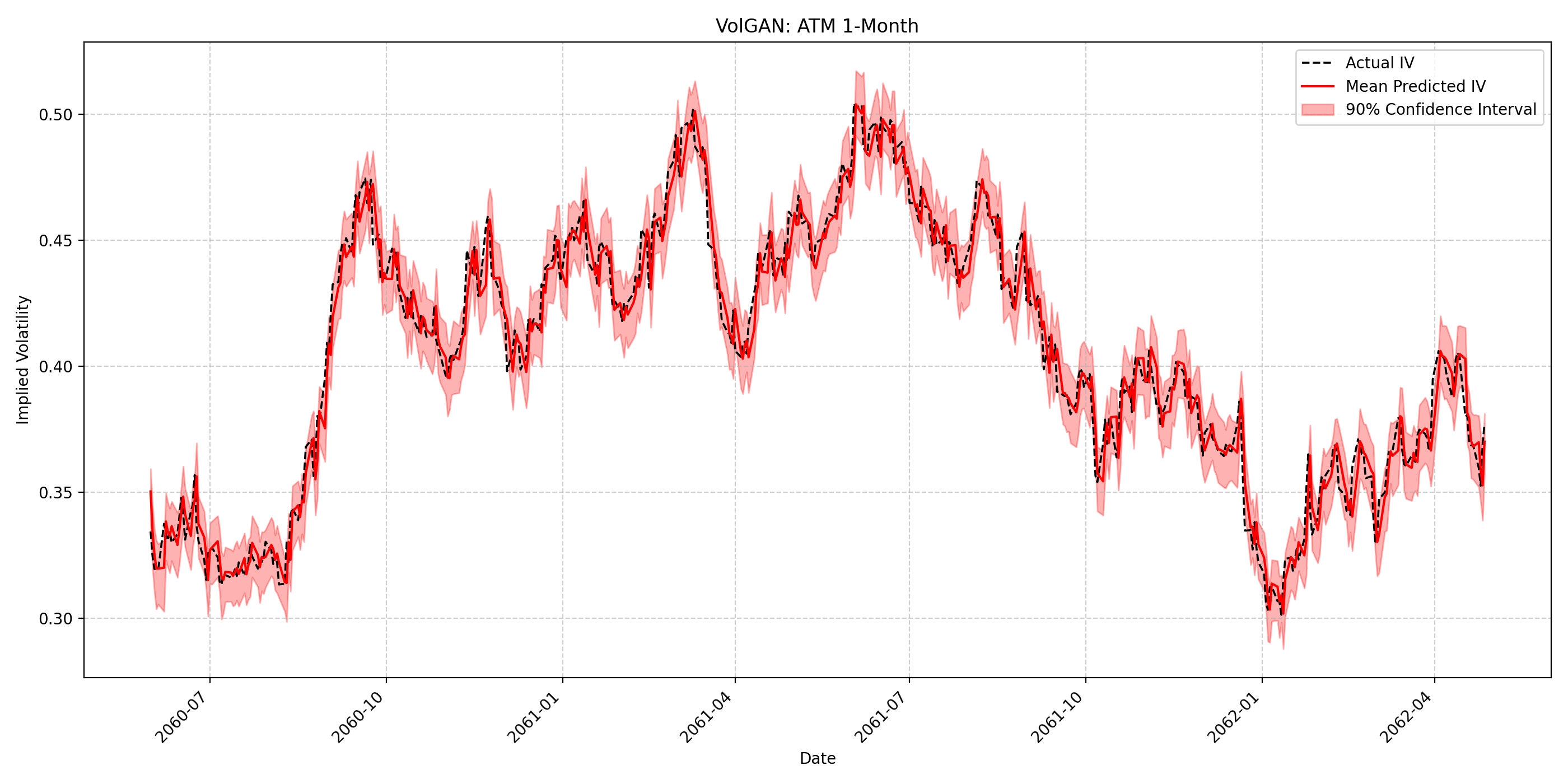}
        \caption{VolGAN Model with Heston Data (ATM 1-Month).}
    \end{subfigure}

    \vspace{0.5cm}

    \begin{subfigure}[b]{0.49\textwidth}
        \centering
        \includegraphics[width=\textwidth]{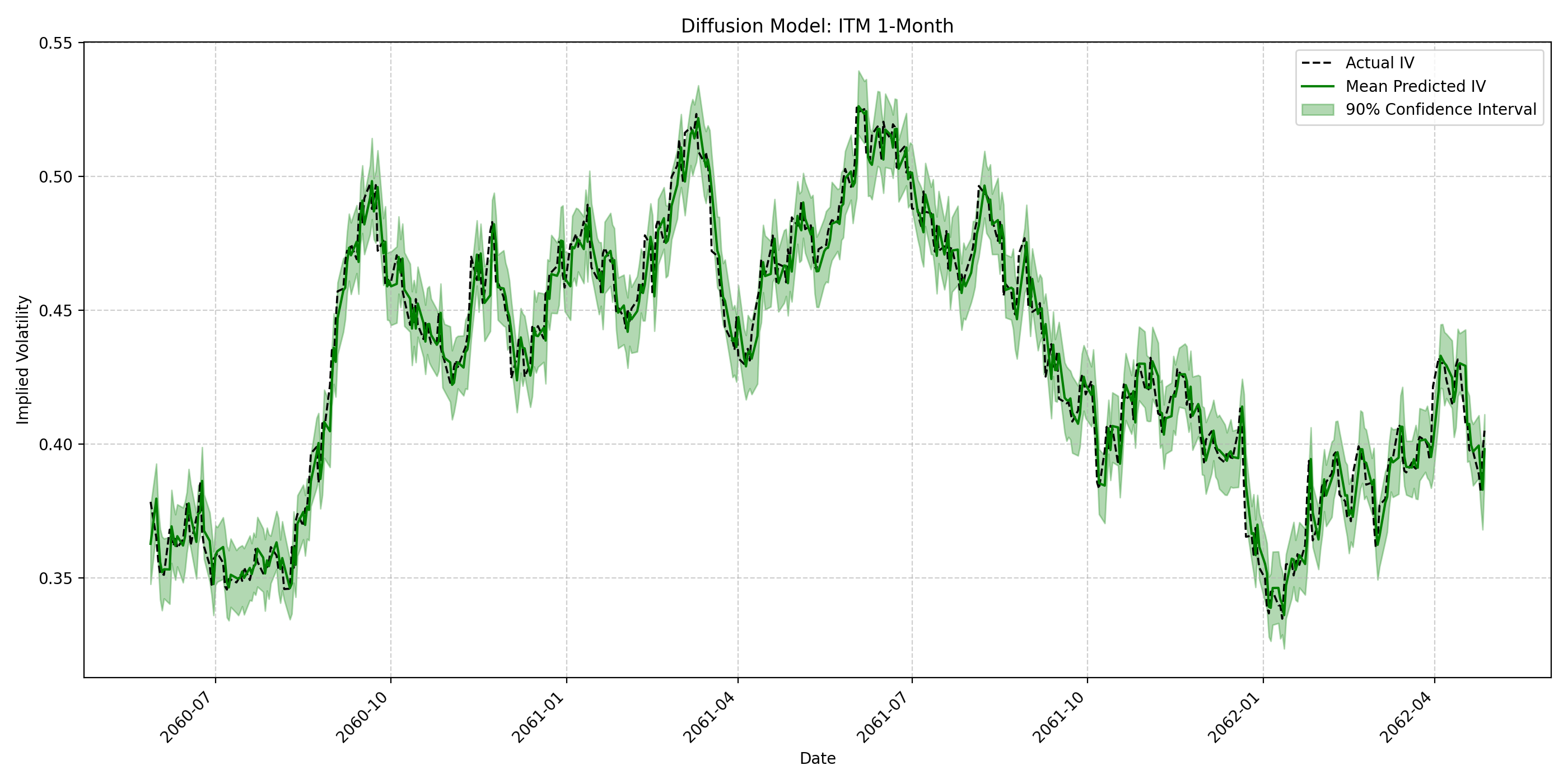}
        \caption{Diffusion Model with Heston Data (ITM 1-Month).}
    \end{subfigure}
    \hfill
    \begin{subfigure}[b]{0.49\textwidth}
        \centering
        \includegraphics[width=\textwidth]{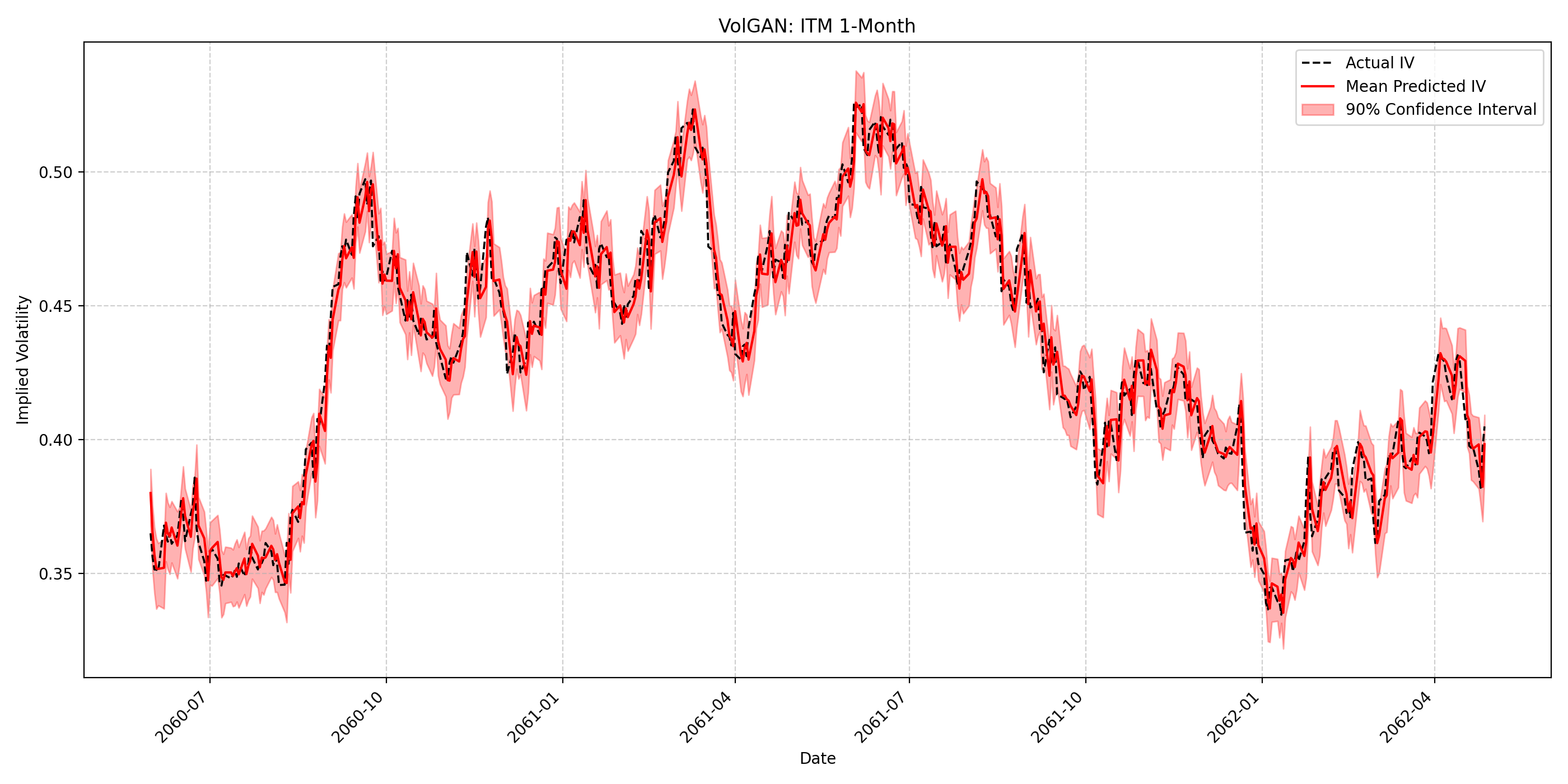}
        \caption{VolGAN Model with Heston Data (ITM 1-Month).}
    \end{subfigure}
    
    \vspace{0.5cm}

    \begin{subfigure}[b]{0.49\textwidth}
        \centering
        \includegraphics[width=\textwidth]{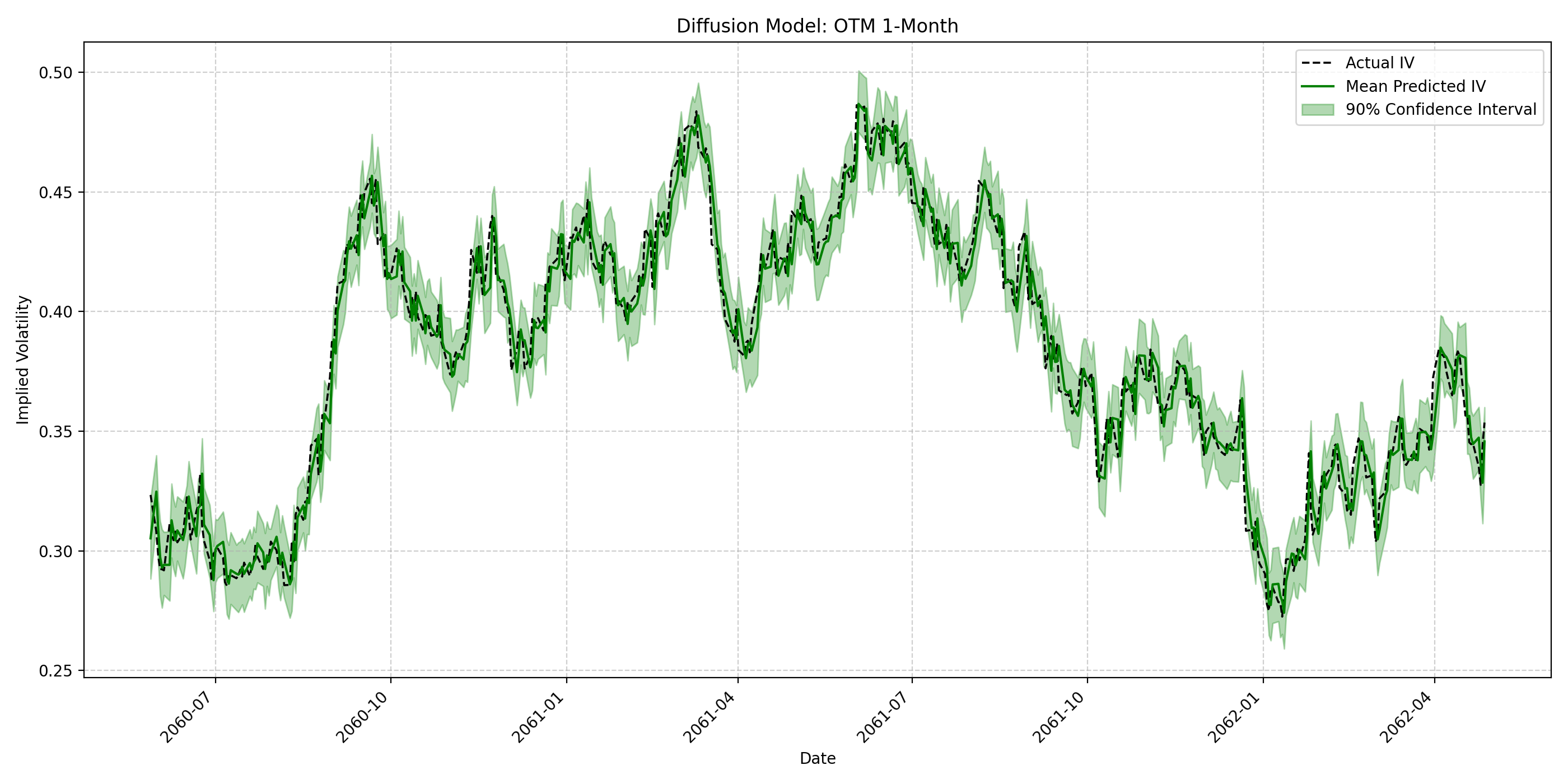}
        \caption{Diffusion Model with Heston Data (OTM 1-Month).}
    \end{subfigure}
    \hfill
    \begin{subfigure}[b]{0.49\textwidth}
        \centering
        \includegraphics[width=\textwidth]{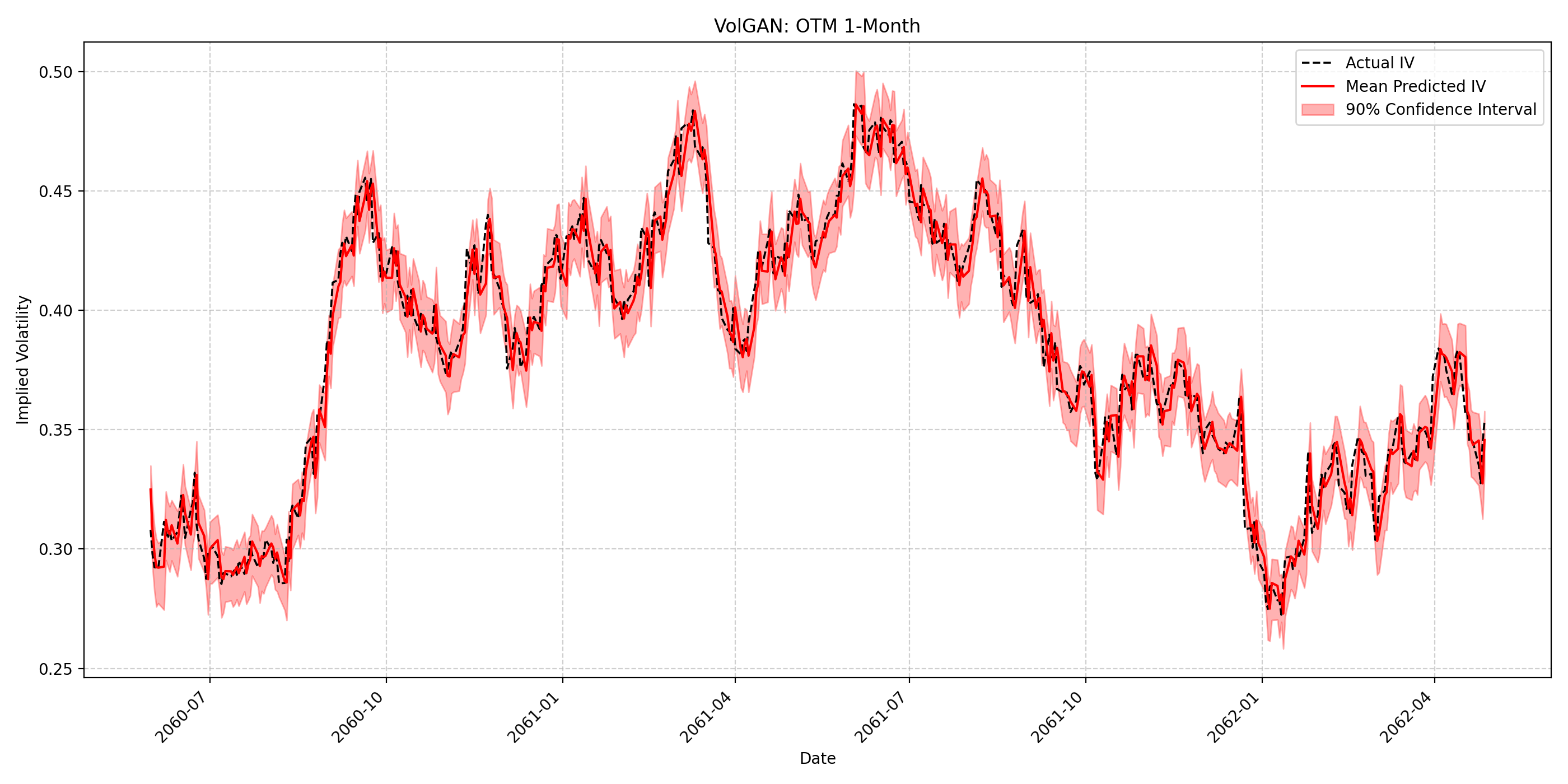}
        \caption{VolGAN Model with Heston Data (OTM 1-Month).}
    \end{subfigure}
    
\end{figure}

This simulation study serves as a sanity check for the learning capability of the diffusion model and to demonstrate that the methodology does not suffer from any issues when learning the underlying distribution of vol-surface in the ITM, ATM and OTM regions. In this controlled environment, the simulated vol-surfaces generated by the Heston model are strictly arbitrage-free, and the data distribution is known. 
We also test whether the model can learn with a truncated dataset by first generating a synthetic dataset of 10,000 vol-surfaces, then truncating the training sample to 70\% of its size, i.e.\ keeping 7,000 of the 10,000 simulated days to evaluate if the model still maintains its learning efficacy. Our expected result is that the conditional diffusion model is still able to capture the vol-surface dynamics with the truncated synthetic dataset albeit with a lower accuracy. The Heston model \citep{Heston_1993} is given by the following SDEs:
\eq{*}{
    dS_t = \mu S_t dt + \sqrt{h_t} S_t dW_t^S, \quad
    dh_t = \kappa(b - h_t) dt + \sigma_v \sqrt{h_t} dW_t^v,
}
where $S_t$ denotes the asset price and $h_t$ its variance. The processes $W_t^S$ and $W_t^v$ are standard Brownian motions with instantaneous correlation $\rho$ such that $dW_t^S dW_t^v = \rho dt$, with $\rho \in [-1, 1]$. The parameters are fixed at $S_0 = 100,$ $h_0 = 0.2,$ $\mu=0.05, \kappa=3, b=0.15, \sigma_v=0.3,$ and $\rho=-0.7$. To generate the vol-surfaces from the Heston model, we adopt the Euler-Maruyama discretization method and simulate a single path of the asset price and its corresponding variance for 10,000 trading days using the fixed parameters above. For each simulated trading day, we apply the Fast Fourier Transform \citep{Carr_Madan_1999} to compute theoretical option prices across the moneyness and time-to-maturity grid. Once these theoretical option prices are available, we convert them into an arbitrage-free vol-surface by inverting the Black-Scholes formula. Since no VIX series exists for the simulated data, the scalar conditioning vector consists of the four return-based EWMAs and the 20-day realized volatility, and the model is trained without the arbitrage penalty.
The moneyness $\mathbf{m}$ and time-to-maturity $\boldsymbol{\tau}$ used in the simulation study are defined respectively as:
\eq{*}{ 
\mathbf{m} = \{0.7, 0.8, 0.9, 0.95, 1.0, 1.05, 1.1, 1.2, 1.3\}, \quad
\boldsymbol{\tau} = \left\{\frac{1}{12}, \frac{1}{6}, \frac{1}{4}, \frac{1}{3}, \frac{1}{2}, \frac{3}{4}, 1.0, 1.5, 2.0\right\}.
}
We split the synthetic dataset into training, validation, and test sets, allocating 90\% of the data for training, 5\% for validation, and 5\% for out-of-sample testing. With our diffusion model being trained on the synthetic data, we compare the performance of our diffusion model against the popular approach, VolGAN \citep{Vuletic_Cont_2024}. In Figure \ref{fig:heston_ts}, we show the one-day-ahead forecasts for the 1-month time-to-maturity and different moneyness over the test window from both the diffusion model and VolGAN. The At-the-Money (ATM), Out-of-the-Money (OTM), In-the-Money (ITM) results in the figure correspond to moneyness levels $m=1, 1.3, 0.7,$ respectively. For each trading day, we generate 100 vol-surfaces using the two models and construct 90\% confidence intervals (CIs) (5\% to 95\%). Both the diffusion model and VolGAN achieve comparable point-wise and overall accuracy  (overall MAPE of 1.007\% for the diffusion model and 1.008\% for VolGAN) and produce smooth, arbitrage-free vol-surfaces, thereby successfully reflecting that both models can learn the distribution and evolution of vol-surfaces under the Heston model. However, we observe that VolGAN suffers from instability, with its CIs varying significantly even when different training instances use identical hyperparameters and data. In contrast, our diffusion model does not suffer from such instability and demonstrates consistent calibration across different training instances.

Next, we trained the model on the synthetic dataset truncated as described above, while keeping the test set identical. As reported in Table \ref{tab:master_sim_metrics}, the out-of-sample performance of the model on the truncated dataset remains highly comparable to the full-dataset baseline, with the overall mean absolute percentage error (MAPE) increasing marginally from 1.007\% to 1.019\%. This result shows that our conditional diffusion model is able to learn the vol-surface dynamics within this controlled arbitrage-free environment, even with data limitations. 

\begin{table}[htbp]
\centering
\resizebox{\textwidth}{!}{%
\begin{tabular}{l cccc cccc cccc c}
\toprule
& \multicolumn{4}{c}{\textbf{ATM}} & \multicolumn{4}{c}{\textbf{OTM}} & \multicolumn{4}{c}{\textbf{ITM}} & \\
\cmidrule(lr){2-5} \cmidrule(lr){6-9} \cmidrule(lr){10-13}
\textbf{Metric} & \textbf{1-Month} & \textbf{2-Month} & \textbf{3-Month} & \textbf{4-Month} & \textbf{1-Month} & \textbf{2-Month} & \textbf{3-Month} & \textbf{4-Month} & \textbf{1-Month} & \textbf{2-Month} & \textbf{3-Month} & \textbf{4-Month} & \textbf{Overall} \\
\midrule
\multicolumn{14}{l}{\textbf{Model Result of Full Synthetic Dataset (10,000 Samples)}} \\
\midrule
\textbf{MAPE (\%)}        & 1.718 & 1.532 & 1.239 & 0.790 & 1.919 & 1.691 & 1.340 & 0.832 & 1.511 & 1.367 & 1.129 & 0.742 & 1.007 \\
\textbf{Std. of APE (\%)} & 1.350 & 1.197 & 0.961 & 0.612 & 1.523 & 1.327 & 1.041 & 0.644 & 1.182 & 1.064 & 0.875 & 0.575 & -- \\
\textbf{Mean CI Width}    & 0.0269 & 0.0241 & 0.0193 & 0.0122 & 0.0283 & 0.0251 & 0.0199 & 0.0124 & 0.0253 & 0.0227 & 0.0186 & 0.0119 & -- \\
\textbf{Std. CI Width}    & 0.0024 & 0.0022 & 0.0019 & 0.0015 & 0.0025 & 0.0023 & 0.0020 & 0.0015 & 0.0023 & 0.0022 & 0.0019 & 0.0014 & -- \\
\textbf{CI Breach \%}     & 12.6 & 13.2 & 13.0 & 12.6 & 13.0 & 12.8 & 12.6 & 12.4 & 12.8 & 13.0 & 12.6 & 13.0 & -- \\
\midrule
\multicolumn{14}{l}{\textbf{Model Result of Truncated Synthetic Dataset (70\% data truncation)}} \\
\midrule
\textbf{MAPE (\%)}        & 1.737 & 1.548 & 1.253 & 0.799 & 1.941 & 1.710 & 1.355 & 0.841 & 1.528 & 1.382 & 1.141 & 0.750 & 1.019 \\
\textbf{Std. of APE (\%)} & 1.342 & 1.189 & 0.956 & 0.608 & 1.515 & 1.320 & 1.036 & 0.641 & 1.175 & 1.058 & 0.870 & 0.572 & -- \\
\textbf{Mean CI Width}    & 0.0262 & 0.0233 & 0.0188 & 0.0119 & 0.0276 & 0.0244 & 0.0194 & 0.0121 & 0.0247 & 0.0222 & 0.0181 & 0.0116 & -- \\
\textbf{Std. CI Width}    & 0.0029 & 0.0027 & 0.0023 & 0.0016 & 0.0030 & 0.0027 & 0.0023 & 0.0017 & 0.0027 & 0.0026 & 0.0022 & 0.0016 & -- \\
\textbf{CI Breach \%}     & 13.0 & 13.2 & 13.4 & 13.0 & 13.2 & 13.4 & 13.4 & 12.8 & 13.4 & 13.4 & 13.4 & 13.6 & -- \\
\bottomrule
\end{tabular}%
}
\caption{Out-of-sample performance metrics for the conditional diffusion model. The Overall column reports the mean MAPE of all trading days of the test set. The CI Breach \% quantifies the proportion of true data that fall outside the predicted 90\% CIs.}
\label{tab:master_sim_metrics}
\end{table}

\section{Hyperparameters}
\label{sec:hyperp}
\begin{table}[H]
    \centering
    \caption{Summary of Model Hyperparameters}
    \label{tab:hyperparameters}
    \begin{tabular}{llc}
        \toprule
        \textbf{Notation} & \textbf{Description} & \textbf{Value} \\
        \midrule

        $N_{\text{epochs}}$ & Total number of training epochs & 2000 \\
        $B$ & Training batch size & 64 \\
        $T$ & Number of discrete diffusion timesteps & 500 \\
        $\eta_{\text{min}}$ & Minimum learning rate bound for plateau scheduler & 1e-6 \\
        $P_{es}$ & Early stopping patience & 200 \\
        $T_{es}$ & Early stopping threshold & 0.01 \\
        $\gamma_{\text{lr}}$ & Learning rate decay factor & 0.8 \\
        $P_{\text{lr}}$ & Patience (consecutive epochs) for learning rate scheduler & 300 \\
        $c_{\text{grad}}$ & Maximum $L_2$-norm for gradient clipping & 0.15 \\
        $\lambda$ & Strength of arbitrage regularization & 13.67 \\
        $\gamma$ & Maximum threshold for SNR capping & 20 \\
        $\beta_{\text{EWMA}}$ & Exponential moving average (EWMA) decay rate for network weights & 0.995 \\
        $\alpha_{\text{short}}$ & Short-term decay factor for historical vol-surfaces & 0.3333 \\
        $\alpha_{\text{long}}$ & Long-term decay factor for historical vol-surfaces & 0.0952 \\
        $\alpha_r^S$ & Short-term decay factor for underlying asset returns & 0.156 \\
        $\alpha_r^L$ & Long-term decay factor for underlying asset returns & 0.118 \\
        $\alpha_\sigma^S$ & Short-term decay factor for squared asset returns & 0.3 \\
        $\alpha_\sigma^L$ & Long-term decay factor for squared asset returns & 0.15 \\
        $h_m$ & Smoothing bandwidth across moneyness & 0.002 \\
        $h_\tau$ & Smoothing bandwidth across time-to-maturity (maturity) & 0.1 \\
        $d_{\text{emb}}$ & Sinusoidal time embedding dimension ($d_{\text{emb}} \geq 2$) & 24 \\
        $\eta_0$ & Initial learning rate & 5.3e-4 \\
        -- & AdamW weight decay & 0.01 \\
        -- & Hidden width  (number of hidden layers) of the context MLP & 10 (2) \\
        \bottomrule
    \end{tabular}
\end{table}

\begin{table}[htbp]
\centering
\caption{Hyperparameters of the benchmark models.}
\label{tab:benchmark_hyperparameters}
\begin{tabular}{lll}
\toprule
Model & Hyperparameter & Value \\
\midrule
PCA-AR(1) & Number of factors & 10 \\
SABR & Smoothing factor $\kappa_{\text{S}}$; $\beta$ & 0.5; 1.0 \\
LSTM+Attention & Window; layers; units; dropout & 3; 2; 243; 0.2 \\
\bottomrule
\end{tabular}
\end{table}

\section{Full Performance Metrics for All Trained Models}
\label{sec:appendix_full_table}
\begin{table}[H]
\centering
\caption{Full comparison of performance metrics for all 10 trained VolGAN models and the proposed Diffusion model (continued in Table~\ref{tab:full_performance_metrics_part2}). VolGAN 3 was selected as the benchmark for comparison in the main text due to having the lowest Overall MAPE. Each VolGAN is trained for 10,000 epochs and the model is included if no exploding or vanishing gradients are observed and the gradient norm oscillation is stable at the end of the training. VolGAN 9 was not stable at the end of the training.
}
\label{tab:full_performance_metrics_part1}
\resizebox{\textwidth}{!}{%
\begin{tabular}{l rrrrrrrrrrrrr}
\toprule

& \multicolumn{4}{c}{ATM} & \multicolumn{4}{c}{OTM} & \multicolumn{4}{c}{ITM} & \\
\cmidrule(lr){2-5} \cmidrule(lr){6-9} \cmidrule(lr){10-13}

Metric & 1-Day & 1-Week & 1-Month & 3-Month & 1-Day & 1-Week & 1-Month & 3-Month & 1-Day & 1-Week & 1-Month & 3-Month & Overall \\
\midrule

\multicolumn{14}{l}{\textbf{VolGAN 1}} \\
MAPE (\%) & 5.4848 & 5.4079 & 5.2264 & 6.6949 & 7.3509 & 7.3813 & 7.9716 & 6.0041 & 3.1493 & 3.1892 & 3.0474 & 3.7308 & 4.6804 \\
Std. of APE (\%) & 4.3854 & 4.8749 & 4.1418 & 4.7774 & 5.4913 & 5.3441 & 5.2115 & 4.0811 & 2.7181 & 2.7836 & 2.4780 & 2.5913 & \\
Mean CI Width & 0.0069 & 0.0074 & 0.0070 & 0.0078 & 0.0105 & 0.0118 & 0.0112 & 0.0091 & 0.0061 & 0.0051 & 0.0055 & 0.0040 & \\
Std. CI Width & 0.0020 & 0.0021 & 0.0017 & 0.0018 & 0.0018 & 0.0017 & 0.0016 & 0.0016 & 0.0013 & 0.0011 & 0.0012 & 0.0008 & \\
CI Breach \% & 71.8795 & 75.5140 & 82.2095 & 83.5007 & 74.7489 & 73.1707 & 77.7618 & 74.6055 & 84.2181 & 86.0832 & 84.7920 & 92.1091 & \\
\addlinespace

\multicolumn{14}{l}{\textbf{VolGAN 2}} \\
MAPE (\%) & 4.9045 & 4.8584 & 4.7725 & 4.2313 & 5.7716 & 6.5407 & 5.7373 & 3.4782 & 3.1053 & 3.0444 & 3.0586 & 2.5059 & 4.1946 \\
Std. of APE (\%) & 4.1963 & 4.1376 & 4.0179 & 3.6930 & 4.1804 & 4.7538 & 4.2727 & 2.4844 & 2.7093 & 2.7118 & 2.6850 & 2.3113 & \\
Mean CI Width & 0.0355 & 0.0385 & 0.0326 & 0.0295 & 0.0488 & 0.0428 & 0.0388 & 0.0271 & 0.0516 & 0.0505 & 0.0528 & 0.0411 & \\
Std. CI Width & 0.0209 & 0.0208 & 0.0200 & 0.0192 & 0.0167 & 0.0163 & 0.0154 & 0.0110 & 0.0262 & 0.0247 & 0.0247 & 0.0203 & \\
CI Breach \% & 16.2124 & 16.3558 & 15.6385 & 15.4950 & 12.6255 & 12.6255 & 10.7604 & 0.4692 & 19.0818 & 18.9383 & 17.5036 & 17.0732 & \\
\addlinespace

\multicolumn{14}{l}{\textbf{VolGAN 3}} \\
MAPE (\%) & 4.8300 & 4.8265 & 4.6083 & 4.1221 & 4.8253 & 4.7117 & 4.6687 & 4.2101 & 2.9745 & 2.8706 & 2.9407 & 2.7524 & 3.8425 \\
Std. of APE (\%) & 4.3137 & 4.1934 & 4.2649 & 3.7653 & 3.5718 & 3.5140 & 3.4992 & 3.0106 & 2.4690 & 2.3908 & 2.4150 & 2.2559 & \\
Mean CI Width & 0.0265 & 0.0270 & 0.0245 & 0.0225 & 0.0417 & 0.0410 & 0.0382 & 0.0273 & 0.0197 & 0.0222 & 0.0178 & 0.0143 & \\
Std. CI Width & 0.0104 & 0.0106 & 0.0100 & 0.0089 & 0.0168 & 0.0167 & 0.0157 & 0.0111 & 0.0065 & 0.0066 & 0.0066 & 0.0075 & \\
CI Breach \% & 23.8506 & 23.2759 & 26.7241 & 26.0057 & 14.5115 & 13.7931 & 14.2241 & 22.5575 & 51.7241 & 46.5517 & 55.0287 & 62.2126 & \\
\addlinespace

\multicolumn{14}{l}{\textbf{VolGAN 4}} \\
MAPE (\%) & 5.1090 & 5.0465 & 4.9615 & 4.5056 & 3.4758 & 9.6190 & 9.9897 & 9.3729 & 3.1169 & 3.0795 & 3.2810 & 2.6945 & 4.8997 \\
Std. of APE (\%) & 4.5452 & 4.3856 & 4.3040 & 4.0932 & 5.5650 & 5.8262 & 5.4114 & 4.5096 & 2.6147 & 2.5816 & 2.5417 & 2.2061 & \\
Mean CI Width & 0.0248 & 0.0249 & 0.0235 & 0.0242 & 0.0292 & 0.0266 & 0.0284 & 0.0224 & 0.0387 & 0.0345 & 0.0330 & 0.0307 & \\
Std. CI Width & 0.0055 & 0.0056 & 0.0057 & 0.0062 & 0.0119 & 0.0112 & 0.0121 & 0.0103 & 0.0040 & 0.0035 & 0.0032 & 0.0030 & \\
CI Breach \% & 26.5423 & 25.9685 & 28.2639 & 24.3902 & 58.6586 & 55.0933 & 62.2669 & 73.1707 & 22.0947 & 20.5165 & 27.1162 & 24.1033 & \\
\addlinespace

\multicolumn{14}{l}{\textbf{VolGAN 5}} \\
MAPE (\%) & 5.8201 & 5.7422 & 5.6428 & 5.0018 & 7.0458 & 6.8858 & 6.5098 & 4.9647 & 2.7388 & 2.6938 & 2.7529 & 2.3215 & 4.5160 \\
Std. of APE (\%) & 4.4566 & 4.3180 & 4.3644 & 3.8236 & 4.6511 & 4.8184 & 4.4095 & 3.4965 & 2.4420 & 2.4070 & 2.4130 & 2.1106 & \\
Mean CI Width & 0.0229 & 0.0225 & 0.0236 & 0.0245 & 0.0639 & 0.0529 & 0.0482 & 0.0365 & 0.0374 & 0.0372 & 0.0347 & 0.0297 & \\
Std. CI Width & 0.0128 & 0.0126 & 0.0128 & 0.0121 & 0.0240 & 0.0238 & 0.0200 & 0.0149 & 0.0149 & 0.0146 & 0.0134 & 0.0112 & \\
CI Breach \% & 52.0803 & 51.0760 & 50.2152 & 45.7675 & 13.7733 & 18.8235 & 16.2124 & 14.3472 & 25.5380 & 25.6815 & 28.5509 & 25.2511 & \\
\bottomrule
\end{tabular}
}
\end{table}

\newpage

\begin{table}[H]
\centering
\caption{Full comparison of performance metrics for all 10 trained VolGAN models and the proposed Diffusion model.}
\label{tab:full_performance_metrics_part2}
\resizebox{\textwidth}{!}{%
\begin{tabular}{l rrrrrrrrrrrrr}
\toprule

& \multicolumn{4}{c}{ATM} & \multicolumn{4}{c}{OTM} & \multicolumn{4}{c}{ITM} & \\
\cmidrule(lr){2-5} \cmidrule(lr){6-9} \cmidrule(lr){10-13}

Metric & 1-Day & 1-Week & 1-Month & 3-Month & 1-Day & 1-Week & 1-Month & 3-Month & 1-Day & 1-Week & 1-Month & 3-Month & Overall \\
\midrule

\multicolumn{14}{l}{\textbf{VolGAN 6}} \\
MAPE (\%) & 4.9009 & 4.8532 & 4.7405 & 4.1738 & 5.0297 & 5.8872 & 5.4906 & 4.3717 & 2.7423 & 2.6738 & 3.3214 & 4.1283 & 4.2254 \\
Std. of APE (\%) & 4.3810 & 4.2428 & 4.2006 & 3.7055 & 4.5995 & 4.4943 & 4.1858 & 3.2465 & 2.3548 & 2.3079 & 2.6845 & 2.7560 & \\
Mean CI Width & 0.0148 & 0.0143 & 0.0148 & 0.0139 & 0.0365 & 0.0355 & 0.0331 & 0.0236 & 0.0290 & 0.0292 & 0.0250 & 0.0240 & \\
Std. CI Width & 0.0076 & 0.0073 & 0.0080 & 0.0072 & 0.0210 & 0.0191 & 0.0196 & 0.0129 & 0.0086 & 0.0085 & 0.0077 & 0.0069 & \\
CI Breach \% & 48.6413 & 51.3630 & 49.0674 & 46.9154 & 34.5768 & 34.4333 & 31.9943 & 39.1679 & 26.3989 & 24.9641 & 42.6112 & 61.4060 & \\
\addlinespace

\multicolumn{14}{l}{\textbf{VolGAN 7}} \\
MAPE (\%) & 4.7346 & 4.6830 & 4.5588 & 4.1578 & 4.5364 & 4.4698 & 4.2965 & 3.5937 & 2.5437 & 2.5627 & 2.9411 & 3.4803 & 3.8615 \\
Std. of APE (\%) & 4.1708 & 4.2017 & 4.0521 & 3.7747 & 3.5830 & 3.5084 & 3.3572 & 3.2544 & 2.4621 & 2.5217 & 2.8368 & 2.8927 & \\
Mean CI Width & 0.0154 & 0.0145 & 0.0144 & 0.0097 & 0.0328 & 0.0323 & 0.0282 & 0.0185 & 0.0400 & 0.0383 & 0.0431 & 0.0287 & \\
Std. CI Width & 0.0044 & 0.0043 & 0.0043 & 0.0057 & 0.0104 & 0.0102 & 0.0106 & 0.0072 & 0.0054 & 0.0053 & 0.0065 & 0.0045 & \\
CI Breach \% & 44.6198 & 47.0588 & 46.4849 & 56.9584 & 14.0603 & 12.7690 & 15.0646 & 39.1679 & 12.9125 & 13.6298 & 14.5841 & 38.8809 & \\
\addlinespace

\multicolumn{14}{l}{\textbf{VolGAN 8}} \\
MAPE (\%) & 6.1914 & 5.9409 & 5.6571 & 5.3297 & 5.8692 & 6.9996 & 7.2199 & 6.4400 & 3.9085 & 3.9522 & 3.6069 & 3.1119 & 4.4841 \\
Std. of APE (\%) & 4.9130 & 4.8142 & 4.5157 & 4.1559 & 4.7578 & 4.7563 & 4.7586 & 4.4847 & 2.9786 & 3.0048 & 2.9723 & 2.5855 & \\
Mean CI Width & 0.0121 & 0.0121 & 0.0103 & 0.0100 & 0.0615 & 0.0604 & 0.0571 & 0.0422 & 0.0256 & 0.0259 & 0.0259 & 0.0236 & \\
Std. CI Width & 0.0038 & 0.0039 & 0.0028 & 0.0026 & 0.0368 & 0.0361 & 0.0334 & 0.0244 & 0.0068 & 0.0072 & 0.0072 & 0.0058 & \\
CI Breach \% & 64.5624 & 64.7059 & 68.7231 & 70.0143 & 19.6556 & 20.6599 & 26.2554 & 34.5768 & 47.7762 & 49.3544 & 45.1937 & 41.4634 & \\
\addlinespace

\multicolumn{14}{l}{\textbf{VolGAN 9}} \\
MAPE (\%) & 5.1526 & 5.2093 & 5.3758 & 8.9616 & 12.0051 & 11.9373 & 11.1095 & 4.8830 & 22.6214 & 22.4364 & 14.8209 & 12.6213 & 9.0688 \\
Std. of APE (\%) & 4.2558 & 4.2584 & 4.2575 & 5.9750 & 5.1833 & 5.0434 & 4.6224 & 3.5354 & 9.2836 & 9.3468 & 6.7452 & 7.1482 & \\
Mean CI Width & 0.0008 & 0.0008 & 0.0011 & 0.0011 & 0.0042 & 0.0040 & 0.0038 & 0.0029 & 0.0066 & 0.0068 & 0.0058 & 0.0065 & \\
Std. CI Width & 0.0002 & 0.0002 & 0.0003 & 0.0003 & 0.0007 & 0.0007 & 0.0007 & 0.0005 & 0.0011 & 0.0012 & 0.0010 & 0.0012 & \\
CI Breach \% & 97.7045 & 96.9871 & 95.2654 & 98.5653 & 98.7088 & 98.7088 & 98.8522 & 88.9526 & 100.0000 & 100.0000 & 98.2783 & 98.7088 & \\
\addlinespace

\multicolumn{14}{l}{\textbf{VolGAN 10}} \\
MAPE (\%) & 4.8474 & 4.8295 & 4.6663 & 4.2226 & 7.7716 & 7.5443 & 6.7917 & 4.2674 & 3.2081 & 3.1988 & 3.0942 & 2.4420 & 4.1273 \\
Std. of APE (\%) & 4.2239 & 4.1958 & 4.0956 & 3.7226 & 4.9421 & 4.8144 & 4.3522 & 3.1403 & 2.5716 & 2.5842 & 2.5012 & 2.0886 & \\
Mean CI Width & 0.0122 & 0.0125 & 0.0116 & 0.0088 & 0.0312 & 0.0301 & 0.0264 & 0.0153 & 0.0194 & 0.0188 & 0.0166 & 0.0107 & \\
Std. CI Width & 0.0053 & 0.0056 & 0.0059 & 0.0036 & 0.0169 & 0.0163 & 0.0138 & 0.0059 & 0.0119 & 0.0108 & 0.0085 & 0.0037 & \\
CI Breach \% & 59.8277 & 59.8346 & 62.5538 & 66.5710 & 26.1206 & 27.5466 & 25.1076 & 55.9540 & 57.6753 & 58.8235 & 60.2582 & 63.5581 & \\
\bottomrule

\multicolumn{14}{l}{\textbf{Diffusion}} \\
MAPE (\%) & 4.6907 & 4.6884 & 4.5146 & 4.0202 & 3.5264 & 3.4573 & 3.1542 & 2.4555 & 2.5129 & 2.4992 & 2.4184 & 2.2024 & 3.0026 \\
Std. of APE (\%) & 4.0338 & 4.0091 & 3.8795 & 3.4909 & 2.6588 & 2.5785 & 2.3190 & 1.8739 & 2.3984 & 2.3844 & 2.2909 & 2.0393 &  \\
Mean CI Width & 0.0372 & 0.0369 & 0.0357 & 0.0327 & 0.0468 & 0.0457 & 0.0412 & 0.0311 & 0.0472 & 0.0463 & 0.0438 & 0.0384 &  \\
Std. CI Width & 0.0160 & 0.0159 & 0.0152 & 0.0134 & 0.0098 & 0.0095 & 0.0082 & 0.0057 & 0.0139 & 0.0137 & 0.0132 & 0.0124 & \\
CI Breach \% & 10.9195 & 10.4885 & 11.0632 & 10.0575 & 0.5747 & 0.7184 & 0.5747 & 1.0057 & 7.9023 & 7.9023 & 8.4770 & 7.9023 &  \\
\bottomrule

\end{tabular}
}
\end{table}

\begin{table}[H]
\centering
\caption{Seed comparison of performance metrics for the diffusion model for nine additional random seeds; the model reported in the main text is the tenth.}
\label{tab:seed_comparison}
\resizebox{\textwidth}{!}{%
\begin{tabular}{l rrrrrrrrrrrrr}
\toprule

& \multicolumn{4}{c}{ATM} & \multicolumn{4}{c}{OTM} & \multicolumn{4}{c}{ITM} & \\
\cmidrule(lr){2-5} \cmidrule(lr){6-9} \cmidrule(lr){10-13}

Metric & 1-Day & 1-Week & 1-Month & 3-Month & 1-Day & 1-Week & 1-Month & 3-Month & 1-Day & 1-Week & 1-Month & 3-Month & Overall \\
\midrule

\multicolumn{14}{l}{\textbf{Seed 1}} \\
MAPE (\%) & 4.6981 & 4.6546 & 4.4943 & 4.0106 & 3.4188 & 3.3467 & 3.0416 & 2.3692 & 2.4510 & 2.4259 & 2.3464 & 2.1512 & 2.9643 \\
Std. of APE (\%) & 3.9449 & 3.9133 & 3.8013 & 3.4205 & 2.7140 & 2.6422 & 2.3479 & 1.9101 & 2.3010 & 2.2883 & 2.1937 & 1.9508 & \\
Mean CI Width & 0.0370 & 0.0367 & 0.0358 & 0.0333 & 0.0469 & 0.0460 & 0.0417 & 0.0313 & 0.0471 & 0.0461 & 0.0433 & 0.0382 & \\
Std. CI Width & 0.0144 & 0.0143 & 0.0139 & 0.0122 & 0.0091 & 0.0089 & 0.0077 & 0.0055 & 0.0126 & 0.0125 & 0.0121 & 0.0112 & \\
CI Breach \% & 9.9138 & 10.2011 & 10.0575 & 9.3391 & 0.4310 & 0.5747 & 0.4310 & 1.0057 & 8.0460 & 8.0460 & 8.0460 & 7.1839 & \\
\addlinespace

\multicolumn{14}{l}{\textbf{Seed 2}} \\
MAPE (\%) & 4.6637 & 4.6226 & 4.4680 & 3.9468 & 3.6245 & 3.5527 & 3.2002 & 2.3907 & 2.5441 & 2.5114 & 2.3956 & 2.1465 & 2.9901 \\
Std. of APE (\%) & 4.0029 & 3.9681 & 3.8564 & 3.4938 & 2.8249 & 2.7602 & 2.4539 & 1.8902 & 2.3981 & 2.3598 & 2.2559 & 1.9918 & \\
Mean CI Width & 0.0360 & 0.0356 & 0.0350 & 0.0325 & 0.0512 & 0.0502 & 0.0450 & 0.0332 & 0.0480 & 0.0477 & 0.0441 & 0.0380 & \\
Std. CI Width & 0.0131 & 0.0130 & 0.0126 & 0.0115 & 0.0095 & 0.0093 & 0.0083 & 0.0058 & 0.0127 & 0.0126 & 0.0120 & 0.0111 & \\
CI Breach \% & 10.4885 & 10.0575 & 9.3391 & 8.4770 & 0.2874 & 0.2874 & 0.4310 & 1.1494 & 7.4713 & 7.1839 & 7.4713 & 7.1839 & \\
\addlinespace

\multicolumn{14}{l}{\textbf{Seed 3}} \\
MAPE (\%) & 4.6491 & 4.6082 & 4.4516 & 3.9434 & 3.7511 & 3.6679 & 3.3232 & 2.4559 & 2.5326 & 2.5085 & 2.4085 & 2.1717 & 3.0954 \\
Std. of APE (\%) & 4.0587 & 4.0396 & 3.9192 & 3.5633 & 2.8289 & 2.7539 & 2.4744 & 1.9342 & 2.3646 & 2.3396 & 2.2319 & 1.9693 & \\
Mean CI Width & 0.0369 & 0.0367 & 0.0356 & 0.0330 & 0.0501 & 0.0490 & 0.0435 & 0.0301 & 0.0466 & 0.0460 & 0.0432 & 0.0375 & \\
Std. CI Width & 0.0163 & 0.0163 & 0.0155 & 0.0140 & 0.0111 & 0.0108 & 0.0095 & 0.0062 & 0.0150 & 0.0151 & 0.0145 & 0.0132 & \\
CI Breach \% & 11.0632 & 10.9195 & 10.4885 & 9.4828 & 0.7184 & 0.7184 & 1.2931 & 2.0115 & 9.1954 & 9.1954 & 8.4770 & 7.6149 & \\
\addlinespace

\multicolumn{14}{l}{\textbf{Seed 4}} \\
MAPE (\%) & 4.6436 & 4.6048 & 4.4370 & 3.9435 & 3.6474 & 3.5541 & 3.2555 & 2.4345 & 2.5163 & 2.5042 & 2.4135 & 2.1964 & 2.9742 \\
Std. of APE (\%) & 4.0689 & 4.0483 & 3.9245 & 3.5430 & 2.9443 & 2.8628 & 2.5514 & 1.9070 & 2.3683 & 2.3519 & 2.2558 & 2.0144 & \\
Mean CI Width & 0.0350 & 0.0348 & 0.0339 & 0.0315 & 0.0463 & 0.0452 & 0.0409 & 0.0306 & 0.0457 & 0.0450 & 0.0424 & 0.0368 & \\
Std. CI Width & 0.0139 & 0.0137 & 0.0134 & 0.0121 & 0.0102 & 0.0099 & 0.0087 & 0.0060 & 0.0139 & 0.0137 & 0.0132 & 0.0118 & \\
CI Breach \% & 12.6437 & 12.3563 & 11.7816 & 11.0632 & 0.7184 & 1.0057 & 0.8621 & 1.7241 & 9.4828 & 9.4828 & 9.3391 & 9.4828 & \\
\addlinespace

\multicolumn{14}{l}{\textbf{Seed 5}} \\
MAPE (\%) & 4.6579 & 4.6117 & 4.4649 & 3.9823 & 3.6020 & 3.5185 & 3.1735 & 2.3795 & 2.4558 & 2.4403 & 2.3618 & 2.1399 & 2.9449 \\
Std. of APE (\%) & 4.0931 & 4.0611 & 3.9447 & 3.5338 & 2.7711 & 2.6926 & 2.4005 & 1.8955 & 2.3676 & 2.3489 & 2.2623 & 2.0097 & \\
Mean CI Width & 0.0345 & 0.0342 & 0.0333 & 0.0309 & 0.0523 & 0.0513 & 0.0462 & 0.0334 & 0.0459 & 0.0451 & 0.0427 & 0.0376 & \\
Std. CI Width & 0.0136 & 0.0135 & 0.0131 & 0.0117 & 0.0122 & 0.0119 & 0.0105 & 0.0070 & 0.0141 & 0.0138 & 0.0132 & 0.0118 & \\
CI Breach \% & 11.7816 & 11.6379 & 11.6379 & 10.7759 & 0.2874 & 0.2874 & 0.2874 & 0.4310 & 8.1897 & 8.4770 & 8.4770 & 7.4713 & \\
\addlinespace

\multicolumn{14}{l}{\textbf{Seed 6}} \\
MAPE (\%) & 4.5671 & 4.5329 & 4.3735 & 3.9003 & 3.4283 & 3.3388 & 3.0173 & 2.3142 & 2.4514 & 2.4372 & 2.3473 & 2.1295 & 2.9199 \\
Std. of APE (\%) & 4.0194 & 3.9895 & 3.8760 & 3.4944 & 2.7341 & 2.6468 & 2.3365 & 1.8154 & 2.3283 & 2.3046 & 2.1999 & 1.9227 & \\
Mean CI Width & 0.0347 & 0.0343 & 0.0335 & 0.0312 & 0.0409 & 0.0396 & 0.0355 & 0.0268 & 0.0440 & 0.0433 & 0.0412 & 0.0365 & \\
Std. CI Width & 0.0149 & 0.0148 & 0.0144 & 0.0130 & 0.0085 & 0.0083 & 0.0073 & 0.0056 & 0.0151 & 0.0149 & 0.0144 & 0.0129 & \\
CI Breach \% & 11.9253 & 12.0690 & 11.6379 & 10.6322 & 1.2931 & 1.1494 & 1.1494 & 1.4368 & 9.9138 & 9.7701 & 9.3391 & 7.4713 & \\
\addlinespace

\multicolumn{14}{l}{\textbf{Seed 7}} \\
MAPE (\%) & 4.5831 & 4.5555 & 4.4104 & 3.9438 & 3.9669 & 3.8833 & 3.4694 & 2.5592 & 2.6542 & 2.6195 & 2.4981 & 2.1939 & 3.0781 \\
Std. of APE (\%) & 4.2193 & 4.1883 & 4.0651 & 3.6953 & 3.0132 & 2.9251 & 2.5278 & 1.9319 & 2.5019 & 2.4707 & 2.3576 & 2.0448 & \\
Mean CI Width & 0.0347 & 0.0345 & 0.0340 & 0.0320 & 0.0519 & 0.0505 & 0.0448 & 0.0318 & 0.0449 & 0.0445 & 0.0422 & 0.0375 & \\
Std. CI Width & 0.0153 & 0.0153 & 0.0150 & 0.0137 & 0.0114 & 0.0110 & 0.0097 & 0.0067 & 0.0172 & 0.0170 & 0.0164 & 0.0145 & \\
CI Breach \% & 11.0632 & 11.6379 & 11.2069 & 10.6322 & 0.1437 & 0.4310 & 0.7184 & 0.8621 & 11.0632 & 10.4885 & 10.3448 & 9.0517 & \\
\addlinespace

\multicolumn{14}{l}{\textbf{Seed 8}} \\
MAPE (\%) & 4.7813 & 4.7478 & 4.5782 & 4.0746 & 3.8681 & 3.7603 & 3.4178 & 2.4772 & 2.4431 & 2.4166 & 2.3353 & 2.1293 & 3.0141 \\
Std. of APE (\%) & 3.9887 & 3.9569 & 3.8423 & 3.4592 & 3.1398 & 3.0391 & 2.6652 & 1.9129 & 2.3215 & 2.2959 & 2.1892 & 1.9413 & \\
Mean CI Width & 0.0369 & 0.0363 & 0.0356 & 0.0333 & 0.0510 & 0.0495 & 0.0443 & 0.0323 & 0.0467 & 0.0464 & 0.0438 & 0.0387 & \\
Std. CI Width & 0.0148 & 0.0146 & 0.0142 & 0.0130 & 0.0137 & 0.0133 & 0.0119 & 0.0077 & 0.0146 & 0.0144 & 0.0141 & 0.0127 & \\
CI Breach \% & 11.4943 & 11.6379 & 10.6322 & 9.4828 & 0.5747 & 0.4310 & 1.1494 & 1.5805 & 6.6092 & 7.0402 & 6.0345 & 6.6092 & \\
\addlinespace

\multicolumn{14}{l}{\textbf{Seed 9}} \\
MAPE (\%) & 4.7119 & 4.6734 & 4.5146 & 4.0113 & 3.6496 & 3.5659 & 3.2304 & 2.3985 & 2.4596 & 2.4358 & 2.3547 & 2.1552 & 2.9748 \\
Std. of APE (\%) & 4.0505 & 4.0186 & 3.8833 & 3.5022 & 2.7855 & 2.7004 & 2.3929 & 1.8516 & 2.2588 & 2.2384 & 2.1513 & 1.9454 & \\
Mean CI Width & 0.0368 & 0.0365 & 0.0355 & 0.0330 & 0.0481 & 0.0468 & 0.0417 & 0.0298 & 0.0479 & 0.0472 & 0.0449 & 0.0394 & \\
Std. CI Width & 0.0146 & 0.0143 & 0.0139 & 0.0126 & 0.0110 & 0.0106 & 0.0093 & 0.0061 & 0.0149 & 0.0147 & 0.0141 & 0.0123 & \\
CI Breach \% & 10.4885 & 10.6322 & 10.9195 & 8.7644 & 0.5747 & 0.5747 & 0.5747 & 1.0057 & 6.4655 & 5.8908 & 6.3218 & 6.6092 & \\
\bottomrule

\end{tabular}
}
\end{table}

\begin{table}[htbp]
\centering

\caption{Mean $\pm$ standard deviation [minimum, maximum] over ten diffusion seeds (main-text model and Table~\ref{tab:seed_comparison}) and ten VolGAN runs (Tables~\ref{tab:full_performance_metrics_part1}--\ref{tab:full_performance_metrics_part2}). All values in \%.}
\label{tab:seed_summary}
\resizebox{\textwidth}{!}{%
\begin{tabular}{lccc}
\toprule
Metric & Diffusion (10 seeds) & VolGAN (10 runs) & VolGAN excl.\ run 9 \\
\midrule
Overall MAPE & $2.996 \pm 0.055$ [2.920, 3.095] & $4.790 \pm 1.541$ [3.843, 9.069] & $4.315 \pm 0.359$ [3.843, 4.900] \\
ATM 1M breach & $10.88 \pm 0.77$ [9.34, 11.78] & $52.52 \pm 25.25$ [15.64, 95.27] & $47.76 \pm 21.53$ [15.64, 82.21] \\
ITM 1M breach & $8.23 \pm 1.34$ [6.03, 10.34] & $47.39 \pm 27.84$ [14.58, 98.28] & $41.74 \pm 22.63$ [14.58, 84.79] \\
OTM 1M breach & $0.75 \pm 0.35$ [0.29, 1.29] & $37.85 \pm 30.77$ [10.76, 98.85] & $31.07 \pm 23.42$ [10.76, 77.76] \\
\bottomrule
\end{tabular}
}
\par\smallskip\footnotesize Every diffusion seed has a lower overall MAPE than the best VolGAN run.

\end{table}

\section{Diagnostics of OTM Calibration}
\label{sec:otm_diagnostics}
This appendix reports the diagnostics referred to in Section~\ref{subsec:real_results}. Table~\ref{tab:coverage_levels} gives the empirical coverage of central predictive intervals at several nominal levels; a coverage above the nominal level at every level indicates over-dispersion. Table~\ref{tab:shift} compares the OTM data across the sample periods, and Table~\ref{tab:otm_split} splits the OTM breach rate by whether the grid point had to be extrapolated.

Table~\ref{tab:coverage_levels} shows that the OTM intervals are too wide at every nominal level, whereas the ATM and ITM intervals are close to nominal. Table~\ref{tab:shift} shows that the OTM data are far noisier and more often extrapolated in the training period than in the test period. This shift is much larger in the OTM region. Table~\ref{tab:otm_split} shows that the over-dispersion does not depend on whether the test-day value was interpolated or extrapolated. Together, these results indicate that the model carries the dispersion of the noisier training period into the test period with lower standard deviation.

\begin{table}[htbp]
\centering

\caption{Empirical coverage of central predictive intervals at several nominal levels, 1-month points (test set, \%).}
\label{tab:coverage_levels}
\begin{tabular}{lcccc}
\toprule
Point & 50\% & 80\% & 90\% & 95\% \\
\midrule
ATM & 48.42 & 78.88 & 88.94 & 94.11 \\
ITM & 53.02 & 83.48 & 91.52 & 94.83 \\
OTM & 75.00 & 97.13 & 99.43 & 99.71 \\
\bottomrule
\end{tabular}

\end{table}

\begin{table}[htbp]
\centering

\begin{threeparttable}
\caption{Grid point statistics in the training, validation and test periods.}
\label{tab:shift}
\begin{tabular}{lcccc}
\toprule
Statistic & Training & Validation & Test & Train / Test \\
\midrule
\multicolumn{5}{l}{Std.\ of daily changes in log IV ($\times 100$), 1-month} \\
\quad ATM ($m = 1.0$) & 5.538 & 8.023 & 6.200 & 0.89 \\
\quad ITM ($m = 0.7$) & 5.800 & 4.247 & 3.250 & 1.78 \\
\quad OTM ($m = 1.3$) & 23.216 & 12.243 & 4.120 & 5.64 \\
Share of days with OTM 1M point extrapolated (\%) & 88.76 & 70.69 & 50.14 & 1.77 \\
Share of penalty $\Phi$ from OTM grid points (\%) & 79.90 & -- & -- & -- \\
\bottomrule
\end{tabular}
\begin{tablenotes}
\footnotesize
\item \textit{Note:} Train / Test is the ratio of the training-period value to the test-period value. OTM grid points ($m = 1.1$--$1.4$) make up 4 of the 9 moneyness levels (44.4\%).
\end{tablenotes}
\end{threeparttable}

\end{table}

\begin{table}[htbp]
\centering

\caption{OTM 1-month breach rates by whether the grid point was interpolated or extrapolated on that day (test set).}
\label{tab:otm_split}
\begin{tabular}{lcc}
\toprule
Days & Number of days & Breach rate (\%) \\
\midrule
Interpolated & 347 & 0.86 \\
Extrapolated & 349 & 0.29 \\
\bottomrule
\end{tabular}

\end{table}

\section{Full Performance Metrics for All Ablation Studies}
\label{sec:ablation_appendix}

\begin{table}[H]
\centering
\caption{Full results for the ablation studies with mean results reported across the ten seeds. The original model is the representative diffusion model in the main text. }
\label{tab:ablation_full}
\resizebox{0.8\textwidth}{!}{%
\begin{tabular}{l rrrrrrrrrrrrr}
\toprule

& \multicolumn{4}{c}{ATM} & \multicolumn{4}{c}{OTM} & \multicolumn{4}{c}{ITM} & \\
\cmidrule(lr){2-5} \cmidrule(lr){6-9} \cmidrule(lr){10-13}

Metric & 1-Day & 1-Week & 1-Month & 3-Month & 1-Day & 1-Week & 1-Month & 3-Month & 1-Day & 1-Week & 1-Month & 3-Month & Overall \\
\midrule

\multicolumn{14}{l}{\textbf{No Current Surface}} \\
MAPE (\%) & 5.6727 & 5.6232 & 5.4376 & 4.8036 & 3.8686 & 3.7925 & 3.4547 & 2.6606 & 2.8009 & 2.7837 & 2.7042 & 2.4708 & 3.3488 \\
Std. of APE (\%) & 4.7661 & 4.7299 & 4.5750 & 4.1059 & 2.9851 & 2.9089 & 2.5864 & 2.0220 & 2.6258 & 2.6027 & 2.4990 & 2.2196 & \\
Mean CI Width & 0.0451 & 0.0448 & 0.0438 & 0.0407 & 0.0546 & 0.0537 & 0.0478 & 0.0351 & 0.0613 & 0.0606 & 0.0576 & 0.0505 & \\
Std. CI Width & 0.0143 & 0.0142 & 0.0140 & 0.0128 & 0.0106 & 0.0105 & 0.0096 & 0.0065 & 0.0139 & 0.0139 & 0.0134 & 0.0126 & \\
CI Breach \% & 10.4885 & 9.9138 & 9.7701 & 8.1897 & 0.5747 & 0.4310 & 0.7184 & 0.4310 & 4.8851 & 4.8851 & 4.5977 & 4.0230 & \\
\addlinespace

\multicolumn{14}{l}{\textbf{No Long Surface}} \\
MAPE (\%) & 4.7074 & 4.6780 & 4.5202 & 3.9991 & 3.5767 & 3.4996 & 3.1683 & 2.4416 & 2.4764 & 2.4565 & 2.3702 & 2.1467 & 3.0359 \\
Std. of APE (\%) & 4.0441 & 4.0103 & 3.8864 & 3.5061 & 2.7270 & 2.6520 & 2.3563 & 1.8015 & 2.2977 & 2.2671 & 2.1834 & 1.9552 & \\
Mean CI Width & 0.0372 & 0.0370 & 0.0360 & 0.0332 & 0.0516 & 0.0505 & 0.0453 & 0.0327 & 0.0471 & 0.0465 & 0.0440 & 0.0376 & \\
Std. CI Width & 0.0132 & 0.0131 & 0.0126 & 0.0114 & 0.0109 & 0.0107 & 0.0095 & 0.0060 & 0.0134 & 0.0134 & 0.0130 & 0.0114 & \\
CI Breach \% & 8.6207 & 9.1954 & 9.0517 & 8.3333 & 0.1437 & 0.1437 & 0.1437 & 0.4310 & 7.1839 & 7.1839 & 7.4713 & 7.4713 & \\
\addlinespace

\multicolumn{14}{l}{\textbf{No Return}} \\
MAPE (\%) & 4.6248 & 4.6018 & 4.4543 & 3.9670 & 3.4589 & 3.3681 & 3.0443 & 2.4020 & 2.4588 & 2.4449 & 2.3541 & 2.1332 & 3.0098 \\
Std. of APE (\%) & 4.0517 & 4.0121 & 3.8762 & 3.4968 & 2.6415 & 2.5600 & 2.2936 & 1.8830 & 2.3104 & 2.2937 & 2.1970 & 1.9379 & \\
Mean CI Width & 0.0323 & 0.0321 & 0.0314 & 0.0290 & 0.0415 & 0.0404 & 0.0361 & 0.0263 & 0.0414 & 0.0407 & 0.0385 & 0.0338 & \\
Std. CI Width & 0.0123 & 0.0124 & 0.0119 & 0.0107 & 0.0088 & 0.0086 & 0.0077 & 0.0054 & 0.0110 & 0.0111 & 0.0107 & 0.0097 & \\
CI Breach \% & 13.6494 & 13.6494 & 13.5057 & 12.6437 & 0.8621 & 1.0057 & 1.2931 & 1.7241 & 9.9138 & 9.9138 & 10.3448 & 9.9138 & \\
\addlinespace

\multicolumn{14}{l}{\textbf{No Short Surface}} \\
MAPE (\%) & 4.8144 & 4.7736 & 4.6040 & 4.0647 & 3.4190 & 3.3271 & 2.9895 & 2.3070 & 2.4787 & 2.4566 & 2.3750 & 2.1583 & 2.9994 \\
Std. of APE (\%) & 4.0132 & 3.9799 & 3.8564 & 3.4690 & 2.6476 & 2.5688 & 2.2812 & 1.7874 & 2.2724 & 2.2447 & 2.1520 & 1.9327 & \\
Mean CI Width & 0.0349 & 0.0345 & 0.0334 & 0.0310 & 0.0429 & 0.0421 & 0.0376 & 0.0273 & 0.0446 & 0.0442 & 0.0419 & 0.0364 & \\
Std. CI Width & 0.0121 & 0.0121 & 0.0117 & 0.0107 & 0.0089 & 0.0086 & 0.0076 & 0.0052 & 0.0122 & 0.0124 & 0.0119 & 0.0107 & \\
CI Breach \% & 12.9310 & 12.9310 & 12.9310 & 11.3506 & 0.5747 & 0.4310 & 0.8621 & 1.0057 & 8.0460 & 8.1897 & 8.4770 & 8.4770 & \\
\addlinespace

\multicolumn{14}{l}{\textbf{No Square Return}} \\
MAPE (\%) & 4.7354 & 4.6977 & 4.5367 & 4.0290 & 3.5059 & 3.4176 & 3.0579 & 2.3451 & 2.5366 & 2.5295 & 2.4479 & 2.2109 & 3.0152 \\
Std. of APE (\%) & 4.0793 & 4.0582 & 3.9361 & 3.5616 & 2.6796 & 2.6037 & 2.3498 & 1.8905 & 2.3737 & 2.3614 & 2.2744 & 2.0151 & \\
Mean CI Width & 0.0340 & 0.0339 & 0.0328 & 0.0306 & 0.0387 & 0.0377 & 0.0338 & 0.0255 & 0.0417 & 0.0411 & 0.0389 & 0.0338 & \\
Std. CI Width & 0.0177 & 0.0177 & 0.0170 & 0.0153 & 0.0083 & 0.0081 & 0.0075 & 0.0060 & 0.0166 & 0.0165 & 0.0157 & 0.0143 & \\
CI Breach \% & 14.2241 & 14.0805 & 14.6552 & 11.7816 & 1.5805 & 1.8678 & 2.2989 & 2.1552 & 10.3448 & 10.7759 & 10.2011 & 10.7759 & \\
\addlinespace

\multicolumn{14}{l}{\textbf{No VIX}} \\
MAPE (\%) & 4.7037 & 4.6680 & 4.5047 & 3.9864 & 3.5276 & 3.4479 & 3.1039 & 2.3662 & 2.4532 & 2.4374 & 2.3566 & 2.1459 & 2.9492 \\
Std. of APE (\%) & 4.0253 & 3.9988 & 3.8827 & 3.4906 & 2.6742 & 2.5949 & 2.3181 & 1.7929 & 2.2710 & 2.2514 & 2.1564 & 1.9294 & \\
Mean CI Width & 0.0359 & 0.0357 & 0.0348 & 0.0323 & 0.0463 & 0.0451 & 0.0405 & 0.0300 & 0.0449 & 0.0446 & 0.0429 & 0.0378 & \\
Std. CI Width & 0.0157 & 0.0156 & 0.0151 & 0.0137 & 0.0106 & 0.0103 & 0.0091 & 0.0063 & 0.0160 & 0.0160 & 0.0151 & 0.0134 & \\
CI Breach \% & 11.0632 & 10.7759 & 10.7759 & 9.3391 & 0.5747 & 0.5747 & 0.5747 & 1.1494 & 7.0402 & 6.8966 & 7.1839 & 7.1839 & \\
\addlinespace

\multicolumn{14}{l}{\textbf{Only Current Surface}} \\
MAPE (\%) & 4.7346 & 4.6980 & 4.5431 & 4.0357 & 4.0696 & 3.9589 & 3.5182 & 2.5649 & 2.7921 & 2.7516 & 2.5801 & 2.2213 & 3.3368 \\
Std. of APE (\%) & 4.1227 & 4.0964 & 3.9714 & 3.5710 & 3.0145 & 2.9185 & 2.5431 & 1.8894 & 2.5664 & 2.5313 & 2.3833 & 2.0465 & \\
Mean CI Width & 0.0353 & 0.0352 & 0.0342 & 0.0316 & 0.0493 & 0.0481 & 0.0430 & 0.0303 & 0.0457 & 0.0456 & 0.0434 & 0.0379 & \\
Std. CI Width & 0.0133 & 0.0132 & 0.0128 & 0.0115 & 0.0101 & 0.0097 & 0.0084 & 0.0051 & 0.0134 & 0.0136 & 0.0132 & 0.0120 & \\
CI Breach \% & 11.3506 & 11.2069 & 10.9195 & 9.9138 & 0.4310 & 0.4310 & 0.5747 & 0.8621 & 10.2011 & 10.3448 & 9.9138 & 8.0460 & \\
\addlinespace

\multicolumn{14}{l}{\textbf{Only VIX}} \\
MAPE (\%) & 4.7890 & 4.7577 & 4.6080 & 4.0912 & 3.8500 & 3.7428 & 3.3755 & 2.5247 & 2.4665 & 2.4506 & 2.3615 & 2.1497 & 3.0941 \\
Std. of APE (\%) & 4.0711 & 4.0407 & 3.9187 & 3.5230 & 2.9440 & 2.8544 & 2.5369 & 1.8870 & 2.2644 & 2.2423 & 2.1605 & 1.9569 & \\
Mean CI Width & 0.0368 & 0.0366 & 0.0357 & 0.0333 & 0.0520 & 0.0506 & 0.0454 & 0.0331 & 0.0492 & 0.0487 & 0.0460 & 0.0401 & \\
Std. CI Width & 0.0148 & 0.0146 & 0.0143 & 0.0129 & 0.0118 & 0.0113 & 0.0101 & 0.0062 & 0.0147 & 0.0145 & 0.0142 & 0.0128 & \\
CI Breach \% & 11.2069 & 10.9195 & 10.9195 & 9.1954 & 0.2874 & 0.1437 & 0.2874 & 0.5747 & 6.1782 & 6.0345 & 6.1782 & 6.8966 & \\
\addlinespace

\multicolumn{14}{l}{\textbf{No Scalar}} \\
MAPE (\%) & 4.7770 & 4.7544 & 4.6032 & 4.1150 & 3.7556 & 3.6638 & 3.2999 & 2.4601 & 2.4890 & 2.4731 & 2.3994 & 2.1805 & 3.2149 \\
Std. of APE (\%) & 4.0132 & 3.9842 & 3.8588 & 3.4649 & 2.9133 & 2.8277 & 2.4925 & 1.8824 & 2.2744 & 2.2515 & 2.1505 & 1.9160 &  \\
Mean CI Width & 0.0374 & 0.0369 & 0.0360 & 0.0331 & 0.0453 & 0.0441 & 0.0393 & 0.0287 & 0.0449 & 0.0441 & 0.0420 & 0.0366 &  \\
Std. CI Width & 0.0168 & 0.0168 & 0.0161 & 0.0142 & 0.0095 & 0.0092 & 0.0081 & 0.0059 & 0.0153 & 0.0154 & 0.0147 & 0.0133 &  \\
CI Breach \% & 10.9195 & 10.9195 & 11.4943 & 10.2011 & 0.5747 & 0.7184 & 1.0057 & 1.1494 & 7.7586 & 8.0460 & 8.0460 & 7.4713 &  \\
\addlinespace

\multicolumn{14}{l}{\textbf{No Surface Cond}} \\
MAPE (\%) & 11.6843 & 11.6343 & 11.4592 & 10.8039 & 9.8532 & 9.8578 & 9.8315 & 9.7726 & 14.0160 & 13.9794 & 13.8045 & 13.3106 & 11.4483 \\
Std. of APE (\%) & 9.1419 & 9.1029 & 8.9646 & 8.4760 & 6.9216 & 6.9170 & 6.9375 & 6.9825 & 8.5267 & 8.5138 & 8.4653 & 8.3390 &  \\
Mean CI Width & 0.1020 & 0.1018 & 0.1015 & 0.1006 & 0.1066 & 0.1053 & 0.1007 & 0.0896 & 0.1417 & 0.1413 & 0.1395 & 0.1356 &  \\
Std. CI Width & 0.0300 & 0.0296 & 0.0294 & 0.0274 & 0.0150 & 0.0149 & 0.0141 & 0.0138 & 0.0308 & 0.0309 & 0.0308 & 0.0301 &  \\
CI Breach \% & 7.7586 & 8.1897 & 7.9023 & 7.6149 & 1.1494 & 1.0057 & 1.2931 & 1.5805 & 38.0747 & 38.0747 & 36.7816 & 33.3333 &  \\
\addlinespace

\multicolumn{14}{l}{\textbf{Original}} \\
MAPE (\%) & 4.6907 & 4.6884 & 4.5146 & 4.0202 & 3.5264 & 3.4573 & 3.1542 & 2.4555 & 2.5129 & 2.4992 & 2.4184 & 2.2024 & 3.0026 \\
Std. of APE (\%) & 4.0338 & 4.0091 & 3.8795 & 3.4909 & 2.6588 & 2.5785 & 2.3190 & 1.8739 & 2.3984 & 2.3844 & 2.2909 & 2.0393 &  \\
Mean CI Width & 0.0372 & 0.0369 & 0.0357 & 0.0327 & 0.0468 & 0.0457 & 0.0412 & 0.0311 & 0.0472 & 0.0463 & 0.0438 & 0.0384 &  \\
Std. CI Width & 0.0160 & 0.0159 & 0.0152 & 0.0134 & 0.0098 & 0.0095 & 0.0082 & 0.0057 & 0.0139 & 0.0137 & 0.0132 & 0.0124 & \\
CI Breach \% & 10.9195 & 10.4885 & 11.0632 & 10.0575 & 0.5747 & 0.7184 & 0.5747 & 1.0057 & 7.9023 & 7.9023 & 8.4770 & 7.9023 &  \\
\bottomrule

\end{tabular}
}
\end{table}


\end{document}